\documentclass[useAMS,usenatbib]{mn2e}
\usepackage{graphicx}
\usepackage{deluxetable}
\usepackage{cleveref}
\usepackage{subfig}
\usepackage{longtable}
\usepackage{csquotes}
\usepackage{lscape}
\usepackage{hhline}
 \usepackage{epsfig}
\usepackage{color}

\newcommand{\degree}{\mbox{$^{\circ}$}}
\crefname{section}{Section}{Sections}

\newcommand{\x}{\mbox{${}\times{}$}}

\newcommand{\msun}{\mbox{M$_\odot$}}
\newcommand{\rsun}{\mbox{R$_\odot$}}

\title{
Ground-based near-UV observations of 15 transiting exoplanets: Constraints on their atmospheres and no evidence for asymmetrical transits
}
\author[Turner et al.]
  {Jake D. Turner$^{1,2}$,
  Kyle A. Pearson$^3$,
  Lauren I. Biddle$^3$,
  Brianna M. Smart$^{3,4}$,
      \newauthor 
  Robert T. Zellem$^1$, 
  Johanna K. Teske$^{3,5}$,
  Kevin K. Hardegree-Ullman$^{3,6}$,
      \newauthor 
             Caitlin C. Griffith$^1$,
             Robin M. Leiter$^{2}$,
             Ian T. Cates$^3$,
    Megan N. Nieberding$^3$,
    \newauthor 
    Carter-Thaxton W. Smith$^3$,
                Robert M. Thompson$^3$,  
   Ryan Hofmann$^3$,
 \newauthor 
             Michael P. Berube$^3$, 
             Chi H. Nguyen$^3$,
                       Lindsay C. Small$^3$,  
             Blythe C. Guvenen$^7$,
\newauthor 
             Logan Richardson$^8$,
             Allison McGraw$^3$,
                  Brandon Raphael$^3$,       
    Benjamin E. Crawford$^3$,
  \newauthor 
              Amy N. Robertson$^3$,
             Ryan Tombleson$^3$,
                    Timothy M. Carleton$^9$,
   Allison P.M. Towner$^3$,
    \newauthor 
             Amanda M. Walker-LaFollette$^3$,
             Jeffrey R. Hume$^3$,
             Zachary T. Watson$^3$,
           \newauthor 
             Christen K. Jones$^3$,
             Matthew J. Lichtenberger$^3$, 
               Shelby R. Hoglund$^{3}$,
Kendall L. Cook$^{3}$,
\newauthor 
             Cory A. Crossen$^3$, 
             Curtis R. Jorgensen$^3$,
             James M. Romine$^3$
Alejandro R. Thompson$^3$,
 \newauthor 
              Christian F. Villegas$^3$,
              Ashley A. Wilson$^{3}$,
             Brent Sanford$^3$, 
             Joanna M. Taylor$^{10}$,
 \newauthor
Triana N. Henz$^3$\\
   $^1$Lunar and Planetary Laboratory, University of Arizona, Tucson, AZ, 85721, USA	\\
  $^2$Department of Astronomy, University of Virginia, Charlottesville, VA 22904, USA \\
  $^3$Steward Observatory, University of Arizona, Tucson, AZ, 85721, USA  \\
  $^4$Department of Astronomy, University of Wisconsin, Madison, WI 53706, USA \\
 $^5$Currently an Origins Fellow at Carnegie DTM/OCIW\\
  $^6$Department of Physics and Astronomy, University of Toledo, Toledo, OH, 43606, USA \\
 $^7$Kitt Peak National Observatory, National Optical Astronomy Observatory, Tucson, AZ 85719, USA \\
   $^8$Institut f{\"u}r Quantenoptik, Leibniz Universit{\"a}t Hannover, D-30167 Hannover, Germany \\
  $^9$Department of Physics and Astronomy, University of California, Irvine, Irvine, CA, 92687, USA  \\
  $^{10}$Space Telescope Science Institute, Baltimore, MD 21218, USA }
\vspace{-3.5em}
\date{Submitted 2016 March 4}
\pagerange{\pageref{firstpage}--\pageref{lastpage}} \pubyear{2016}

\def\LaTeX{L\kern-.36em\raise.3ex\hbox{a}\kern-.15em
    T\kern-.1667em\lower.7ex\hbox{E}\kern-.125emX}

\begin{document}
\label{firstpage}
\clearpage\thispagestyle{empty}
\maketitle
\maketitle

\begin{abstract}
Transits of exoplanets observed in the near-UV have been used to study the scattering properties of their atmospheres and possible star-planet interactions. We observed the primary transits of 15 exoplanets (CoRoT-1b, GJ436b, HAT-P-1b, HAT-P-13b, HAT-P-16b, HAT-P-22b, TrES-2b, TrES-4b, WASP-1b, WASP-12b, WASP-33b, WASP-36b, WASP-44b, WASP-48b, and WASP-77Ab) in the near-UV and several optical photometric bands to update their planetary parameters, ephemerides, search for a wavelength dependence in their transit depths to constrain their atmospheres, and determine if asymmetries are visible in their light curves. Here we present the first ground-based near-UV light curves for 12 of the targets (CoRoT-1b, GJ436b, HAT-P-1b, HAT-P-13b, HAT-P-22b, TrES-2b, TrES-4b, WASP-1b, WASP-33b, WASP-36b, WASP-48b, and WASP-77Ab). We find that none of the near-UV transits exhibit any non-spherical asymmetries, this result is consistent with recent theoretical predictions by Ben-Jaffel et al. and Turner et al. The multi-wavelength photometry indicates a constant transit depth from near-UV to optical wavelengths in 10 targets (suggestive of clouds), and a varying transit depth with wavelength in 5 targets (hinting at Rayleigh or aerosol scattering in their atmospheres). We also present the first detection of a smaller near-UV transit depth than that measured in the optical in WASP-1b and a possible opacity source that can cause such radius variations is currently unknown. WASP-36b also exhibits a smaller near-UV transit depth at 2.6$\sigma$. Further observations are encouraged to confirm the transit depth variations seen in this study.  
\end{abstract}

\begin{keywords}
 planets and satellites: atmospheres -- techniques:
photometric -- planet-star interactions -- planets and satellites: individual: CoRoT-1b, GJ436b, HAT-P-1b, HAT-P-13b, HAT-P-16b, HAT-P-22b, TrES-2b, TrES-4b, WASP-1b, WASP-12b, WASP-33b, WASP-36b, WASP-44b, WASP-48b, WASP-77Ab
\end{keywords}
\section{Introduction} \label{intro}

Near-ultraviolet (near-UV) transits of short period exoplanets are a great tool to study star-planet interactions (e.g. tidal, gravitational, magnetic) and the scattering properties of their atmospheres (e.g. \citealt{Fossati2015}). The atmospheres of hot Jovian exoplanets in the near-UV (300 -- 450 nm) can be dominated by Rayleigh scattering, other forms of scattering or absorption, or clouds/hazes (\citealt{Seager2000}; \citealt{Brown2001}; \citealt{Benneke2013}; \citealt{Benneke2012}; \citealt{Griffith2014}). Clouds reduce the strength of spectral features thus causing the transit depth from near-UV to optical to be constant (\citealt{Seager2000}; \citealt{Brown2001}; \citealt{Gibson2013b}; \citealt{Kreidberg2014}; \citealt{Knutson2014}), and the Rayleigh scattering signature causes the transit depth to increase in the near-UV (\citealt{Etangs2008}; \citealt{Tinetti2010}; \citealt{deWit2013}; \citealt{Griffith2014}). Additionally, near-UV transits may exhibit asymmetries in their light curves such as ingress/egress timing differences, asymmetric transit shapes, longer durations, or significantly deeper transit depths ($>\sim1\%$) than the optical (e.g. \citealt{Madjar2003}; \citealt{Fossati2010b}; \citealt{Ehrenreich2012}; \citealt{Kulow2014}). The physical interpretations of these abnormalities vary, and include bow shocks, tidal interactions, star-planet magnetic interactions, a plasma torus originating from an active satellite, or escaping planetary atmospheres (e.g. \citealt{Vidotto2010}; \citealt{Lai2010}; \citealt{BenJaffel2014}; \citealt{Matsakos2015}).

There are 19 exoplanets with ground- or space-based observations in the UV (100--450 ${\rm nm}$). These observations can be subdivided into two groups: asymmetric and symmetric light curves. There are 5 exoplanets (55 Cnc b, GJ 436b, HD 189733b, HD 209458b, WASP-12b) where asymmetries in their light curves are observed (\citealt{Ehrenreich2012}; \citealt{Kulow2014}; \citealt{Ehrenreich2015};\citealt{Jaffel2013}; \citealt{Madjar2003}; \citealt{Jaffel2007}; \citealt{Jaffel2008}, \citealt{Madjar2004}; \citealt{Madjar2008}; \citealt{Madjar2013}; \citealt{Jaffel2013}; \citealt{Fossati2010b}; \citealt{Haswell2012}; \citealt{Nichols2015}). For the symmetric transits, 9 hot Jupiters (HAT-P-1b, HAT-P-12b, HAT-P-16b, TrES-3b, WASP-12b, WASP-17b, WASP-19b, WASP-43b, WASP-39b) are observed to have a constant planetary radii from near-UV to optical wavelengths (\citealt{Turner2013a}; \citealt{Copperwheat2013}; \citealt{Pearson2014}; \citealt{Bento2014}; \citealt{Nikolov2014}; \citealt{Mallonn2015}; \citealt{Ricci2015}; \citealt{Sing2016}). Additionally, 11 exoplanets with symmetric light curves (GJ 3470b, HD 189733b, HD 209458b, HAT-P-5b, HATP-12b, WASP-6b, WASP-12b, WASP-17b, WASP-31b, WASP-39b, XO-2b) are observed to have a larger near-UV radii than optical wavelengths (\citealt{Sing2008}, \citealt{Etangs2008}; \citealt{Sing2011}; \citealt{Southworth2012b}, \citealt{Sing2013}; \citealt{Nascimbeni2013}; \citealt{Sing2015}; \citealt{Zellem2015}; \citealt{Sing2016}). There seems to be a wavelength distinction between asymmetric and symmetric lights where asymmetric transits are only observed below 300 ${\rm nm}$. However, recent observations of asymmetric transits at optical wavelengths (\citealt{Rappaport2012}; \citealt{Rappaport2014}; \citealt{Werkhoven2014}; \citealt{Cabrera2015}; \citealt{Cauley2015}) hint that this dichotomy might not be the case.  

In this study, we investigate whether ground-based near-UV observations exhibit asymmetries. Most notably, it was predicted that a transiting exoplanet can potentially show an earlier transit ingress in the UV than in the optical, while the transit egress times will be unaffected due to the early absorption of star light due to a bow shock (\citealt{Vidotto2010,Vidotto2011a,Vidotto2011b}; \citealt{Vidotto2011c}; \citealt{Llama2011,llama2013}). Additionally, the near-UV transit will have a greater drop in flux than the optical transit and will no longer be symmetric about the mid-transit time (\citealt{Vidotto2011b}; \citealt{Llama2011,llama2013}). This effect is explained by the presence of a bow shock on the leading edge of the planet formed by interactions between the planet's magnetosphere and the stellar coronal plasma. If the shocked material in the magnetosheath becomes sufficiently opaque, it will absorb starlight and cause an early ingress in the near-UV light curve \citep[see fig. 6]{Vidotto2011b}. \citet[hereafter VJH11a]{Vidotto2011a} predict that near-UV ingress asymmetries should be common in transiting exoplanets and tabulated a list of the 69 targets that should exhibit this effect.

Is it possible to observe near-UV asymmetries from the ground? Previous observations of an early ingress on WASP-12b and HD 189733b observe a flux drop difference of about $\sim 1 \%$ and a timing difference of $\sim > $30 minutes between the near-UV and optical light curves (\citealt{Fossati2010b}; \citealt{Haswell2012}; \citealt{Jaffel2013}; \citealt{Nichols2015}). Both these properties are well within reach for ground-based meter-sized telescopes (e.g. \citealt{Copperwheat2013}; \citealt{Turner2013a}; \citealt{Pearson2014}), like the Steward Observatory 1.55-m Kuiper Telescope used for the near-UV observations in this study. Additionally, \citealt{Nichols2015} find that summing over the entire NUV band (253.9--281.1 $\rm{nm}$) on \textit{Hubble Space Telescope} (\textit{HST}) still resulted in an early ingress, which they attributed to a blend of thousands of lines of metals (e.g. Mg, Na, Fe, Al, Co, Al Mn). Therefore, ground-based broadband near-UV observations (303--417 nm) might also experience an early near-UV ingress by the blending of lots of lines from the same metal species that exist at \textit{HST} wavelengths (e.g. Na I/II, Ca II/III, Na I, Mg I, Al I, Mn I/II, Fe I/II, Co I/II; \citealt{Morton1991,Morton2000}; \citealt{Sansonetti2005}). Therefore, it is be feasible to observe near-UV asymmetries from the ground by taking all the factors discussed above into consideration.

However, recent studies by \citealt{BenJaffel2014} and \citealt{Turner2016} cast doubt on observing asymmetries in all ground- and space-based UV wavelengths using the VJH11a bow shock model. \citet{BenJaffel2014} use simple recombination and ionization equilibrium calculations for realistic parameters of the stellar corona ($T\sim10^{6} \rm{K}$; \citealt{Aschwanden2005}) to determine that only highly ionized stages of heavy elements can cause any detectable optical depth. Furthermore, \citealt{Turner2016} use the plasma photoionization and microphysics code \texttt{CLOUDY} (\citealt{Ferland1998}; \citealt{Ferland2013}) to investigate all opacity sources at UV and optical wavelengths that could cause an early ingress due to the presence of a bow shock compressing the coronal plasma. \citealt{Turner2016} also find that the optical depths in the compressed stellar wind ($T\sim10^{6} \rm{K}$; \citealt{Aschwanden2005}, $n \sim 10^{4} \rm{cm^{-3}}$; \citealt{McKenzie1997}) are orders of magnitude too small ($> 3\times10^{-7}$) to cause an observable absorption in space- and ground-based UV and optical observations (even for stellar wind densities 10$^4$ times higher than what is expected).

\begin{table*}
\centering
\caption{Comparison of the planetary systems in this study$^{a}$}
\begin{tabular}{ccccccccccccc}
  \hline
  \hline
Planet              & M$_{p}$		&R$_{p}$ 		&a		& P$_{p}$ 	&Spec. 	 	&M$_{*}$ 		&R$_{*}$     	&$[Fe/H]$		& $\log{(R_{hk}^{'})}^{c}$ & $\lambda^{d}$\\
Name		&(M$_{Jup}$)	&(R$_{Jup}$) 	&(au) 	&(d)			& Type 	       & ($\msun$) 	& ($\rsun$) 	&				& 					&    ($\degree$) \\ 
  \hline
  \hline
  CoRoT-1b 	& 1.03 	& 1.49 	& 0.025 & 1.51 & G0V 	& 0.95 & 1.11	& -0.30	    	& -5.132	& 77$^{e}$ \\
 GJ346b 		& 0.072	& 0.38	& 0.029&  2.63 & M2.5V  &  0.45 & 0.46   & -0.32		&-5.298    & --       \\
  HAT-P-1b 	& 0.53 	& 1.24 	& 0.055 & 4.47 & G0V 	& 1.13 & 1.14	&0.13		&-4.984	& 3.7$^{f}$\\
  HAT-P-13b 	& 0.85 	& 1.28 	& 0.043 & 2.92 & G4		& 1.22 & 1.56	& 0.43		&-5.134	&1.9$^{h}$\\
  HAT-P-16b 	& 4.2 	& 1.29 	& 0.041 & 2.78 & F8 	& 1.22 & 1.24	&0.17		&-4.864	&-10$^{i}$	\\
  HAT-P-22b 	& 2.15 	& 1.08 	& 0.041 & 3.21 & G5 	& 0.92 & 1.04	& 0.22		&--		&--\\
  TrES-2b 		& 1.19 	& 1.22 	& 0.036 & 2.47 & G0V 	& 0.98 & 1.0	&-0.15		&-4.949	&-9$^{j}$\\
  TrES-4b 		& 0.91 	& 1.78 	& 0.051 & 3.55 & F 		& 1.39 & 1.82	&0.14		&-5.104	&6.3$^{k}$\\
  WASP-1b	&1.03	&1.49	&0.025 &	1.51	&F7V	&0.95 &	1.11	&-0.30		& -5.114	&-59$^{l}$\\
  WASP-12b 	& 1.35	& 1.79 	& 0.023 & 1.09 & G0 	& 1.28 & 1.63 	&0.30		&-5.500	&63$^{m}$\\
  WASP-33b 	& 1.76 	& 1.50	& 0.026 & 1.22 & A5 	& 1.5 & 1.44	& 0.1		&--		&251.6$^{n}$\\
  WASP-36b 	& 2.26 	& 1.27 	& 0.026 & 1.54 & G2		& 1.02 & 0.94	& -0.31		&--		&--\\
  WASP-44b 	& 0.89 	& 1.14 	& 0.035 & 2.42 & G8V 	& 0.95 & 0.93	& 0.06		&--		&--\\
  WASP-48b 	& 0.97 	& 1.67 	& 0.034 & 2.14 & G 		& 1.19 & 1.75 	& -0.12		&--		&--\\
  WASP-77Ab 	& 1.76 	& 1.21 	& 0.024 & 1.36 & G8V 	& 1.0 & 0.96	& 0.1		&--		&-- \\
  \hline
\end{tabular}
\vspace{-2em}
\tablenotetext{a}{Information about the systems is obtained from the Exoplanet Data Explorer at exoplanets.org \citep{Wright2011exo}}
\tablenotetext{d}{$\lambda$ is the angle between the sky projections of the planetary orbital axis and the stellar rotation axis}
\tablerefs{(c) \citet{Knutson2010}; (e) \citealt{Pont2010}; (f) \citealt{Johnson2008}; (g) \citealt{Hirano2011}; (h) \citealt{Winn2010}; (i) \citealt{Moutou2011}; (j) \citealt{Winn2008b}; (k) \citealt{Narita2010}; (l) \citealt{Albrecht2011}; (m) \citealt{Albrecht2012}; (n) \citealt{Cameron2010} }
\label{tb:compare}
\end{table*}
\begin{table*}
\centering
\caption{Journal of observations}
\begin{tabular}{cccccccc}
\hline 
\hline
Planet 		&Date             & Filter$^{1}$	& Cadence  		&OoT RMS$^{2}$ & Res RMS$^{3}$ 	& Seeing  	& $\chi_{r}^2$$^{a}$ \\
Name 		& (UT)             &			&(s)  			&(mmag) & (mmag) 	& ($\arcsec$)  \\
\hline
\hline 
CoRoT-1b		&2012 December 06	&      U		&    	70					& 3.59	&	3.95			& 1.46--2.95	&0.49	\\
GJ436b		&2012 March 23		&	U		&	60				& 2.96	&	2.85			&0.96--1.99	&0.68	\\
"			&2012 April  07 		&	U		&	61				&2.83	&	2.70			& 1.22--2.10 	&1.37\\
HAT-P-1b		&2012 October 02  		&      U		&	40				&1.44	&	1.45			&1.57--2.00	&1.69	\\
HAT-P-13b	&2013 March 02		&	 U		&	58				& 1.91	&	1.63			& 1.67--2.89	&1.76	\\
HAT-P-16b	& 2013 November 02	&      U		&	55				&2.50	&	2.50			&1.40--3.98	&1.23	\\	
HATP-22b		& 2013 February 22		& U			&	  70				&3.42	 & 	3.17			&1.41--4.12	&0.26	 \\
"			&2013 March 22		&	U		&   71				& 2.07	&  2.12			&1.34--2.26 	&1.16\\
TrES-2b		& 2012 October 29		&	U		&	50				&	3.05	&	2.54			&1.36--2.53	&1.27	\\
TrES-4b		&	 2011 July 26		&	U		&     	116				&	4.42	&	4.08			&1.29--3.05      &3.65		\\
"			&  2011 July 26			& 	R		&	116				&      5.54	&	3.93			&1.29--3.05	&2.09		\\
WASP-1b		&	2013 September 19	&      U 		&	133				&	2.92	&	3.31			&1.10--2.98	&1.56		\\
"			& 2013 September 19	&      B		&	135				&2.80	&	2.36			& 1.10--2.98	&3.61			\\
"			&	2013 October 22	&	U		&	137				&1.58	&	1.79			& 1.21--2.64	&1.06			\\
"			& 2013 October 22		&	B		&	136				&1.23	&	1.25			& 1.21--2.64	&1.60	\\
WASP-12b	&2011 November 15     	&      R      	&        126    	 		& 1.40  	&      1.47			&  1.72--2.10	&2.14	\\  
"			&2011 November 15    	&      U       	&       126    			& 1.67 	&  	1.62			& 1.72--2.10	&0.92	\\
"			& 2012 March 22    		&      U 		&    	   61    			&  2.54  	&	2.23  		& 1.33--2.15	&0.47	\\
	"		& 2012 October 02      	&      U 		&        61    			&   2.53 	&  	2.11			& 2.07--3.18	&0.79	\\
"			& 2012 November 30 	&	U		&	  55 				& 3.30	&	3.61			& 1.45--3.24	&0.94	\\
WASP-33b	& 	2012 October 01	&	U		&	27				&	2.57	&	2.60			&1.12--1.99	&1.54\\
"			&	2012 December 01  	&	U		&	91				&	2.45	&	2.63			&1.75--2.90	&8.60\\
"			&	2012 December 01	&	B		&	91				&	7.17	&	6.68			&1.75--2.90	&1.91	\\
WASP-36b	&	2012 December 29  & 	R		& 	31				&	1.90 &	2.50			& 1.92--2.80	&1.44\\
"			&	2013 March 16 		&	U		&	60				&	3.86	&	5.74			&1.46--2.96	&0.63\\
WASP-44b	&  2012 October 13		&	U		& 	68				& 6.22	& 5.64			&1.77--2.58	&1.08\\
"			&  2013 October 19		&	B		&	116				&2.33	&2.50			&1.07--1.95	&1.19\\
"			&  2013 October 19		&	V		&	120				&2.04	&2.25			&1.07--1.95	&1.85\\
WASP-48b	&  2012 October 09 		&	U		&     71				&1.92	&2.36			&1.54--3.11	&1.40\\
WASP-77Ab	&  2012 December 06	&      U		&     68				& 1.58	& 1.53			& 2.31--2.95    &2.87     \\    
\hline
\end{tabular}
\vspace{-2em}
\tablenotetext{1}{Filter: B is the Harris B (330--550 nm), R is the Harris R (550--900 nm), V is the Harris V (473--686 nm) and U is the Bessell U (303--417 nm) } 
\tablenotetext{2}{Out-of-Transit (OoT) root-mean-squared (RMS) relative flux} 
\tablenotetext{3}{Residual (res) RMS flux after subtracting the EXOplanet MOdeling Package (\texttt{EXOMOP}) best-fitting model from the data}
\tablenotetext{a}{Reduced $\chi^2$ calculated using the \texttt{EXOMOP} best-fitting model}
\label{tb:obs_new}
\end{table*}

The goals of this paper are to study the atmospheres of 15 transiting exoplanet targets and to determine whether ground-based near-UV transit observations are sensitive to light curve asymmetries. Our data can be used to confirm the predictions by \citealt{BenJaffel2014} and \citealt{Turner2016} that an early ingress should not be present in ground-based near-UV transits. Our sample is chosen to contain a wide variety of different system parameters to determine if any system parameters correlate with the existence of a bow shock (Table \ref{tb:compare}). We also perform follow-up ground-based near-UV observations of WASP-12b \citep{Copperwheat2013} and HAT-P-16b \citep{Pearson2014}. Using our data set, we update the planetary system parameters (Section \ref{sec:physical_properites}), present a new ephemeris to aid in future observations (Section \ref{sec:period}), and search for a wavelength dependence in the planetary radii that can be used to constrain their atmospheric compositions (Section \ref{sec:atmo}).


\section{Observations and Data Reduction} \label{sec:obs_redu}

All of our observations were conducted at the University of Arizona's Steward Observatory 1.55-m (61") Kuiper Telescope on Mt. Bigelow near Tucson, Arizona, using the Mont4k CCD. The Mont4k CCD contains a 4096$\x$4096 pixel sensor with a field of view of 9.7'$\x$9.7'. We used 3$\x$3 binning to achieve a resolution of 0.43$\arcsec$/pixel and shorten our read-out time to $\sim$10 s. Our observations were taken with the Bessell U (303--417 nm), Harris B (360--500 nm), Harris V (473--686 nm), and Harris R (550--900 nm) photometric band filters. Specifically, the Bessell U filter is a near-UV filter and has a transmission peak of 70 per cent near 370 nm. To ensure accurate timing in these observations, the clocks were synchronized with an NTP time server every few seconds. In all the data sets, the average shift in the centroid of our targets is less than 1 pixel (0.43$\arcsec$) due to excellent autoguiding (the max is 3 pixels), which minimizes our need to worry about intrapixel sensitivity. Seeing ranged from 0.86--4.12$\arcsec$ throughout our complete set of observations. A summary of all our observations is displayed in Table \ref{tb:obs_new}.

\begin{figure*}
\centering
\begin{tabular}{cc}
\vspace{0.5cm}
\epsfig{file=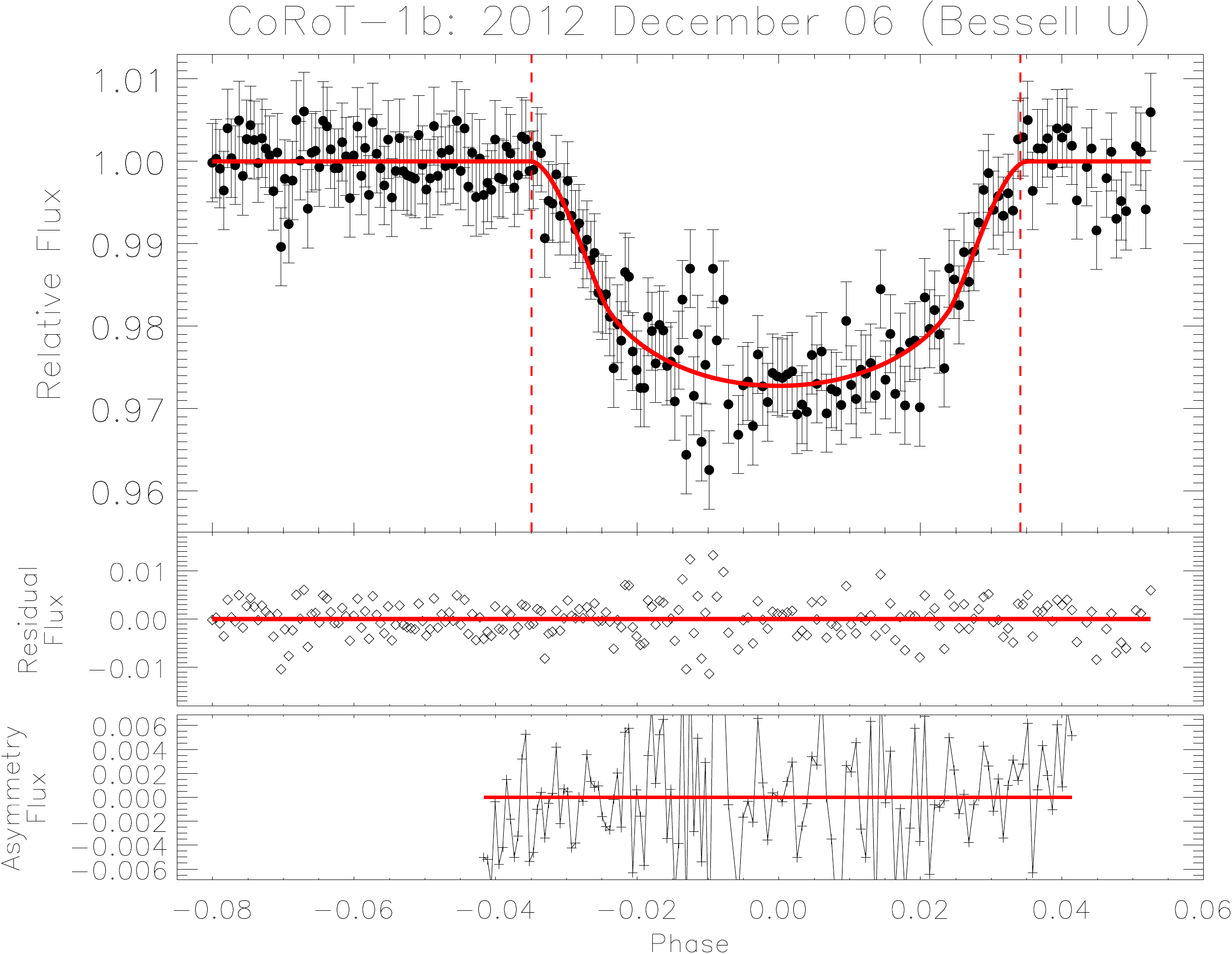,width=0.38\textwidth,angle=90} 	& \hspace{0.3cm} \epsfig{file=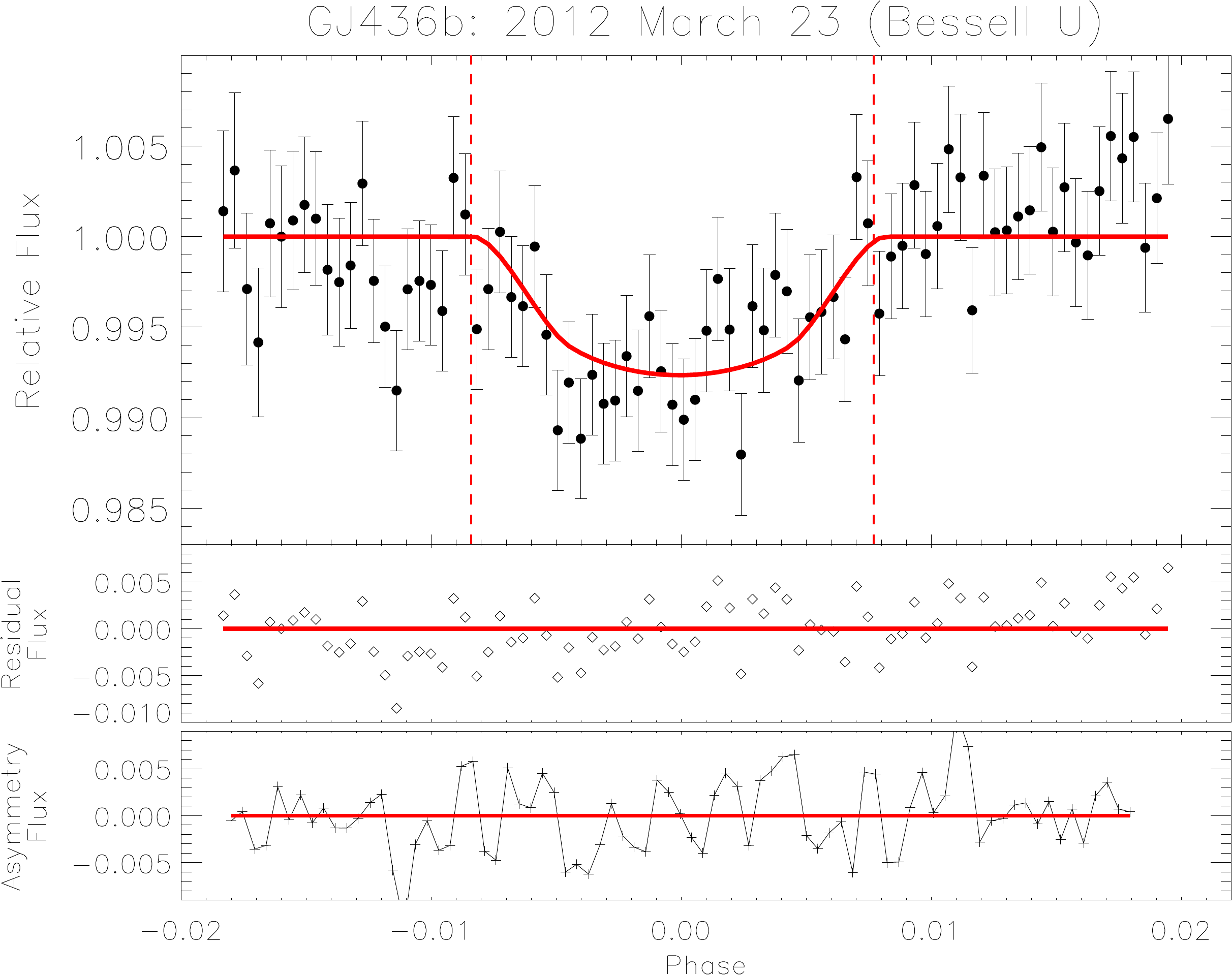,width=0.38\textwidth,angle=90} \\
\vspace{0.5cm}
\epsfig{file=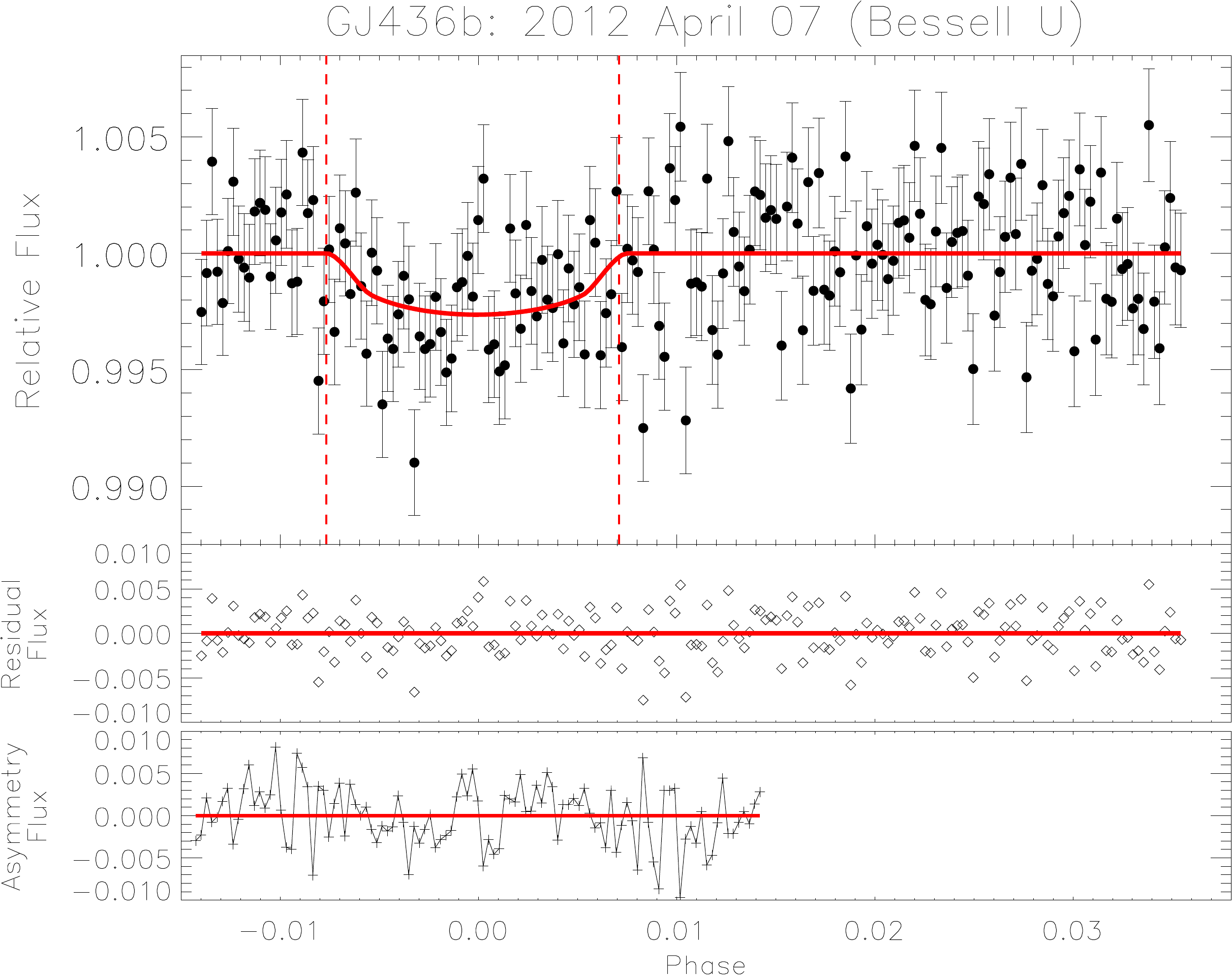,width=0.38\linewidth,angle=90} 		&\hspace{0.3cm} \epsfig{file=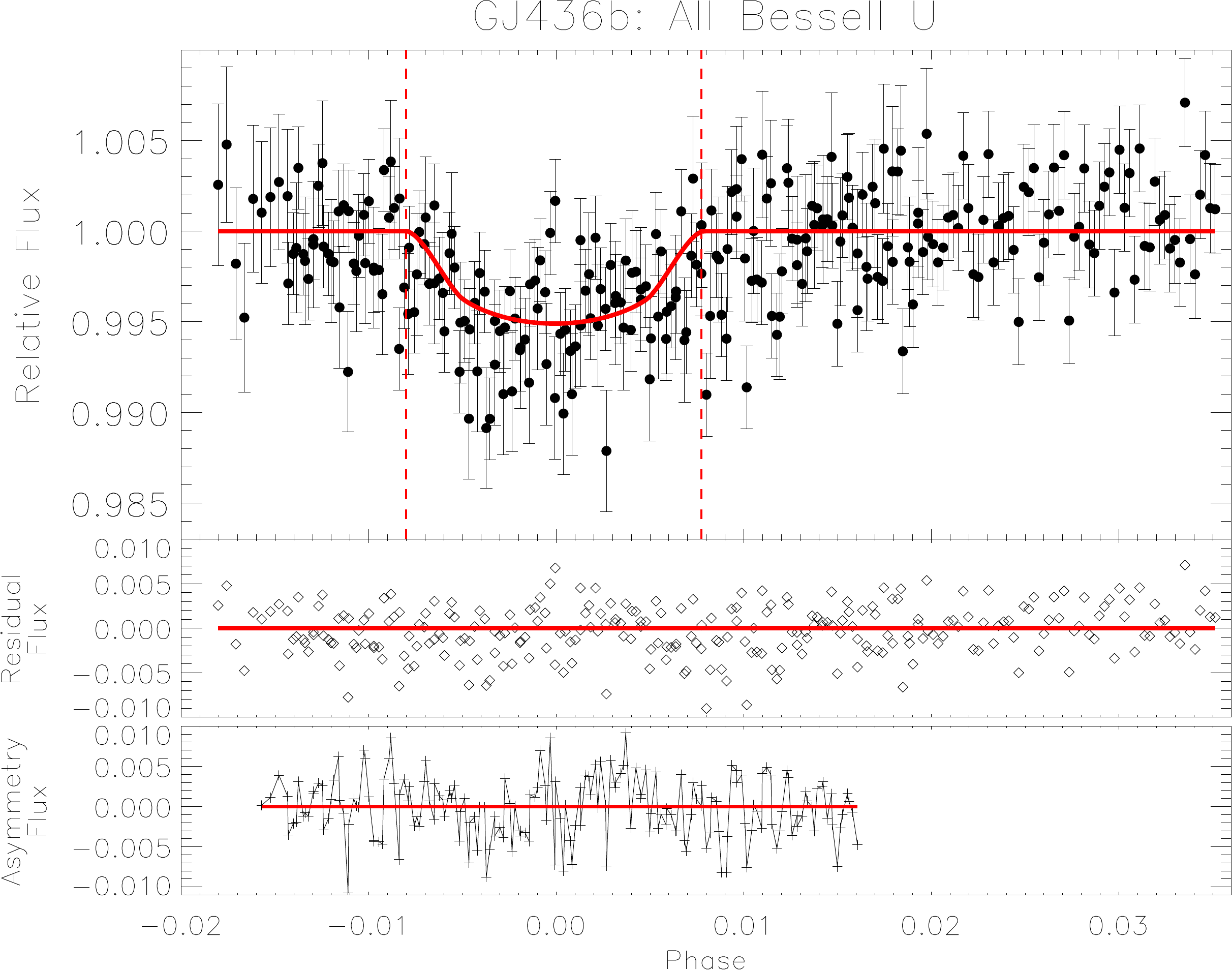,width=0.38\linewidth,angle=90} \\
\vspace{0.1cm}
\epsfig{file=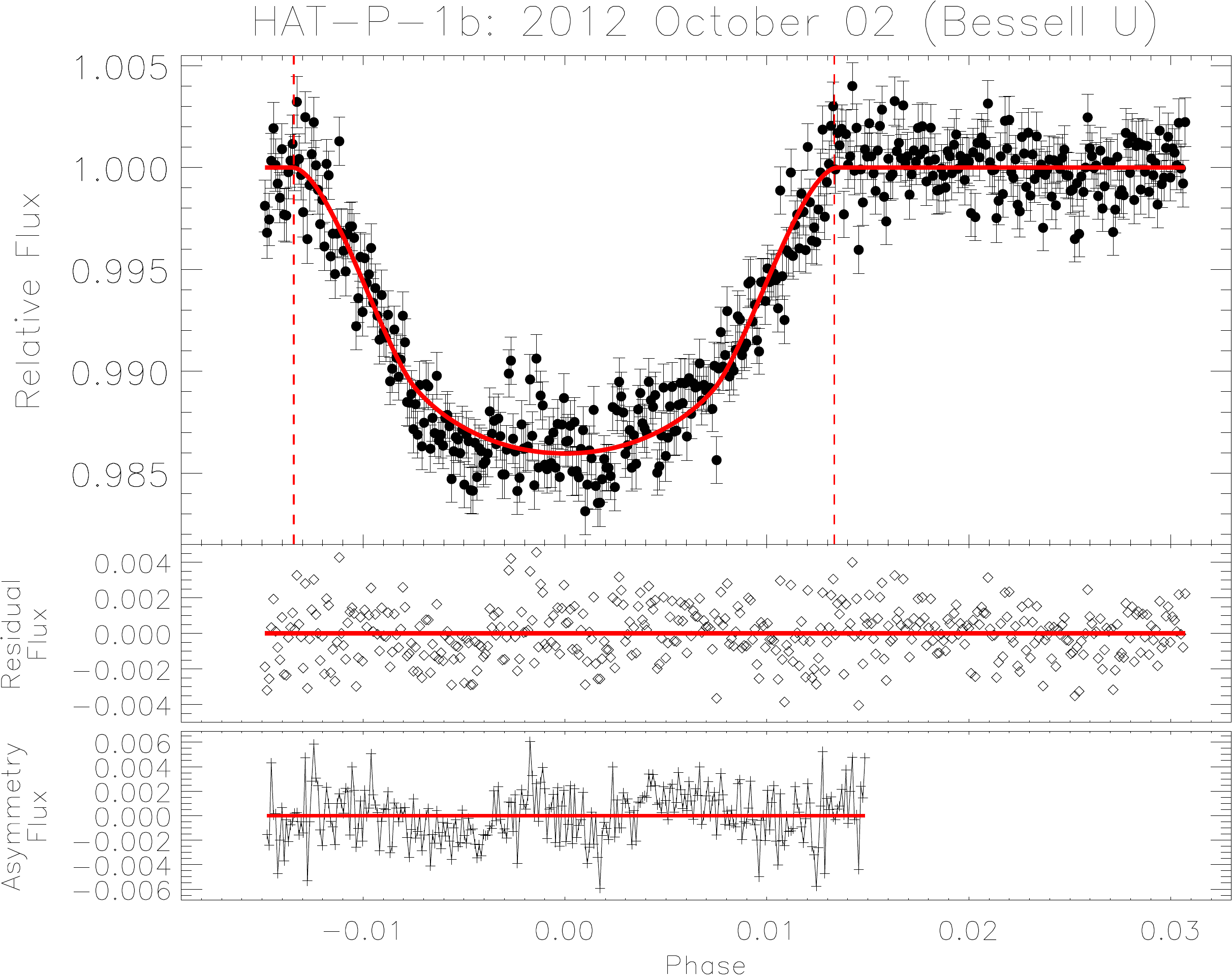,width=0.38\linewidth,angle=90}		& \hspace{0.3cm} \epsfig{file=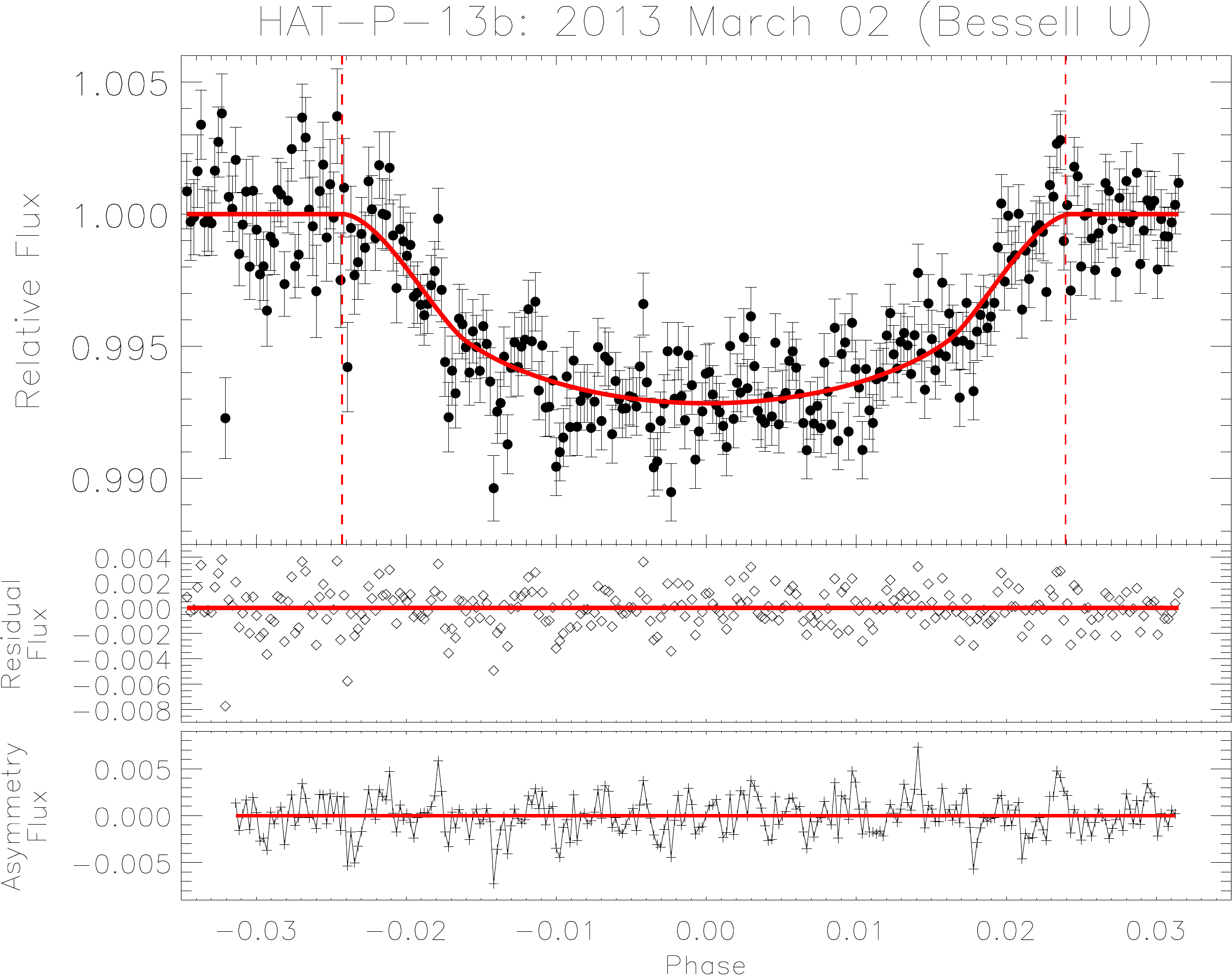,width=0.38\linewidth,angle=90}\\
\end{tabular}
\caption{Light curves of CoRoT-1b, GJ436b, HAT-P-1b, and HAT-P-13b. The 1$\sigma$ error bars include the readout noise, the Poisson noise, and the flat-fielding error. The best-fitting models obtained from the EXOplanet MOdeling Package (\texttt{EXOMOP}) are shown as a solid red line. The \texttt{EXOMOP} best-fitting model predicted ingress and egress points are shown as dashed red vertical lines. The residuals (Light Curve - \texttt{EXOMOP} Model) are shown in the second panel. The third panel shows the residuals of the transit subtracted by the mirror image of itself (Section \ref{sec:add}). See Table \ref{tb:obs_new} for the cadence, Out-of-Transit root-mean-squared (RMS) flux, and residual RMS flux for each light curve. We do not observe an early ingress or any non-spherical asymmetries in any of the near-UV transits. The data points for all the transits are available in electronic form (see Table \ref{tb:mr}).}
\label{fig:light_1}
\end{figure*}

\begin{figure*}
\centering
\begin{tabular}{cc}
\vspace{0.5cm}
\epsfig{file=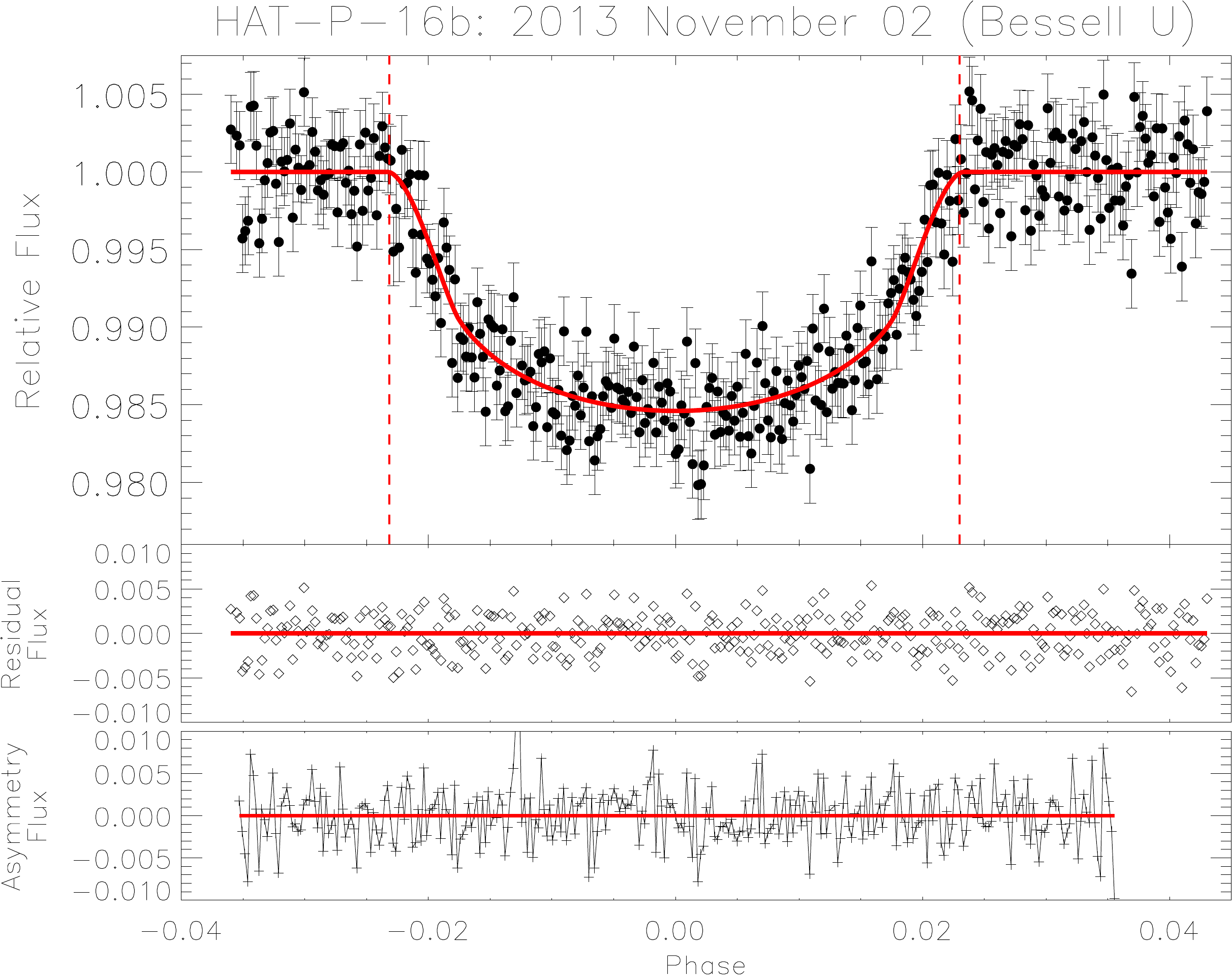,width=0.38\textwidth,angle=90} 	& \hspace{0.3cm} \epsfig{file=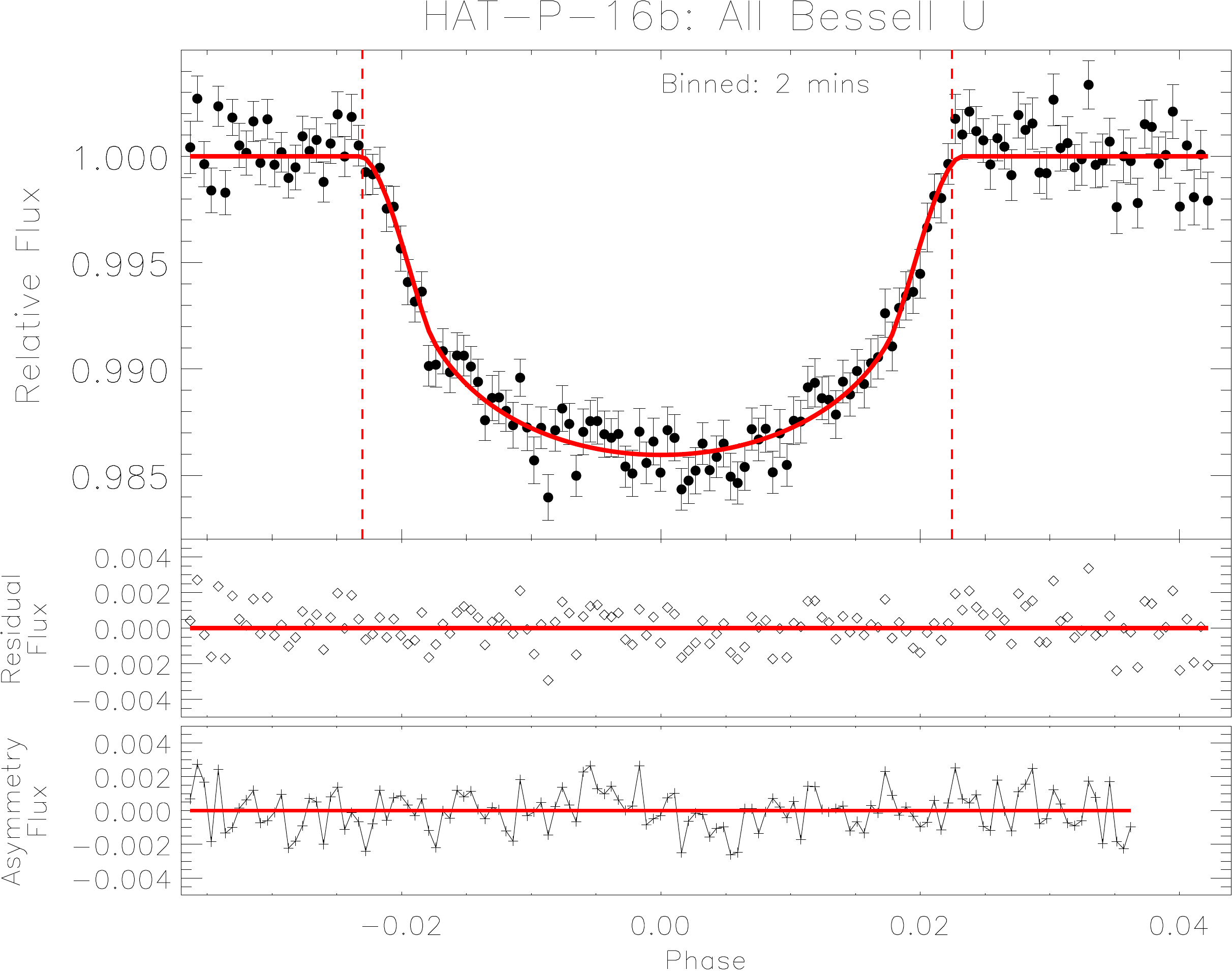,width=0.38\textwidth,angle=90} \\
\vspace{0.5cm}
\epsfig{file=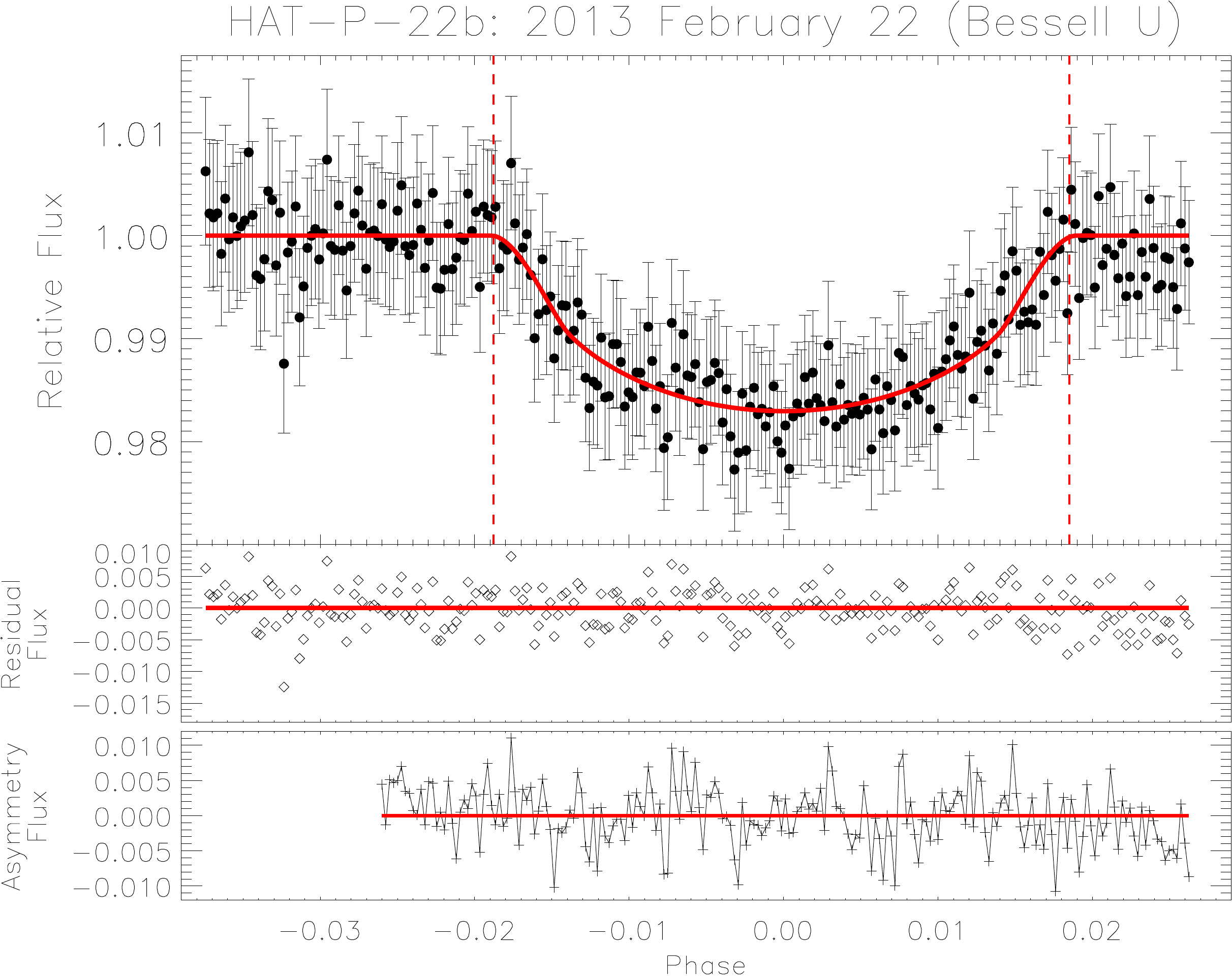,width=0.38\linewidth,angle=90} 	&\hspace{0.3cm} \epsfig{file=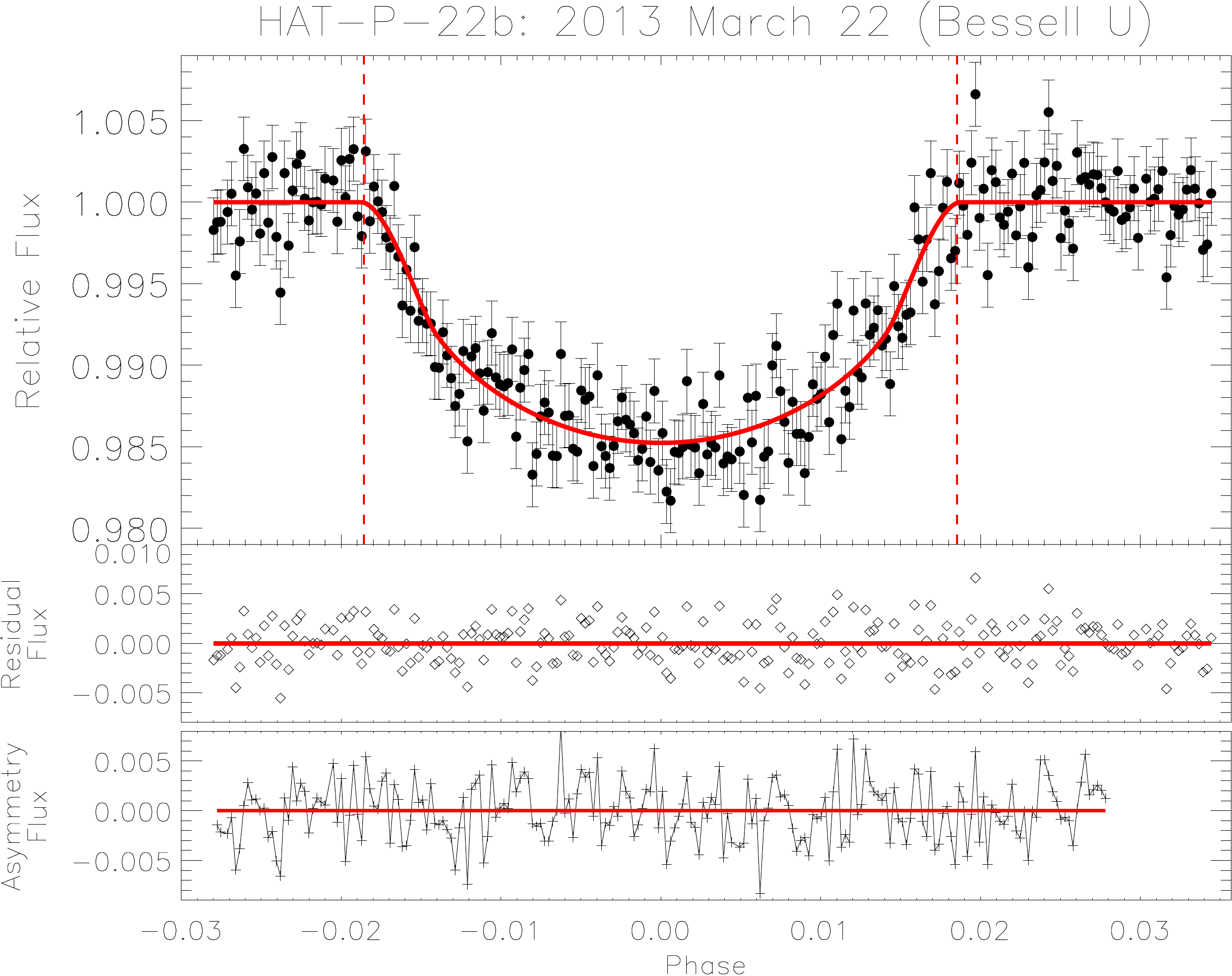,width=0.38\linewidth,angle=90} \\
\vspace{0.1cm}
\epsfig{file=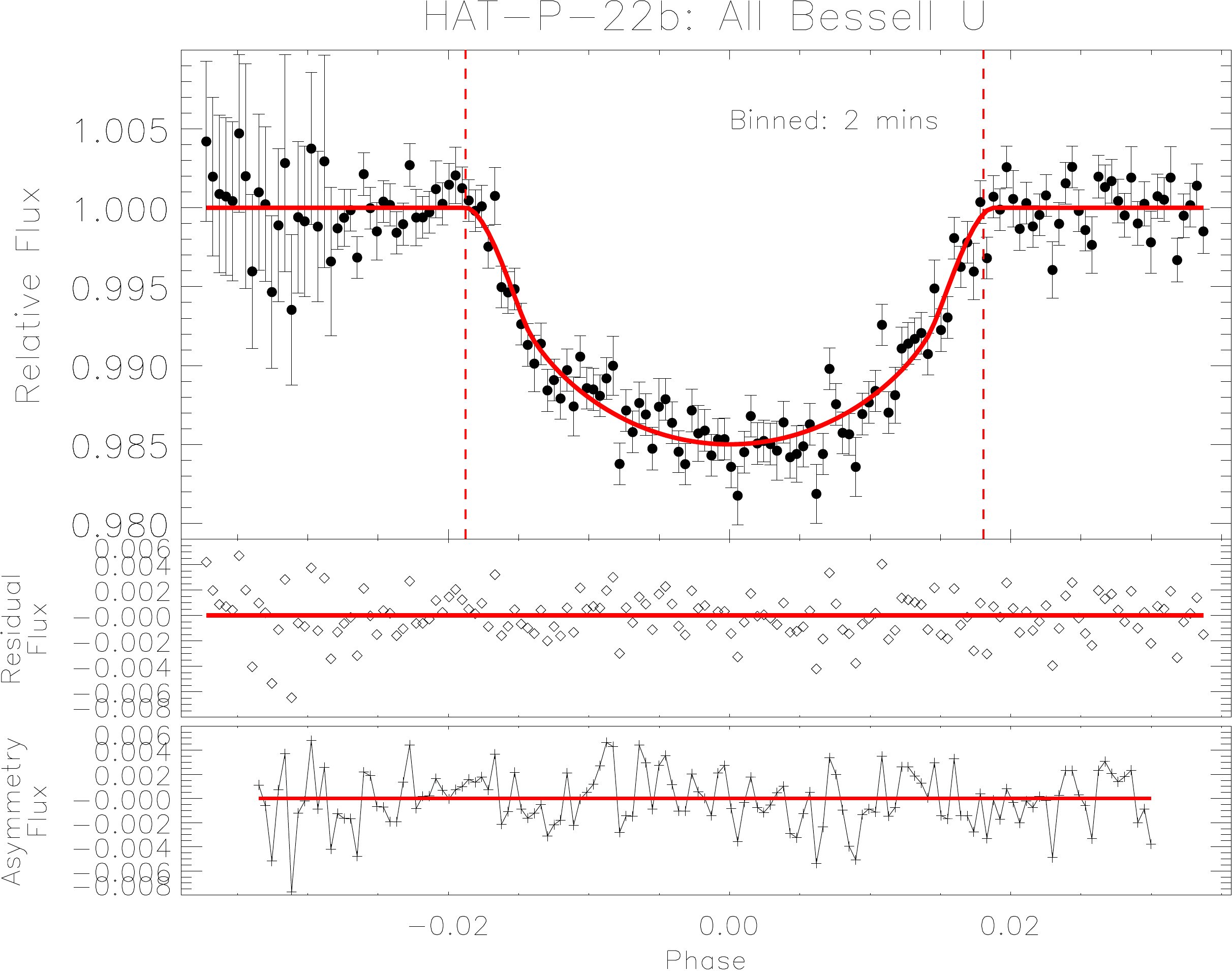	,width=0.38\linewidth,angle=90} 		&\hspace{0.3cm} \epsfig{file=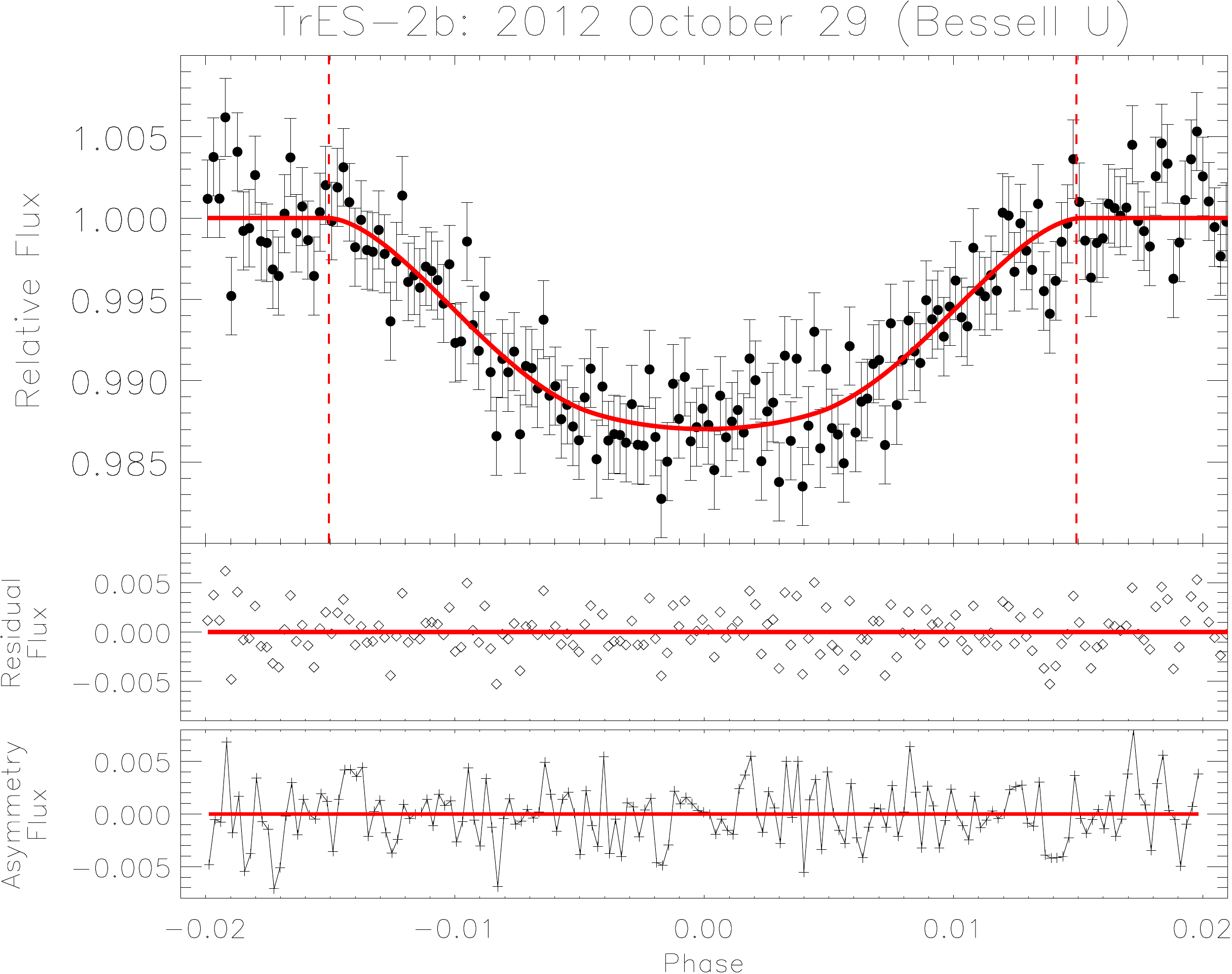,width=0.38\linewidth,angle=90} 
\end{tabular}
\caption{Light curves of HAT-P-16b, HAT-P22b, and TrES-2b. Other comments are the same as Fig. \ref{fig:light_1}.}
\label{fig:light_2}
\end{figure*}

\begin{figure*}
\centering
\begin{tabular}{cc}
\vspace{0.5cm}
\epsfig{file=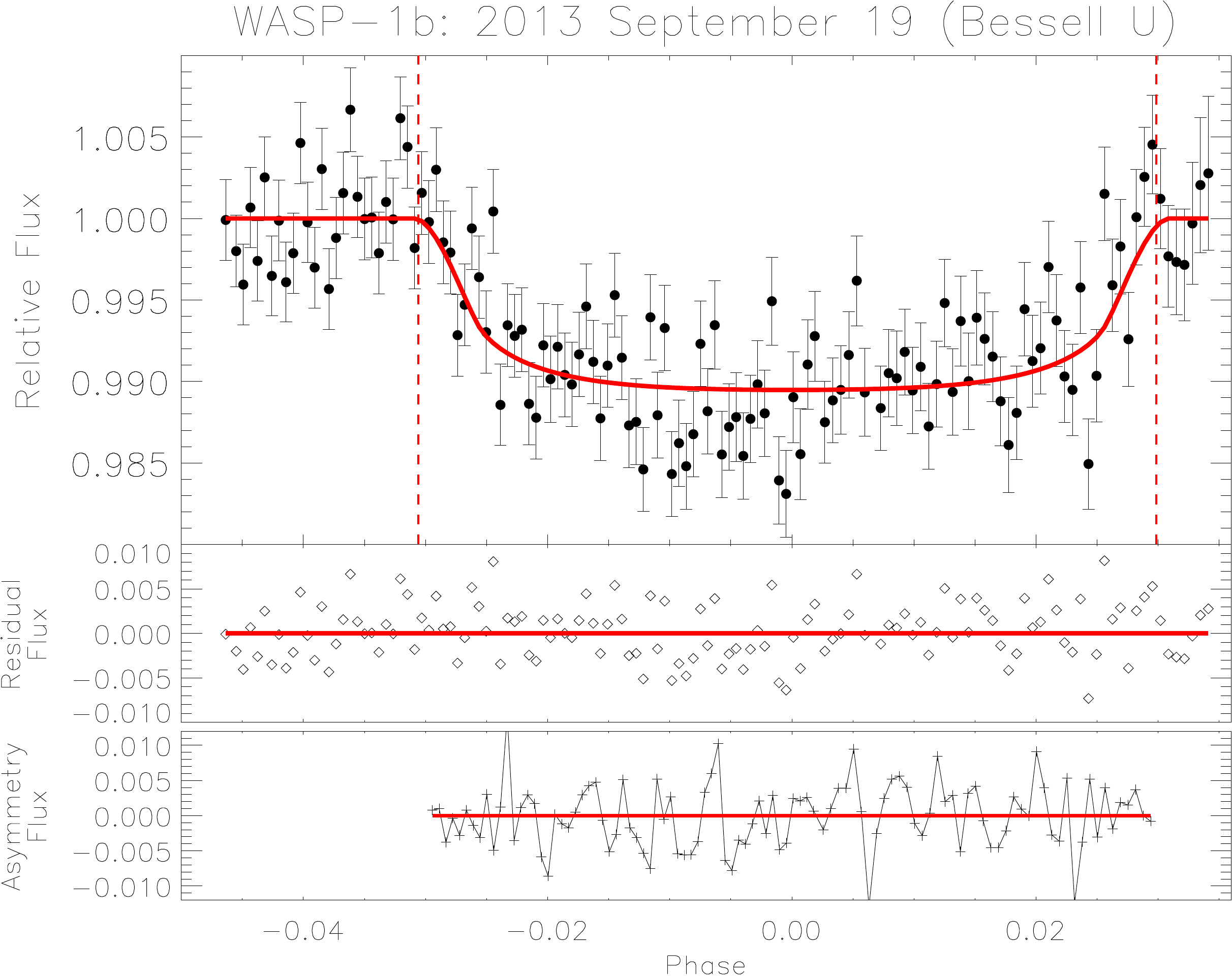,width=0.38\textwidth,angle=90} & \hspace{0.3cm}
\epsfig{file=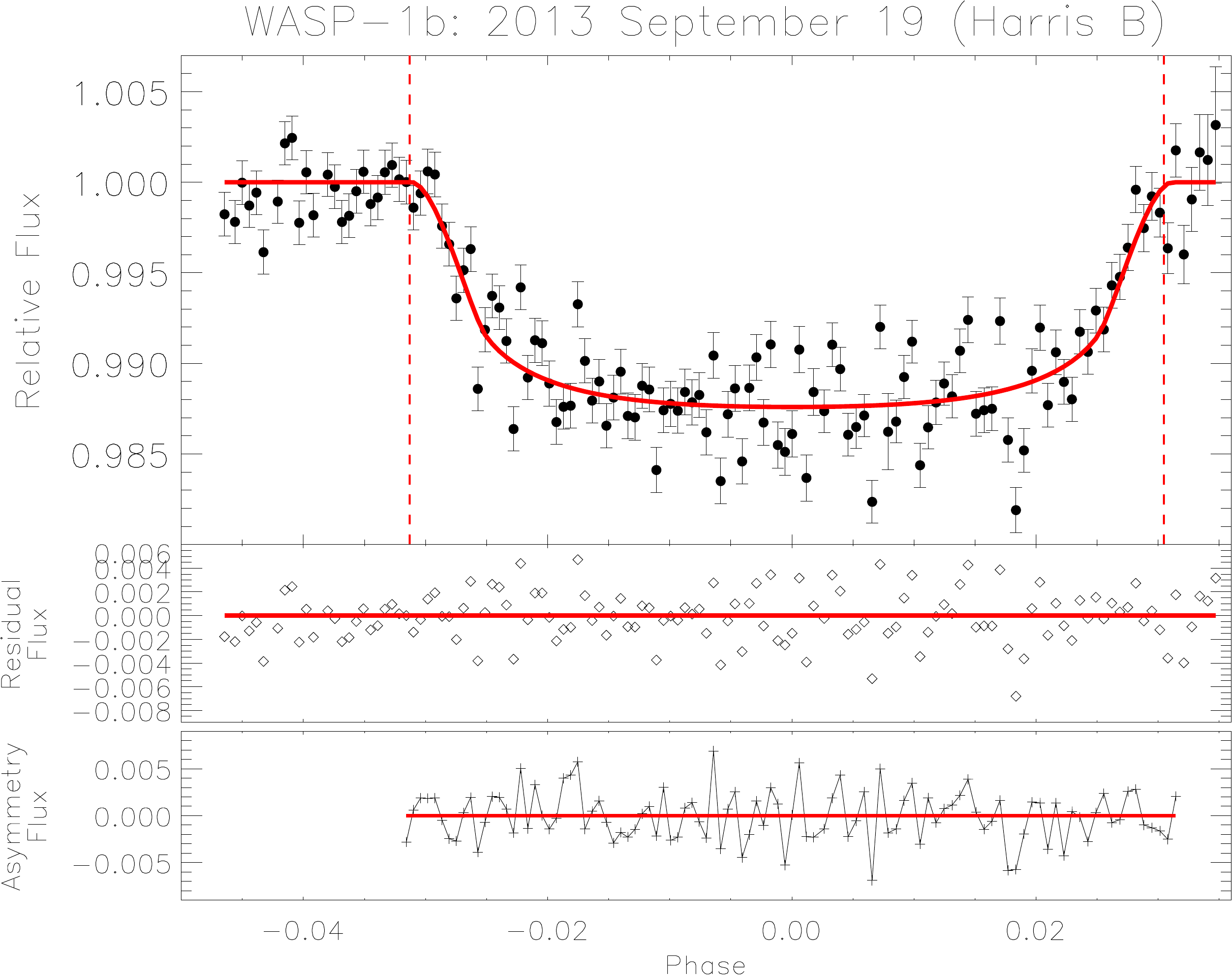,width=0.38\textwidth,angle=90} \\
\vspace{0.5cm}
\epsfig{file=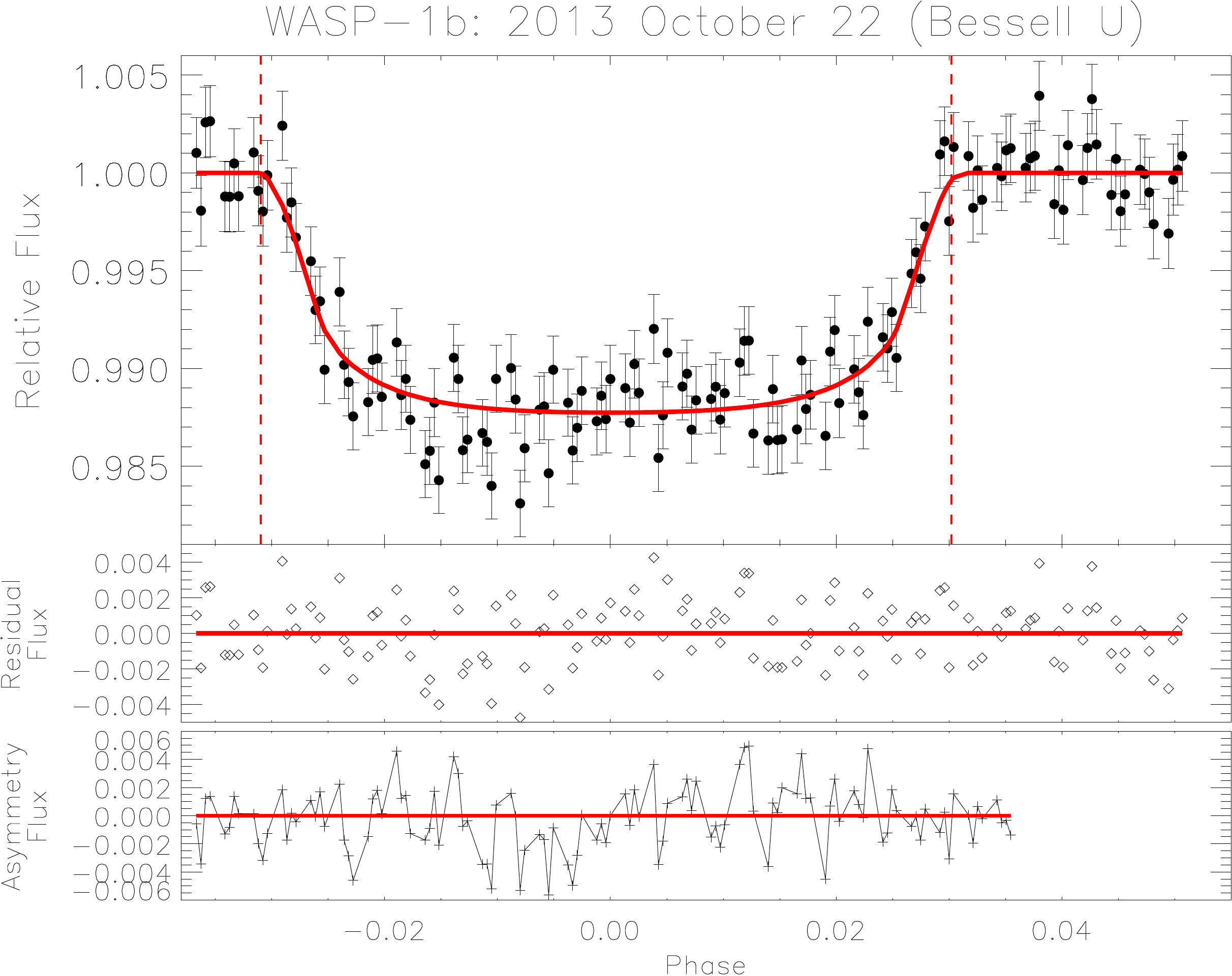,width=0.38\linewidth,angle=90} &\hspace{0.3cm}
\epsfig{file=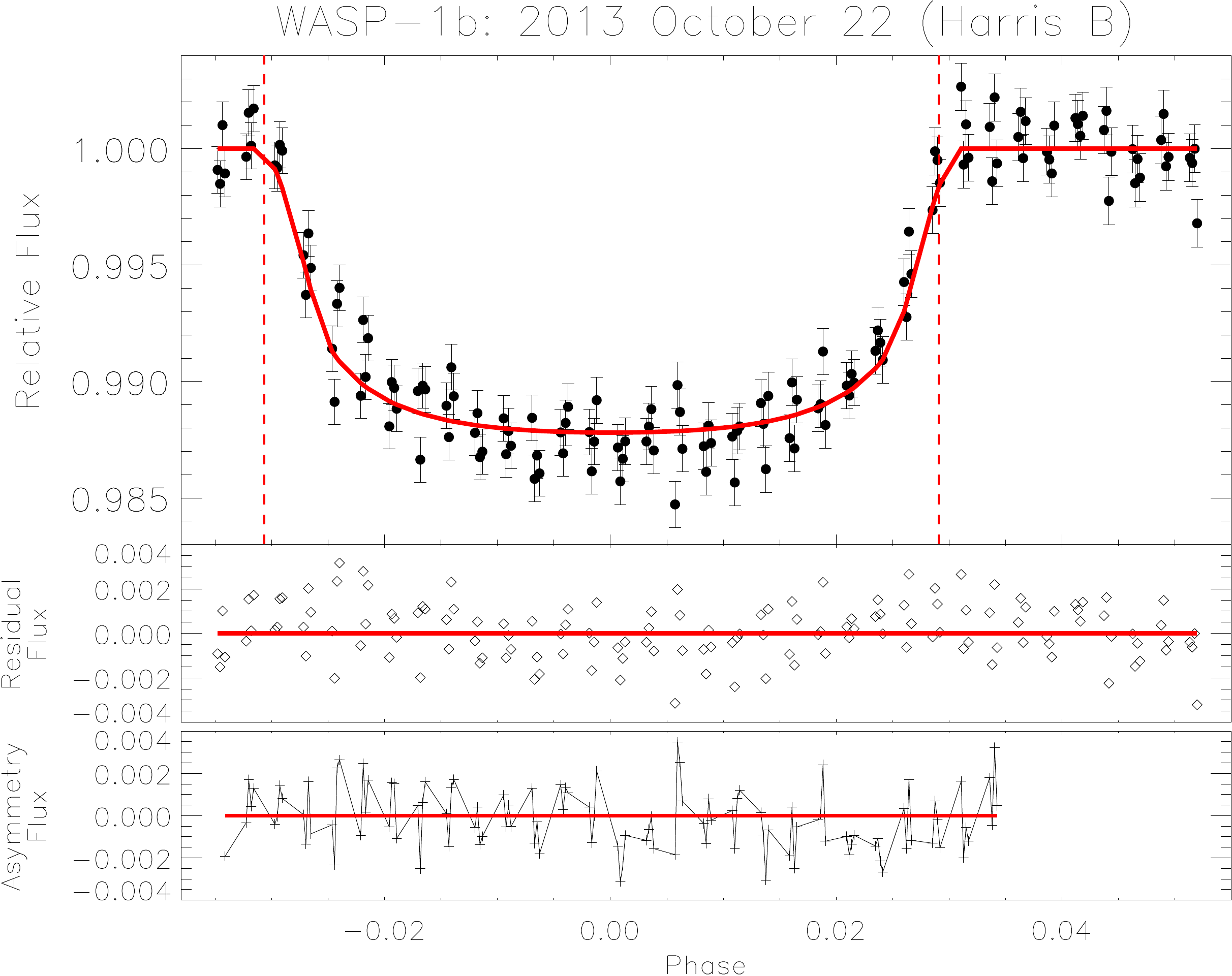,width=0.38\linewidth,angle=90} \\
\vspace{0.1cm}
\epsfig{file=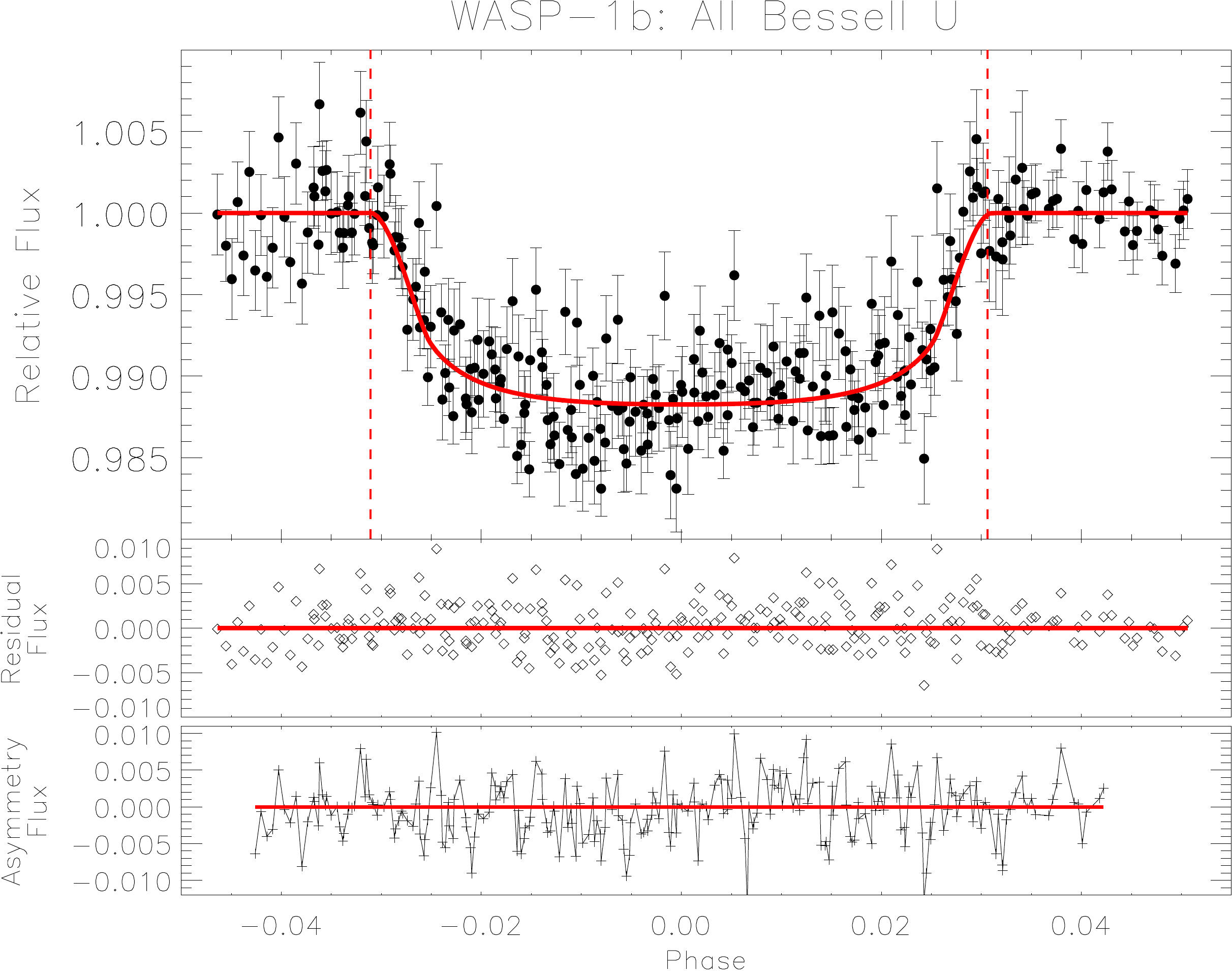,width=0.38\linewidth,angle=90}& \hspace{0.3cm}
\epsfig{file=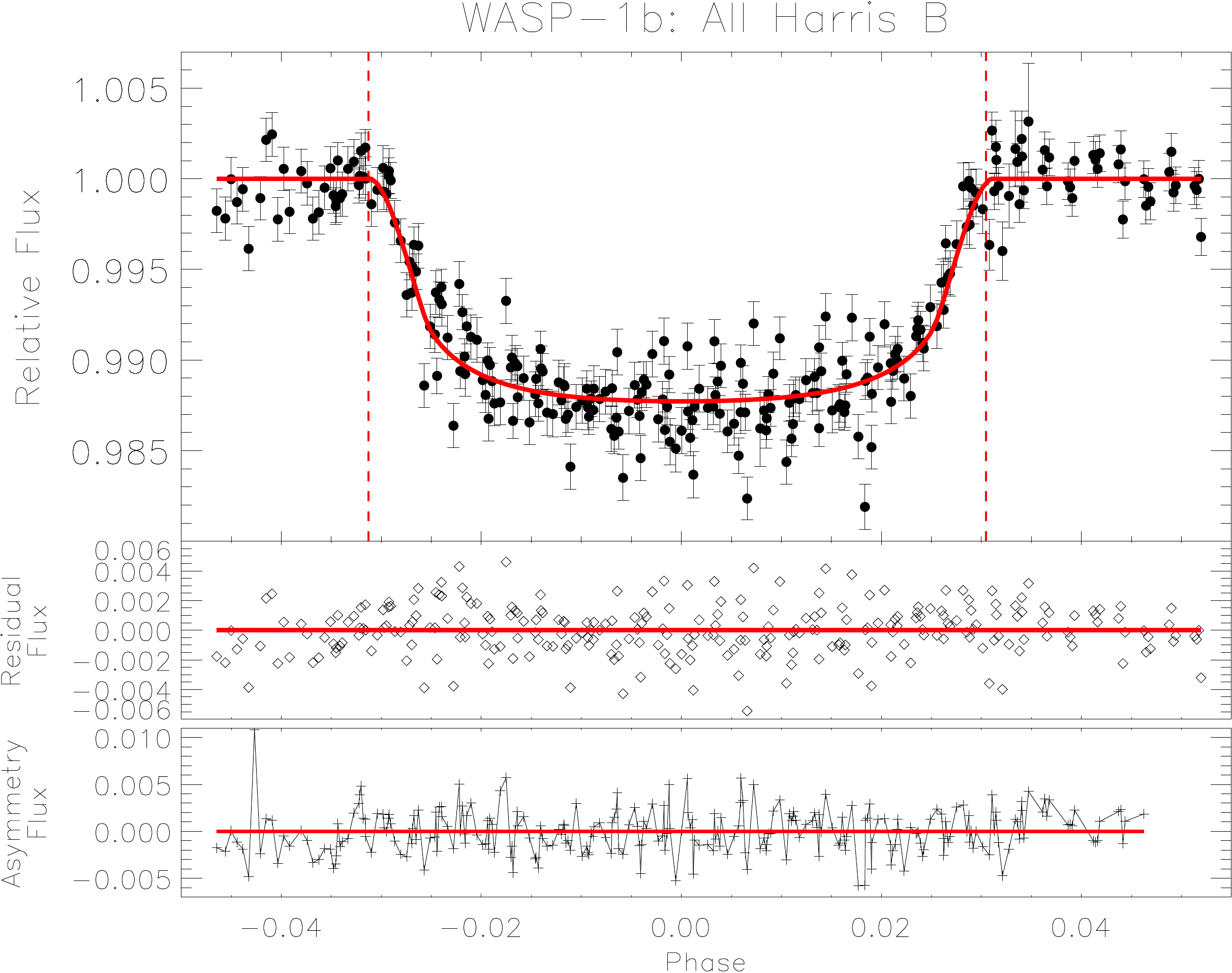,width=0.38\linewidth,angle=90}\\
\end{tabular}
\caption{Light curves of WASP-1b. Other comments are the same as Fig.~\ref{fig:light_1}.}
\label{fig:light_3}
\end{figure*}

\begin{figure*}
\centering
\begin{tabular}{cc}
\vspace{0.5cm}
\epsfig{file=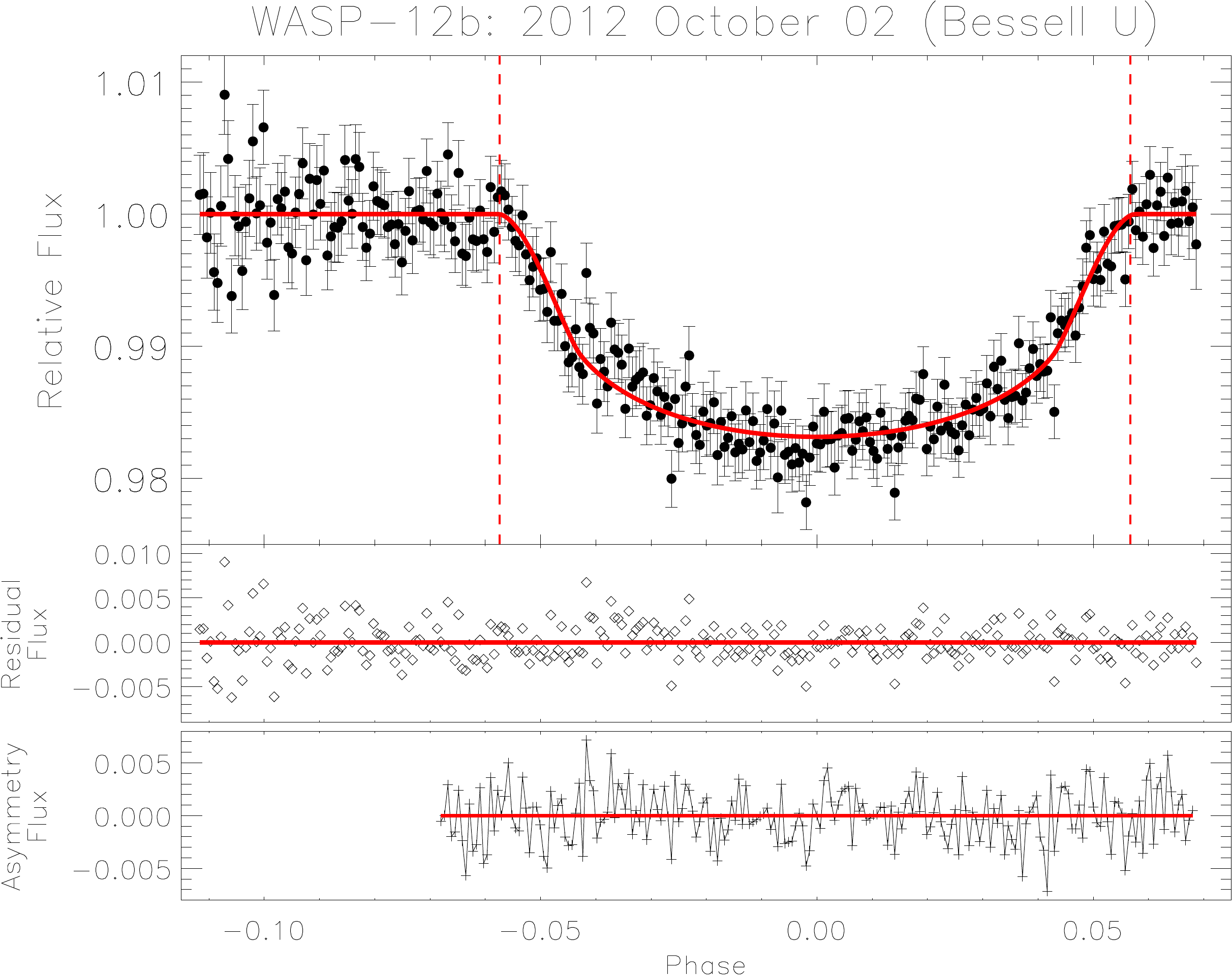,width=0.38\linewidth,angle=90} & \hspace{0.3cm}
\epsfig{file=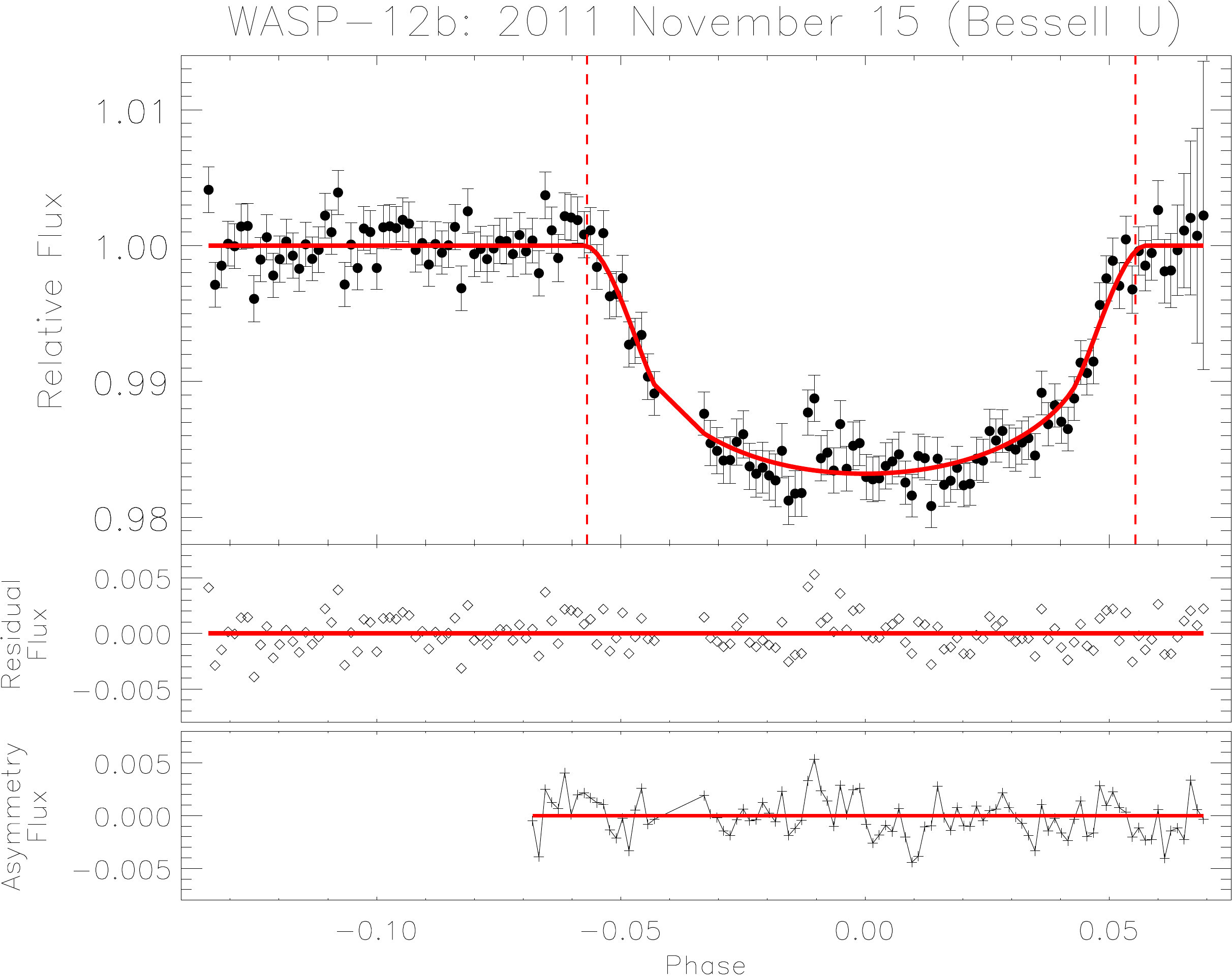,width=0.38\linewidth,angle=90} \\
\vspace{0.5cm}
\epsfig{file=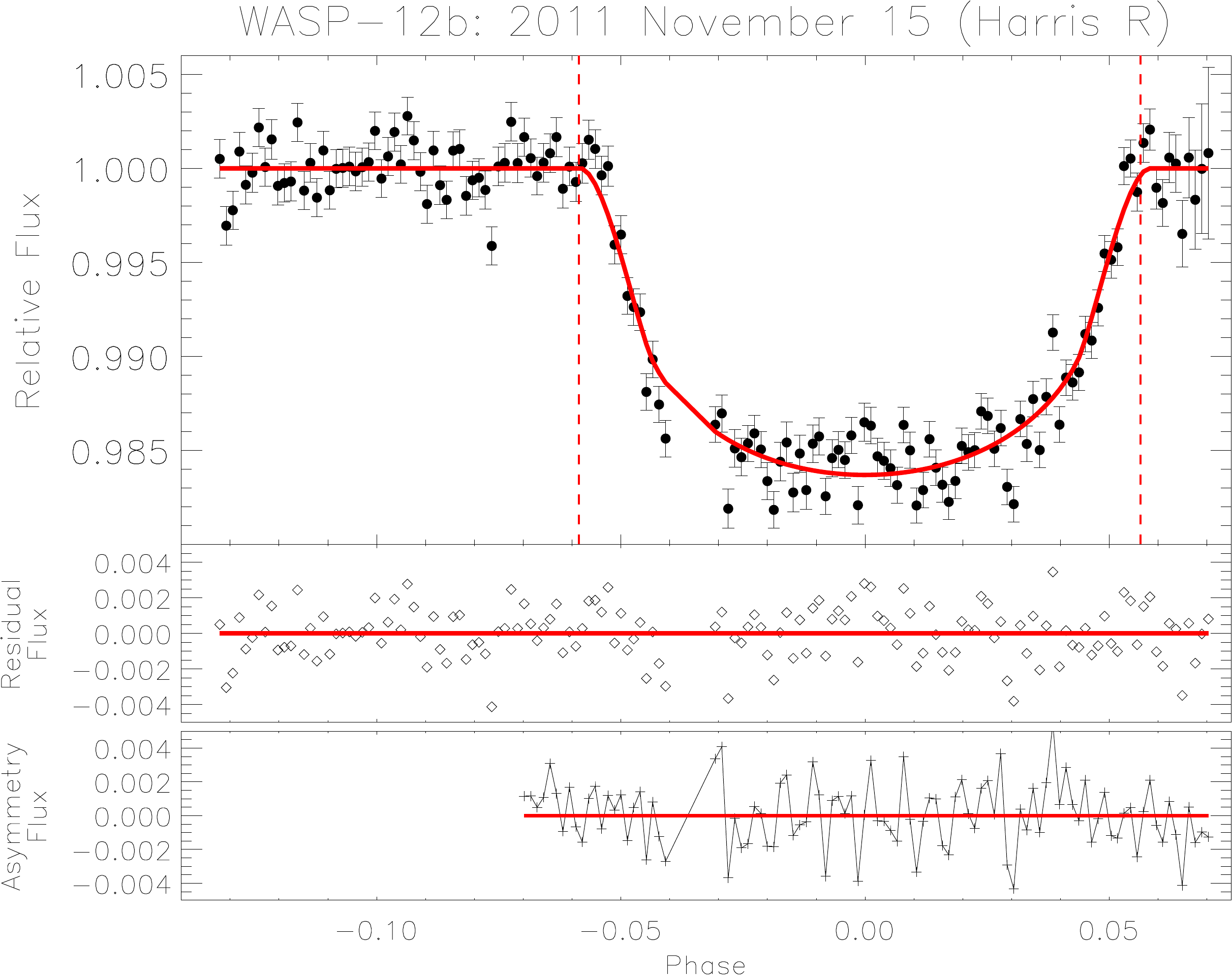,width=0.38\linewidth,angle=90} & \hspace{0.3cm}
\epsfig{file=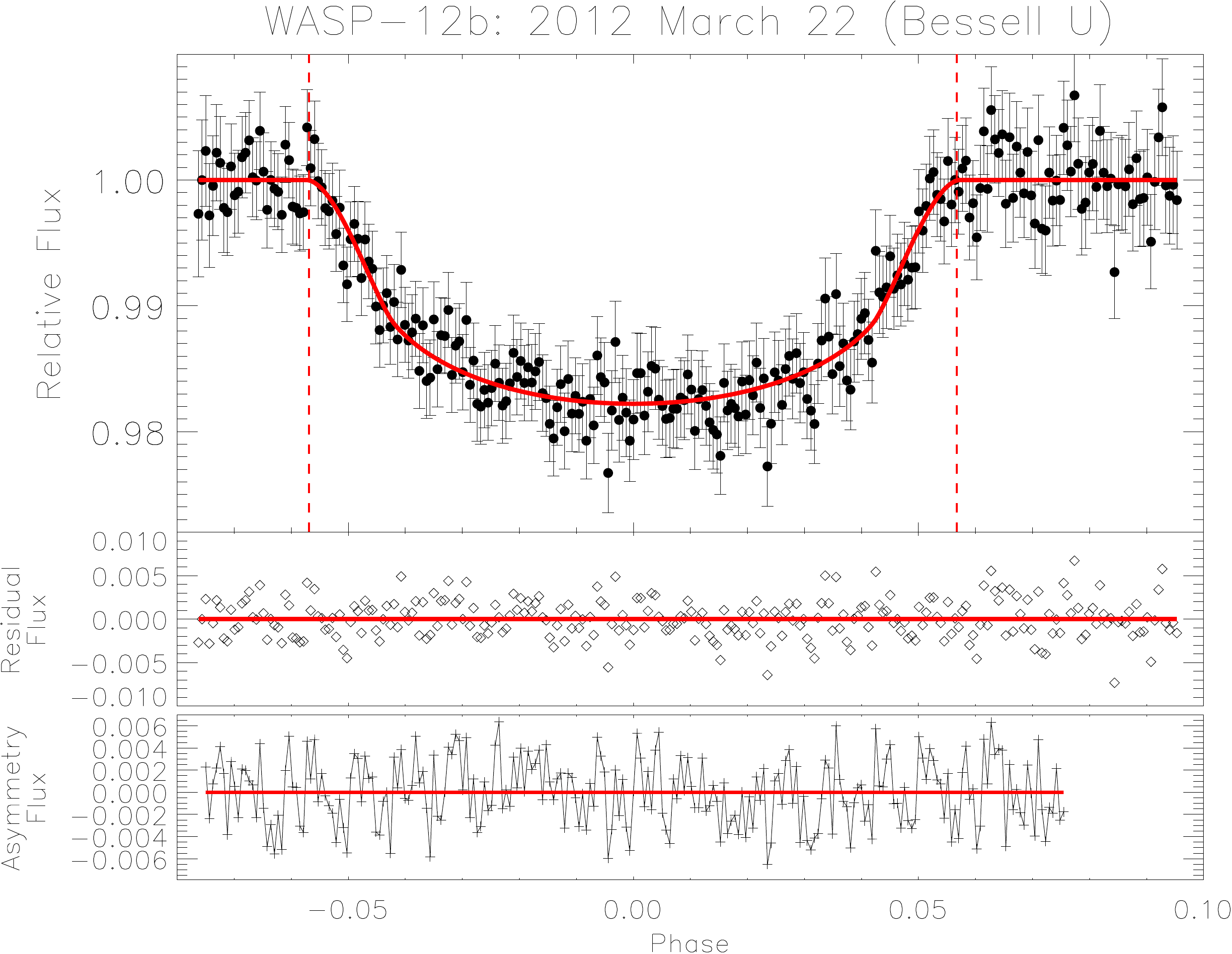,width=0.38\linewidth,angle=90} \\
\vspace{0.1cm}
\epsfig{file=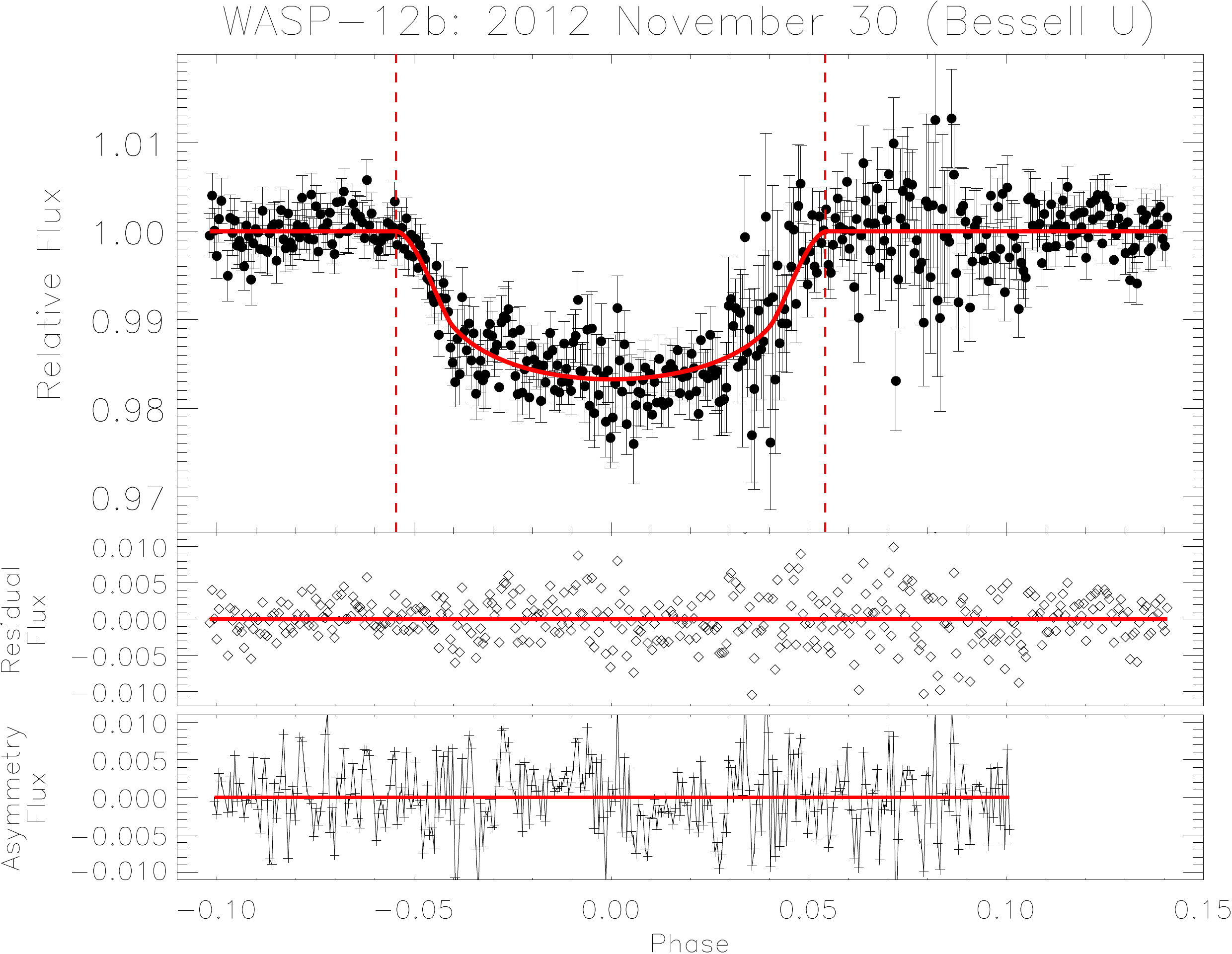,width=0.38\linewidth,angle=90}& \hspace{0.3cm}
\epsfig{file=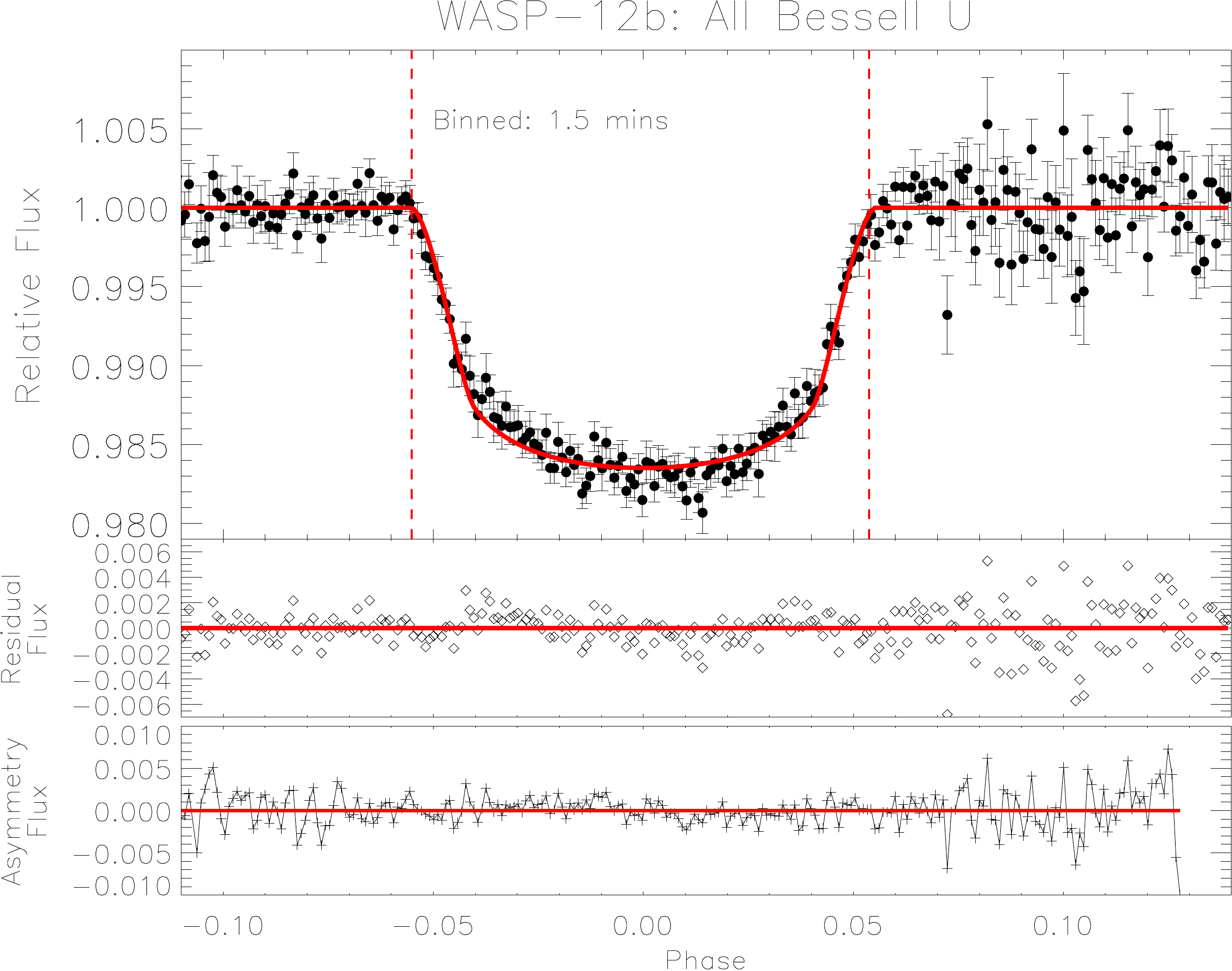,width=0.38\linewidth,angle=90}\\
\end{tabular}
\caption{Light curves of WASP-12b. Other comments are the same as Fig.~\ref{fig:light_1}.}
\label{fig:light_4}
\end{figure*}

\begin{figure*}
\centering
\begin{tabular}{cc}
\vspace{0.5cm}
\epsfig{file=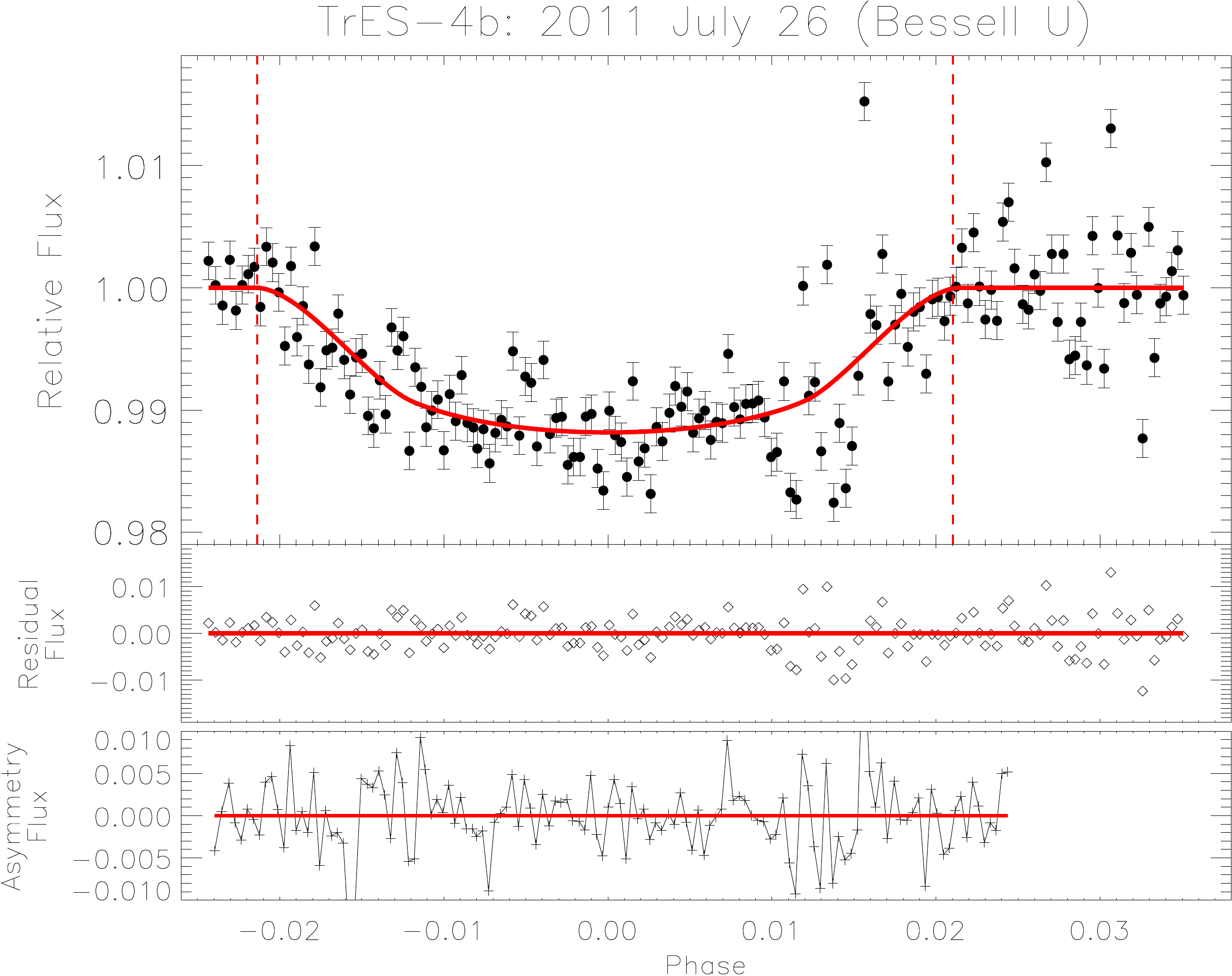,width=0.38\linewidth,angle=90}		& \hspace{0.3cm} \epsfig{file=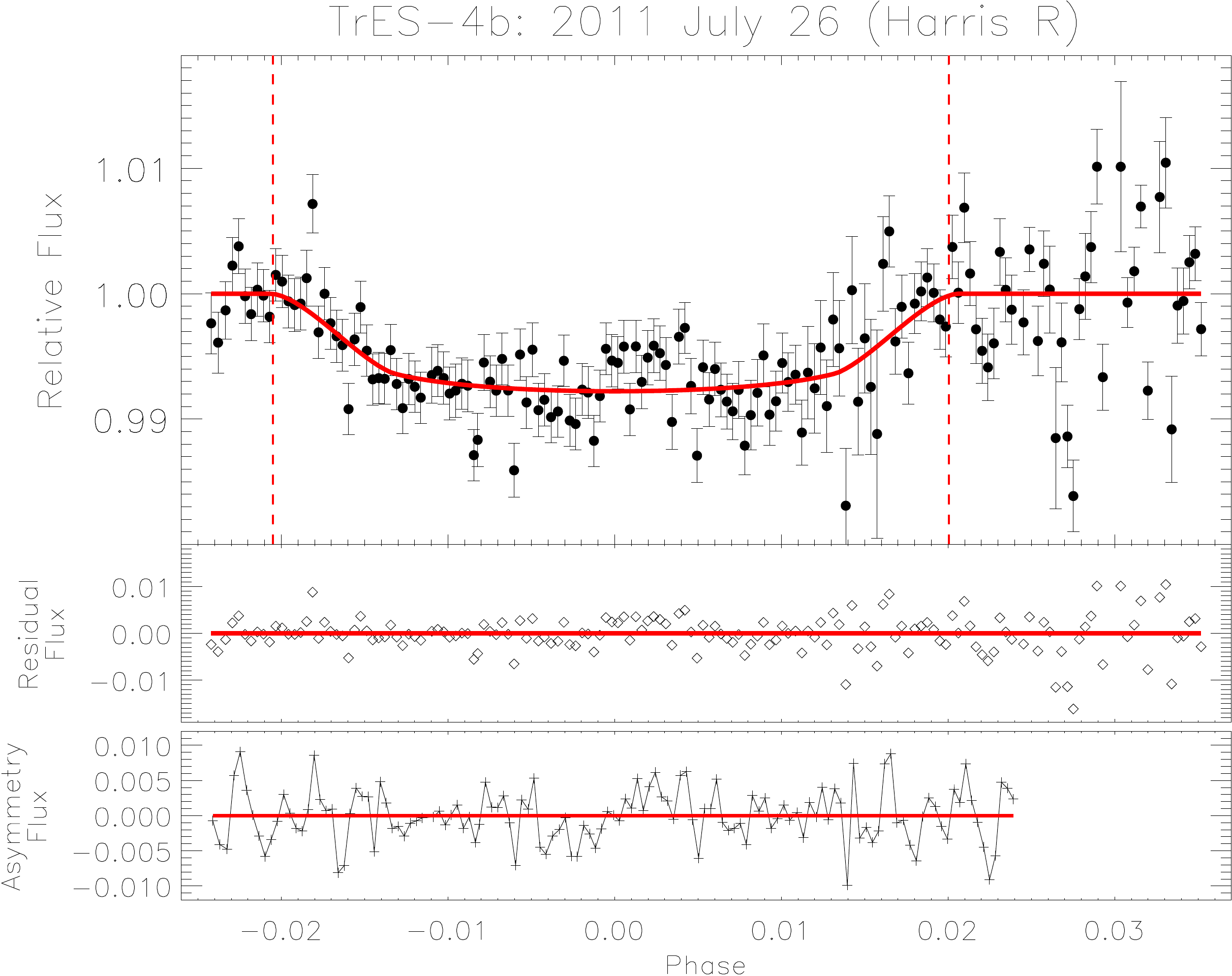,width=0.38\linewidth,angle=90}\\
\vspace{0.5cm}
\epsfig{file=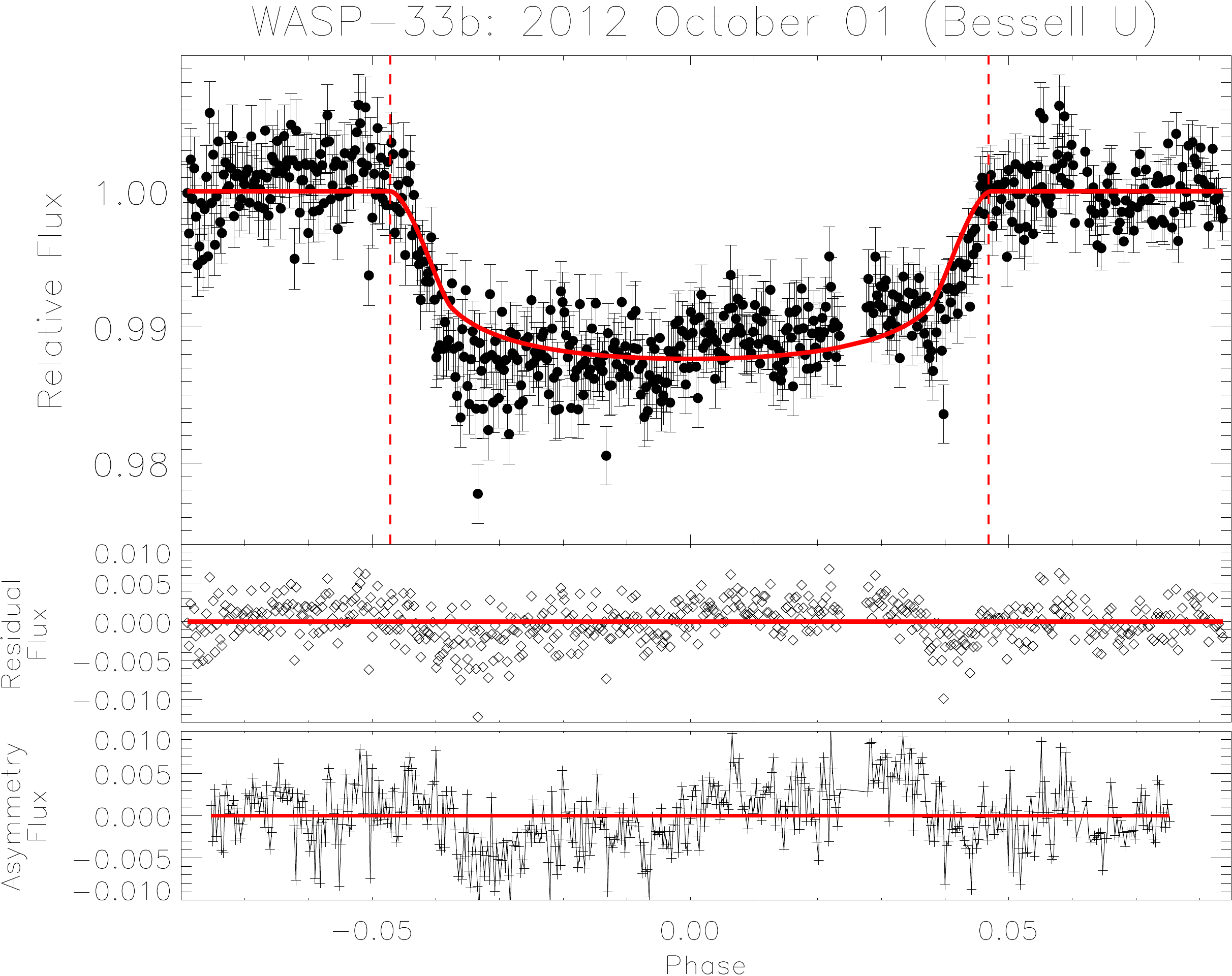,width=0.38\textwidth,angle=90} 		& \hspace{0.3cm} \epsfig{file=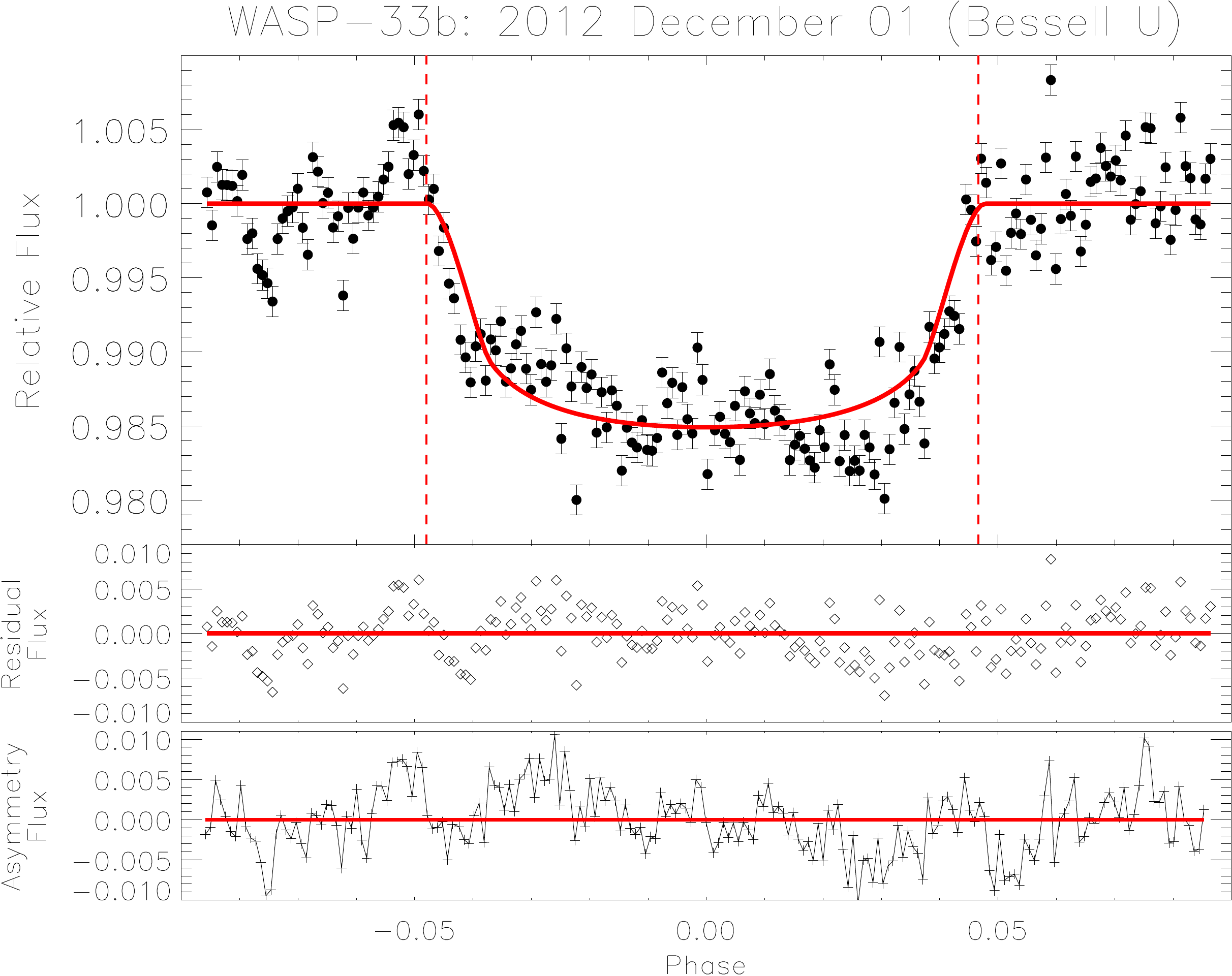,width=0.38\textwidth,angle=90} \\
\vspace{0.1cm}
\epsfig{file=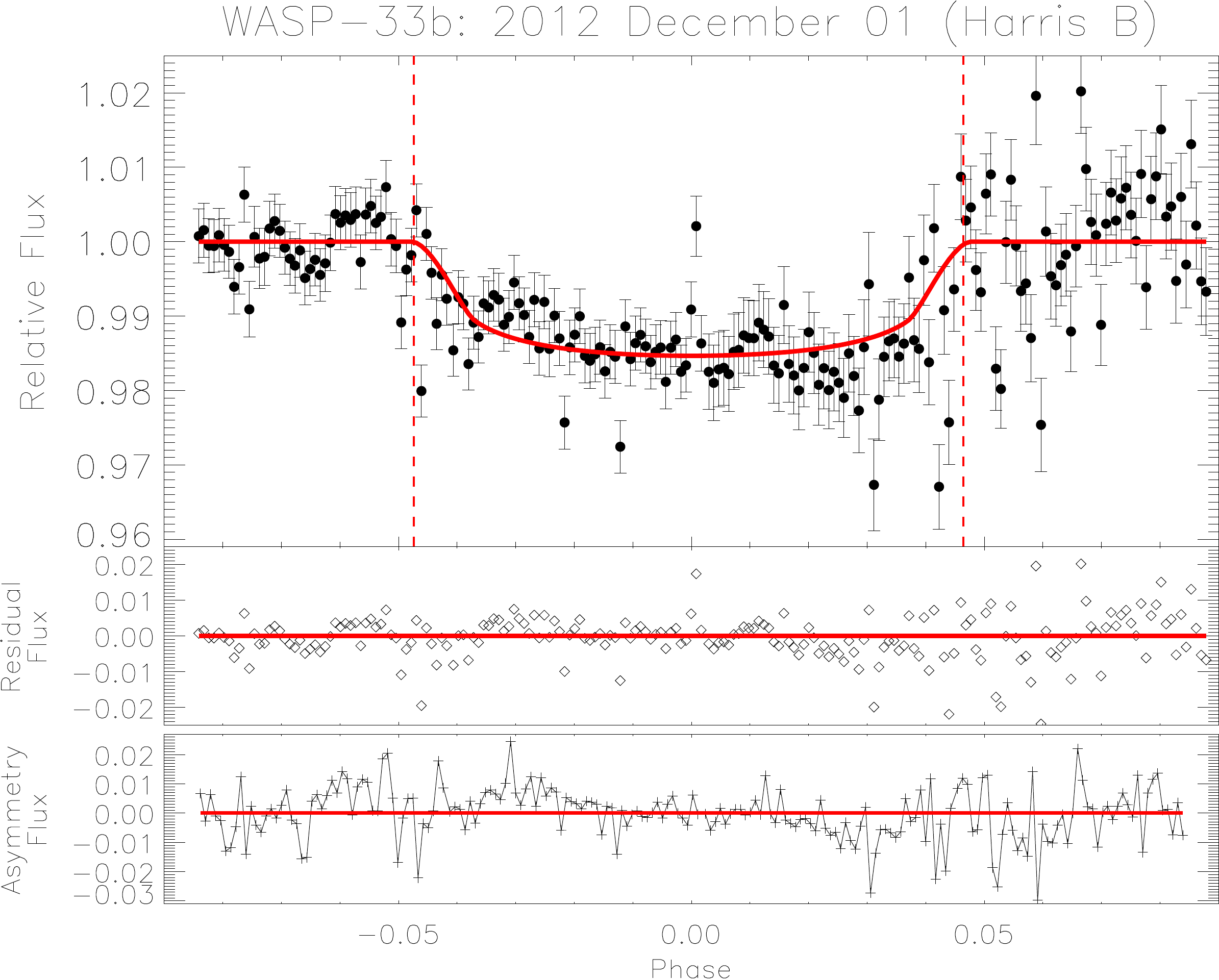,width=0.38\linewidth,angle=90} 		&\hspace{0.3cm} \epsfig{file=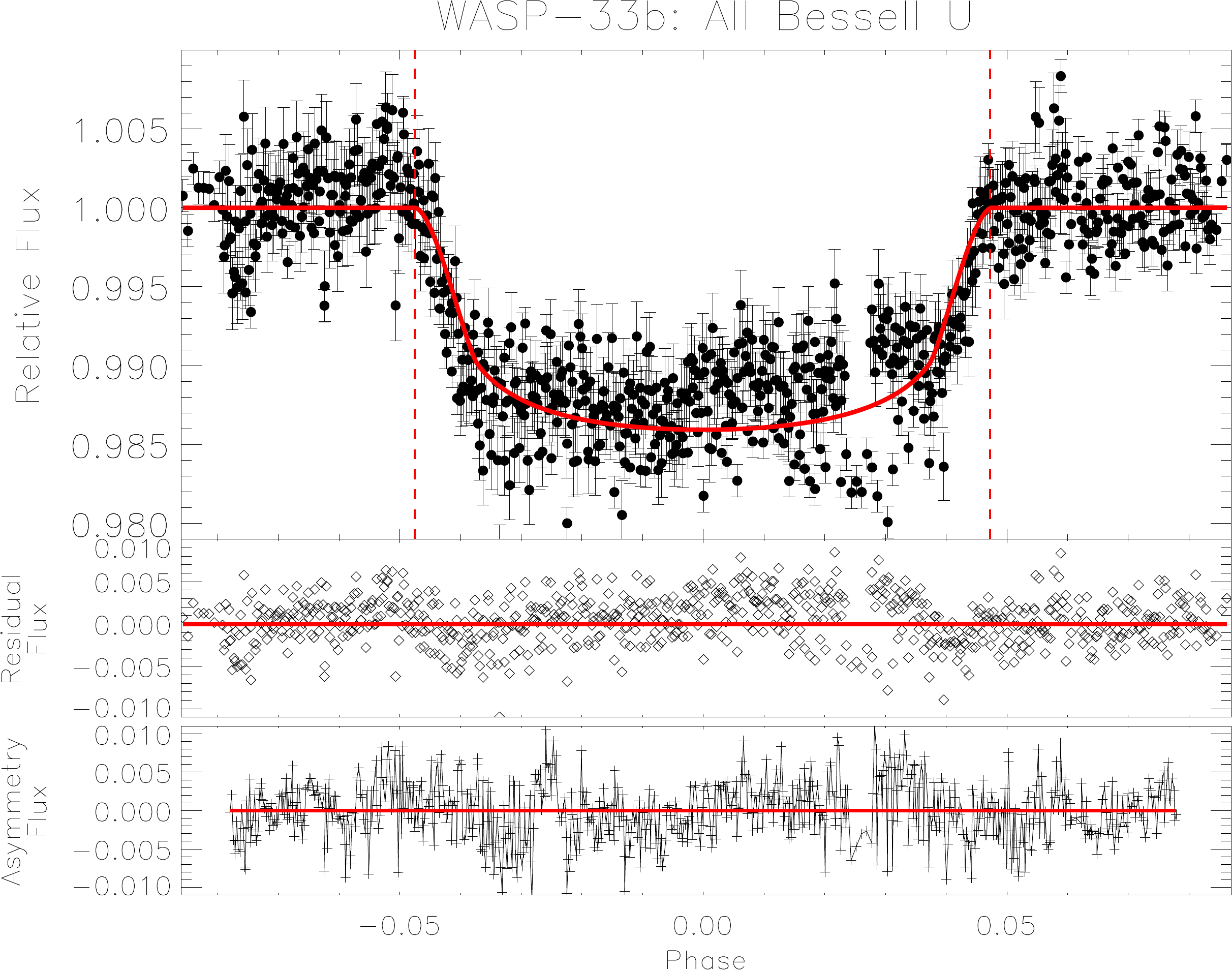,width=0.38\linewidth,angle=90} \\
\end{tabular}
\caption{Light curves of TrES-4b and WASP-33b. Other comments are the same as Fig. \ref{fig:light_1}.}
\label{fig:light_5}
\end{figure*}

To reduce the data we use the automated reduction pipeline \texttt{ExoDRPL}\footnote{https://sites.google.com/a/email.arizona.edu/kyle-pearson/exodrpl} which generates a series of \texttt{IRAF}\footnote{\texttt{IRAF} is distributed by the National Optical Astronomy Observatory, which is operated by the Association of Universities for Research in Astronomy, Inc., under cooperative agreement with the National Science Foundation.} scripts that calibrate images using standard reduction procedures and perform aperture photometry \citep{Pearson2014}. Each of our images are bias-subtracted and flat-fielded. \citet{Turner2013a} determined that using more than 10 flat-field images in the reduction of Kuiper/Mont4k data does not significantly reduce the noise in the resulting images. To optimize telescope time, we use 10 flats and 10 bias frames in our all of our observations and reductions.\\
\indent To produce the light curve for each observation we perform aperture photometry (using the task \texttt{phot} in the \texttt{IRAF DAOPHOT} package) by measuring the flux from our target star as well as the flux from up to eight different reference stars with 110 different circular aperture radii. We insure that each reference star is not a variable star by checking the Aladin Sky Atlas\footnote{http://aladin.u-strasbg.fr/; \citealt{Bonnarel2000}}, the International Variable Star Index\footnote{http://www.aavso.org/vsx}, and by examining their light curves divided by the average of the other reference stars. The aperture radii sizes we explore differ for every observation due to changes in seeing conditions. For the analysis we use a constant sky annulus for every night of observation of each target (a different sky annulus is used depending on the seeing for each date and the crowdedness of the field for each target). The sky analysis is chosen to be a radius greater than the target aperture so that no stray light from the target star is included. We also make sure that no other stars fall in the chosen sky annulus. A synthetic light curve is produced by averaging the light curves of the reference stars. Then, the final transit light curve of each date is normalized by dividing by this synthetic light curve to correct for systematics due to atmospheric variations and airmass differences throughout the observations. Every combination of reference stars and aperture radii are considered. We systematically choose the best reference stars and aperture by minimizing the scatter in the Out-of-Transit (OoT) data points. The 1$\sigma$ error bars on the data points include the readout noise, the Poisson noise, and the flat-fielding errors. The final light curves are presented in Figs. \ref{fig:light_1}--\ref{fig:light_5}. The data points of all our transits are available in electronic form (see Table \ref{tb:mr}). For all the transits, the OoT baselines have a photometric root-mean-squared (RMS) value between 1.23 and 6.22 millimagnitude (mmag), consistent with previous high S/N transit photometry using the Mont4k on the 1.55-m Kuiper telescope (\citealt{Dittmann2009a,Dittmann2009b,Dittmann2010,Dittmann2012}; \citealt{Scuderi2010}; \citealt{Turner2013a}; \citealt{Teske2013};  \citealt{Pearson2014}; \citealt{Zellem2015}).

\section{Light Curve Analysis}

\begin{table*}
\centering
\caption{Photometry of all our light curves$^{1}$}
\begin{tabular}{cccccccc}
\hline
\hline
 Planet Name           & 
Filter            & 
 Time (BJD$_{TBD}$)           & 
Relative flux  &
Error bars &
CCD X-Pos &
CCD Y-Pos &
Median Airmass \\
\hline
\hline
CoRoT-1b &Bessell-U & 2456268.870437 &0.9957161 &0.0047672 &549.708 &768.763 &1.2330450  \\
CoRoT-1b &Bessell-U &2456268.871252 &0.9961978 &0.0047672 &549.505 &768.637 &1.2325040  \\
CoRoT-1b &Bessell-U & 2456268.872066 &0.9950270 &0.0047406 &549.488 &768.543 &1.2319960  \\
CoRoT-1b &Bessell-U &2456268.872881 &0.9924044 &0.0046875 &548.965 &768.693 &1.2315230  \\
\hline
\end{tabular}
\vspace{-2em}
\tablenotetext{1}{This table is available in its entirety in machine-readable form in the online journal.  A portion is shown here for guidance regarding its form and content. }	
\label{tb:mr}	
\end{table*}


\begin{figure*}
\centering
\begin{tabular}{cc}
\vspace{0.5cm}
\epsfig{file=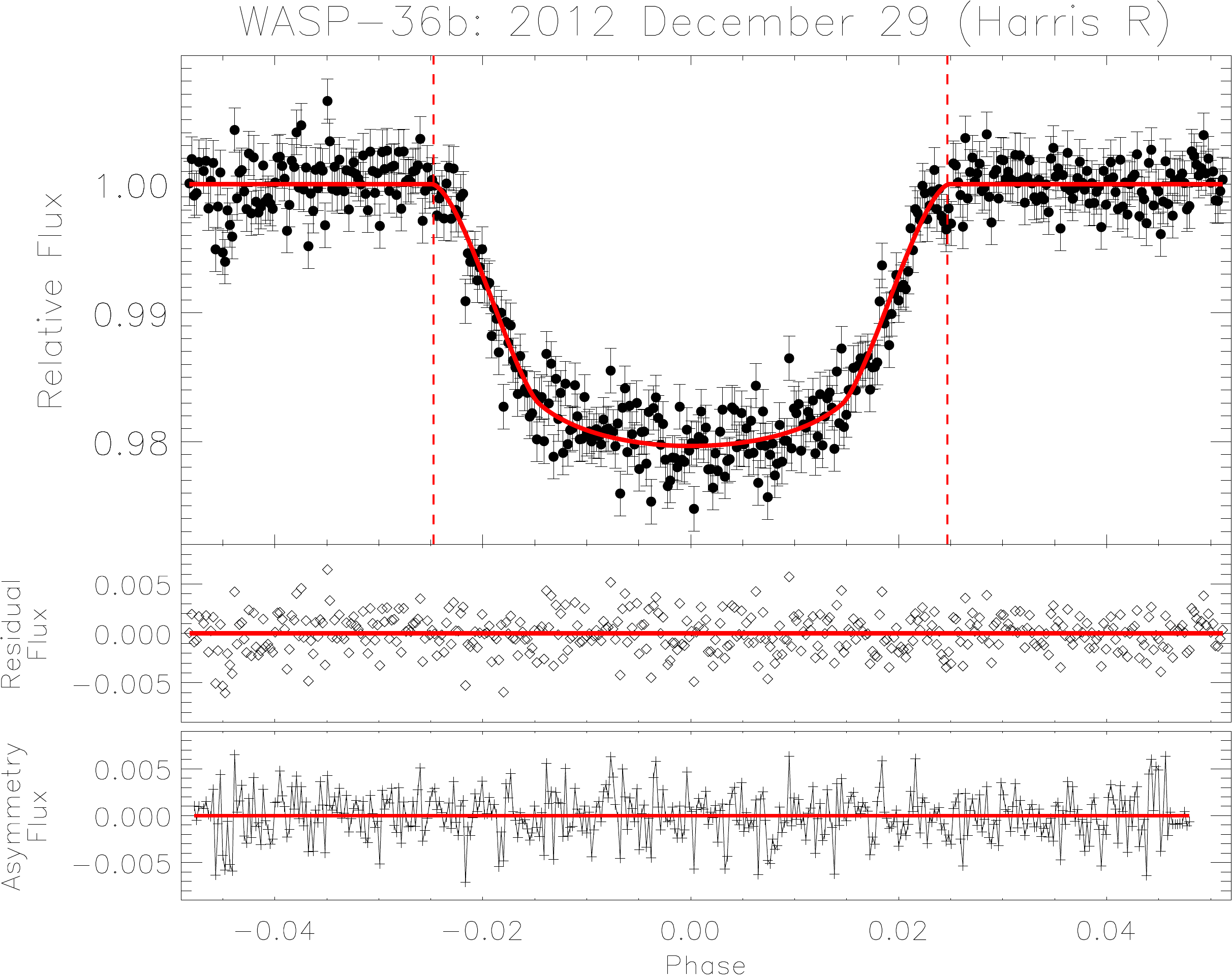,width=0.38\linewidth,angle=90}		& \hspace{0.3cm} \epsfig{file=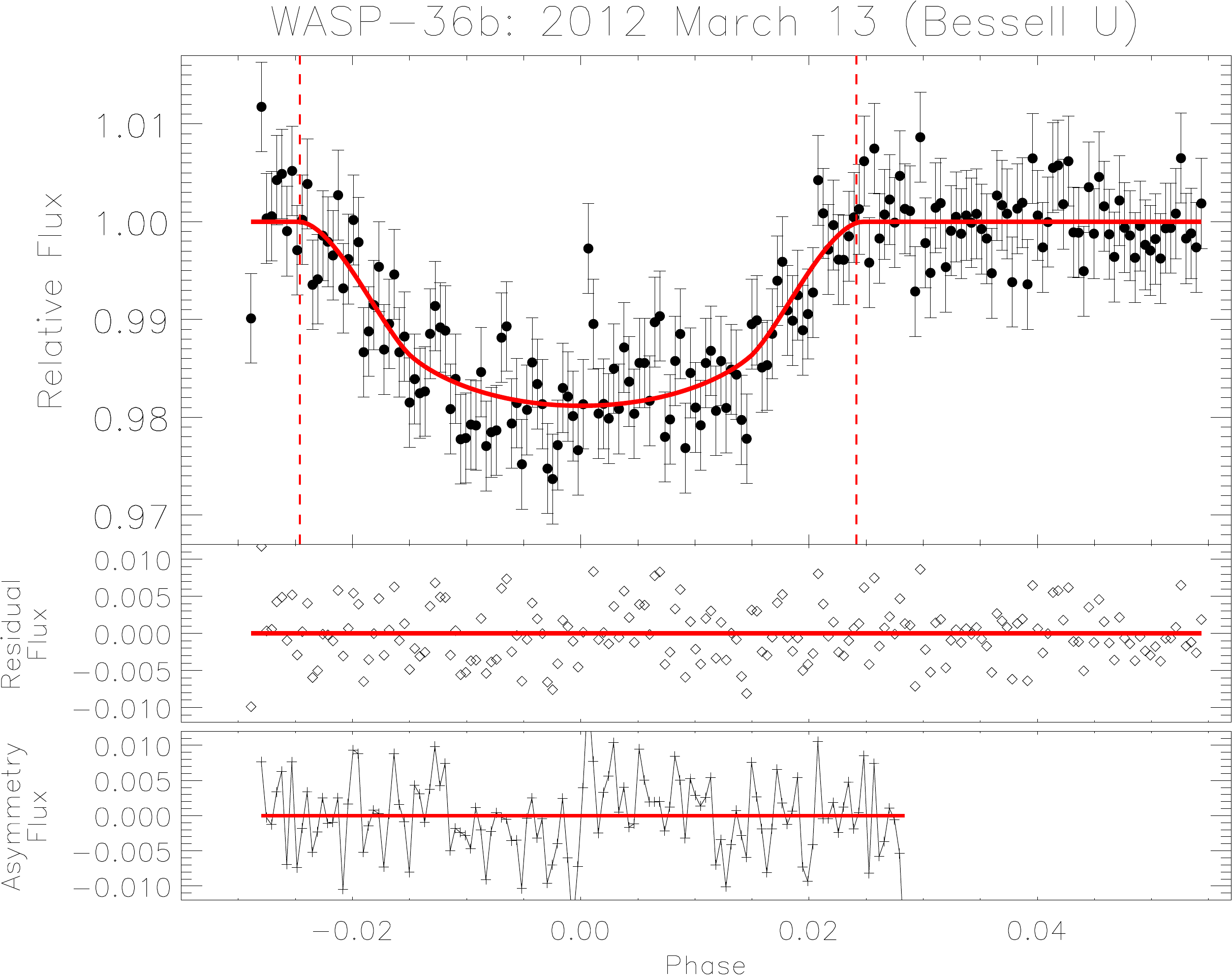,width=0.38\linewidth,angle=90}\\
\vspace{0.5cm}
\epsfig{file=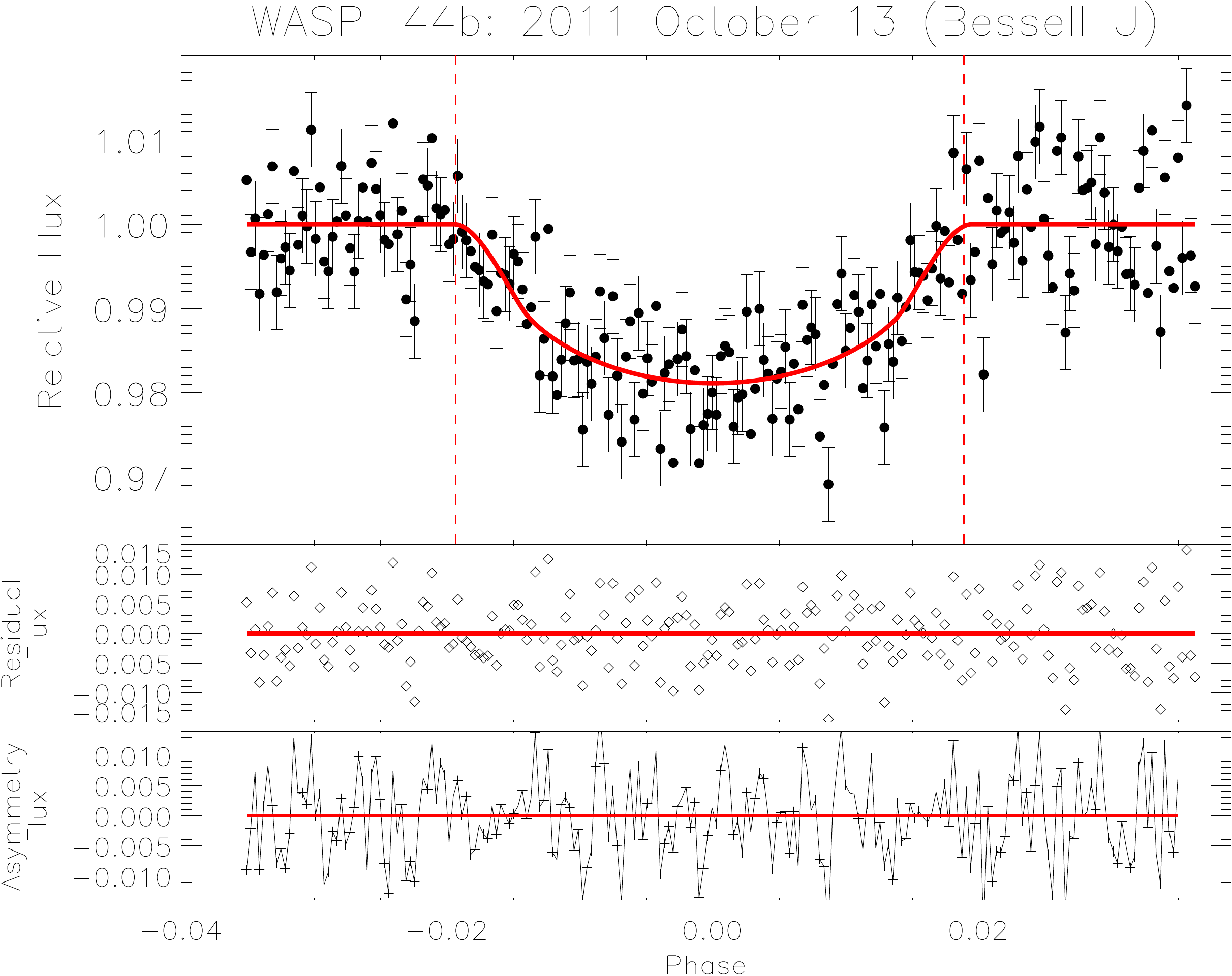,width=0.38\textwidth,angle=90} 	& \hspace{0.3cm} \epsfig{file=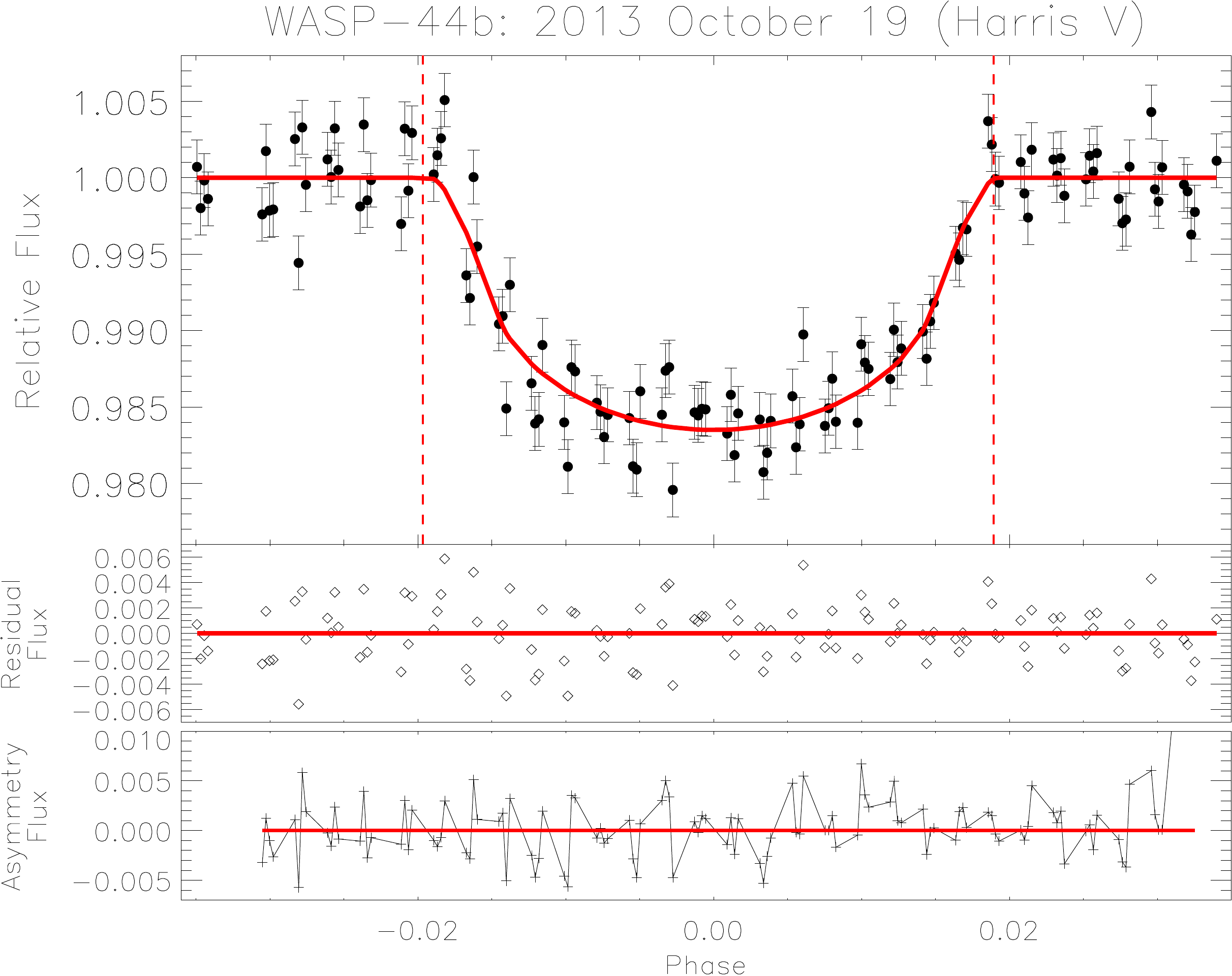,width=0.38\textwidth,angle=90} \\
\vspace{0.1cm}
\epsfig{file=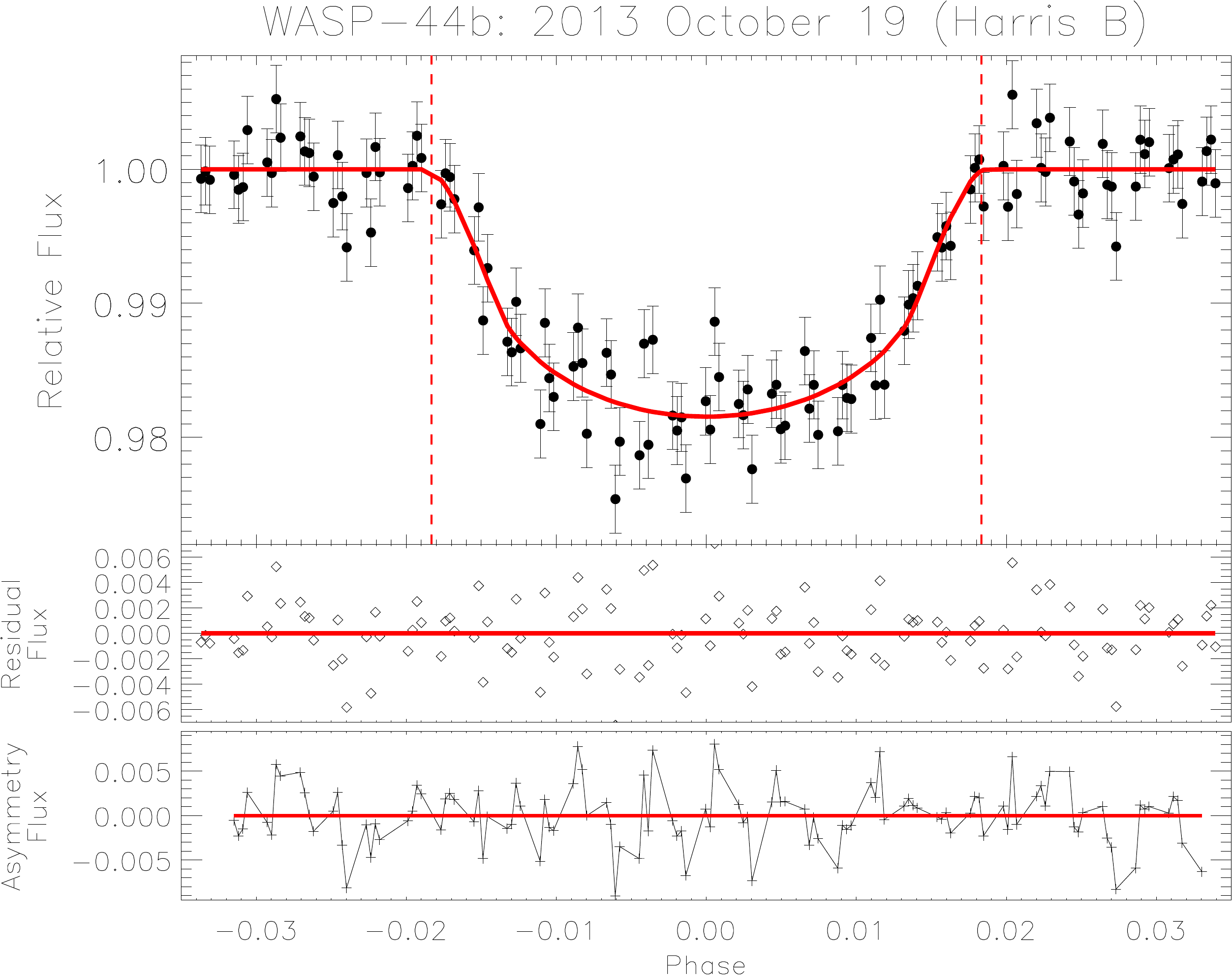,width=0.38\linewidth,angle=90} 	&\hspace{0.3cm} \epsfig{file=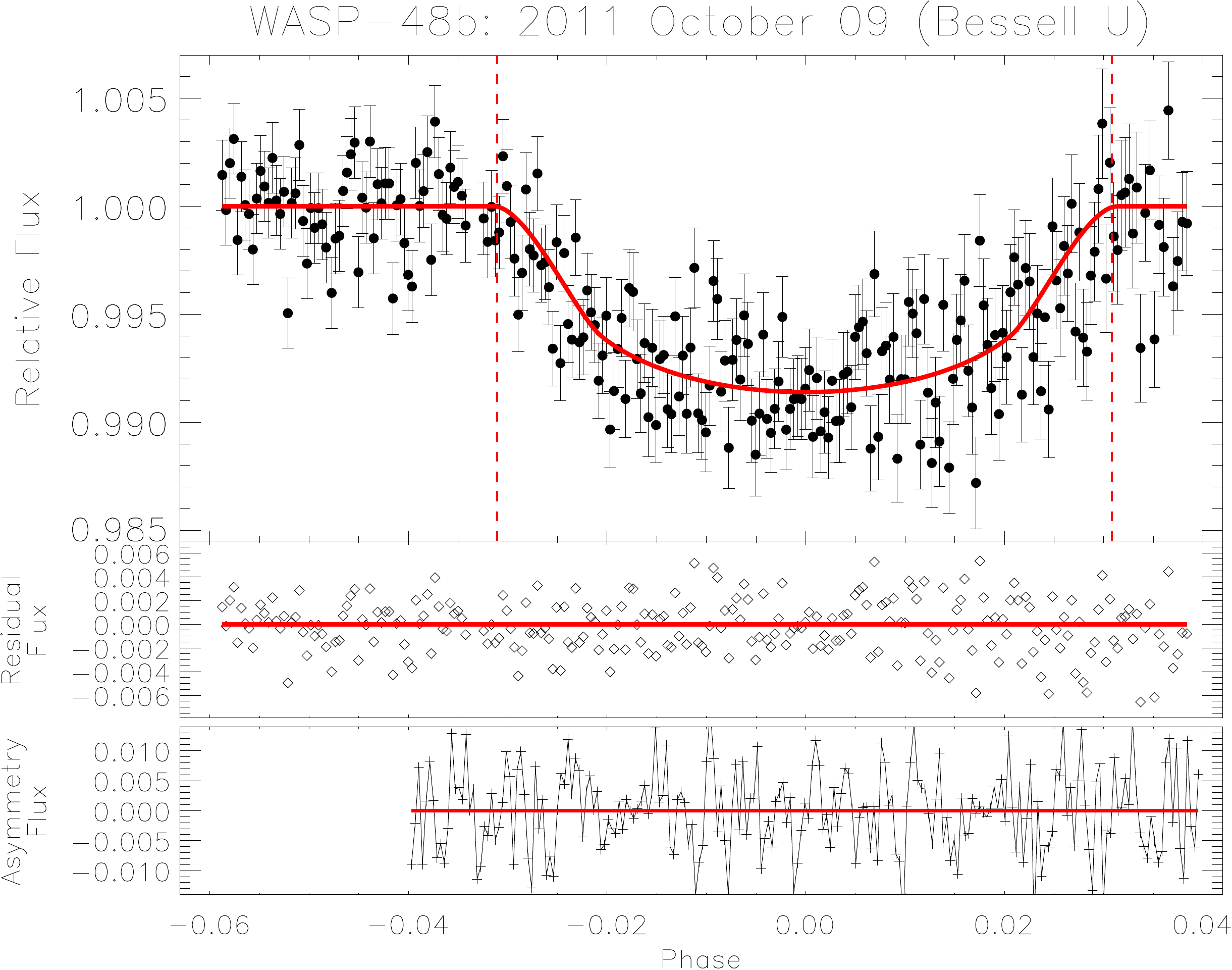,width=0.38\linewidth,angle=90} \\
\end{tabular}
\caption{Light curves of WASP-36b, WASP-44b, and WASP-48b. Other comments are the same as Fig. \ref{fig:light_1}.}
\label{fig:light_6}
\end{figure*}

\begin{figure*}
\centering
\begin{tabular}{c}
\epsfig{file=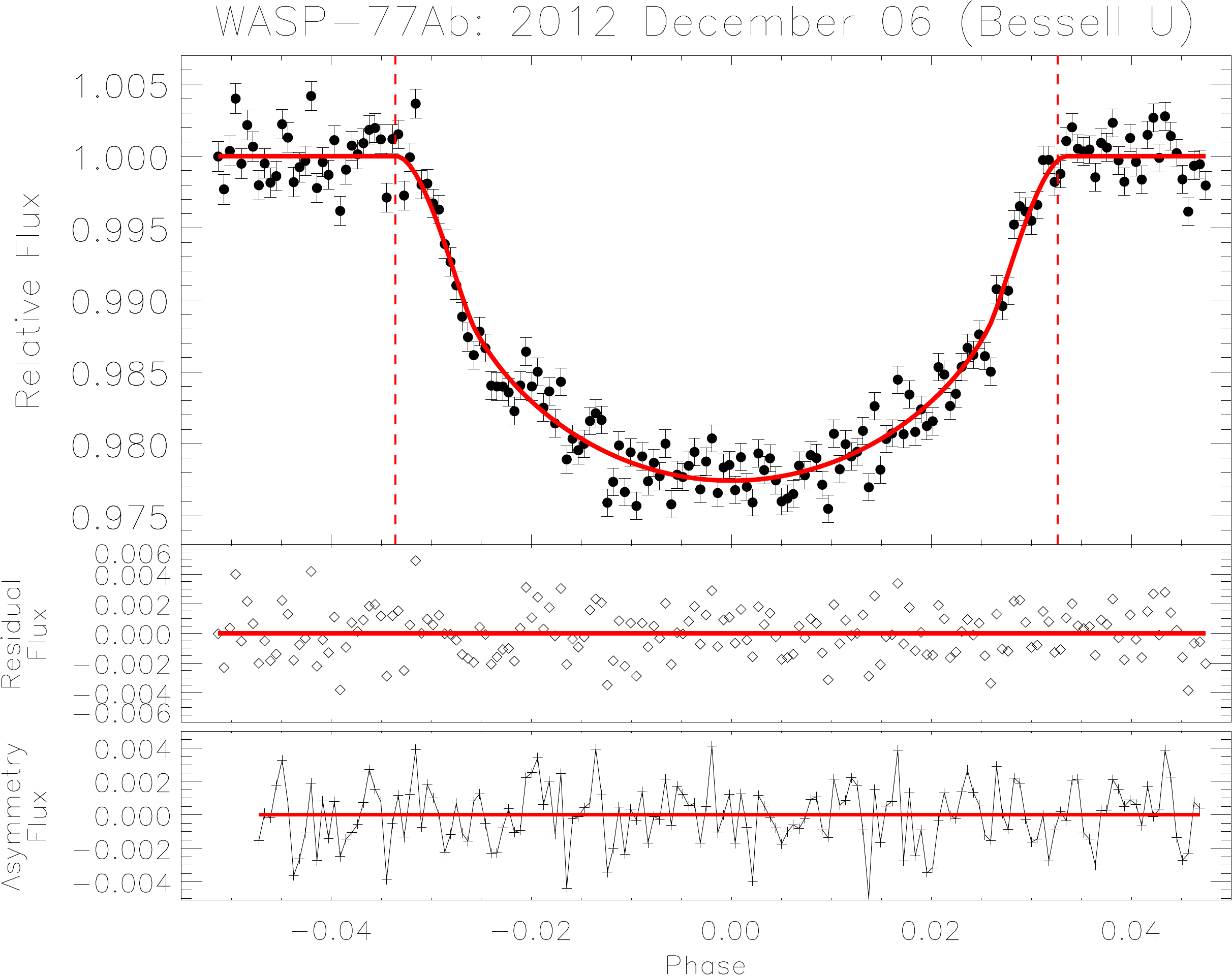,width=0.38\linewidth,angle=90}
\end{tabular}
\caption{Light curve of WASP-77Ab. Other comments are the same as Fig. \ref{fig:light_1}.}
\label{fig:light_7}
\end{figure*}


\subsection{EXOplanet MOdeling Package (EXOMOP)} \label{sec:EXOMOP}

To find the best-fit to the light curves we develop a modeling package called the EXOplanet MOdeling Package (\texttt{EXOMOP}; \citealt{Pearson2014})\footnote{\texttt{EXOMOPv7.0} is used in the analysis and is available at \\ https://sites.google.com/site/astrojaketurner/codes} that uses the analytic equations of \cite{Mandel2002} to generate a model transit. The $\chi^{2}$-fitting statistic for the model light curve is: 
\begin{equation}
\chi^2 = \sum_{i=1}^{n} \left[  \frac{f_{i}( \mbox{obs} )  - f_{i}(\mbox{model})}{\sigma_{i}(\mbox{obs})}    \right]^2, \label{eq:chi2}
\end{equation}
where $n$ is the total number of data points, $f_{i}(\mbox{obs})$ is the observed flux at time $i$, $\sigma_{i}(\mbox{obs})$ is the error in the observed flux, and $f_{i}(\mbox{model})$ is the calculated model flux. The goal of the light curve modeling is to explore the solution-space effectively to determine the $f_{i}(\mbox{model})$ that minimizes $\chi^2$. 

The Bayesian Information Criterion (BIC; \citealt{schwarz1978}) is used to assess over-fitting of the data with \texttt{EXOMOP}. The BIC is defined as 
\begin{equation}
BIC =  \chi^{2} + k \ln{(n)}, \label{eq:BIC}
\end{equation}
where $\chi^{2}$ is the chi-squared calculated for the best-fitting model (equation \ref{eq:chi2}), $k$ is the number of free parameters in the model fit [$f_{i}(\mbox{model})$], and $n$ is the number of data points in the transit. The possible free parameters in the \cite{Mandel2002} model are the planet-to-star radius ($R_{p }/R_{*}$), the scaled semi-major axis ($a/R_{*}$), inclination ($i$), mid-transit time ($T_{c}$), linear limb darkening coefficient ($\mu{_1}$), and quadratic limb darkening coefficient ($\mu{_2}$). The power of the BIC is the penalty for a higher number of fitted model parameters, making it a robust way to compare different best-fit models. The preferred model is the one that produces the lowest BIC value. The BIC has been used extensively in many other exoplanet transit studies (e.g. \citealt{Kipping2010}; \citealt{Croll2011}; \citealt{Sing2011}; \citealt{Gibson2010,Gibson2013}; \citealt{Demory2013}; \citealt{Crossfield2013}; \citealt{Rogers2013}; \citealt{Howard2013}; \citealt{Stevenson2014}; \citealt{Murgas2014}; \citealt{Zellem2014}). 

We perform a Levenberg-Marquardt (LM) non-linear least squares minimization (\texttt{MPFIT}; \citealt{Markwardt2009}; \citealt{Press1992}) to find a best-fit to the data and a bootstrap Monte Carlo technique (\citealt{Press1992}) to calculate robust errors of the LM fitted parameters. In addition, we perform a Differential Evolution Markov Chain Monte Carlo (DE-MCMC; \citealt{terBrrak2002}; \citealt{Eastman2013}) analysis to find a best-fit to the data and associated errors. Both the LM and DE-MCMC methods take into account the photometric error bars on the data points. 

The formal errors in the LM fit can underestimate the parameter uncertainties under strongly correlated parameters (\citealt{Popper1984}; \citealt{Maceroni1997}; \citealt{Southworth2004a,Southworth2004b}; \citealt{Southworth2008}), which is the case for exoplanet transits (\citealt{Carter2009}). Therefore, we determine a robust estimation of the uncertainties using the following Monte Carlo bootstrap procedure. (1) We obtain the best-fit light curves and parameters from the LM non-linear least squares algorithm. 
(2) We find new error bars, $\sigma_{n}$, by the following equation:
\begin{equation}
\sigma_{n} = \sigma_{p} N(\mu,\sigma^2), \label{eq:new}
\end{equation}
where $\sigma_{p}$ are the photometric (observational) error bars for each data point in the light curve, $N(\mu,\sigma^2)$ is a random Gaussian distributed variable ($N$) with a mean $\mu = 0$ and a standard deviation $\sigma = 1$. (3) We add $\sigma_{n}$ to the flux measurements in the light curve. (4) Step (1) is repeated to find a new best-fit light curve (the original photometric error bars, $\sigma_{p}$, are used for the error on the flux measurements). This process is repeated at least 10000 times to avoid biasing the Gaussian fit due to small-number statistics. When all iterations are finished, each fit parameter from step (4) is subtracted from the original best-fit value and a Gaussian function is fit to the distribution. The standard deviations of the distributions are taken as the one sigma uncertainties in the fitted parameters.  

We use the DE-MCMC analysis to find more robust parameter values because the solution is a global minimum in solution-space and $\chi^2$. By default the DE-MCMC in \texttt{EXOMOP} uses 20 chains and 20$^{6}$ links. The Gelman-Rubin statistic (\citealt{Gelman1992}) is used to ensure chain convergence \citep{Ford2006}. We use the DE-MCMC model from \texttt{EXOFAST} (exofast$\_$demc; \citealt{Eastman2013}) in \texttt{EXOMOP}. \texttt{EXOMOP} uses the Metropolis-Hastings sampler and characterizes the uncertainties using a Bayesian inference that accounts for non-Gaussian errors and covariances between parameters \citep{Eastman2013}. The LM solution and errors are used as the seed for the DE-MCMC model. 
   
 \texttt{EXOMOP} is also capable of fitting a function to the OoT baseline to account for any residual curvature due to the atmospheric extinction. Either a linear or quadratic fit can be found in both the LM and DE-MCMC models. The baseline function is fit to the transit simultaneously with the \cite{Mandel2002} model. The BIC is also used to determine whether to include a baseline fit in the best-fit model.  

\subsubsection{Red noise estimation} \label{sec:red_noise}
 
\texttt{EXOMOP} uses the residual permutation (rosary bead; \citealt{Southworth2008}), time-averaging (\citealt{Pont2006}), and wavelet (\citealt{Carter2009}) methods to access the importance of temporally-correlated (red) noise in both fitting methods. Red noise is accounted for in our analysis because the errors in the fitted parameter values can be underestimated if we don't account for red noise (\citealt{Pont2006}; \citealt{Carter2009}). In order to be conservative, the red noise method that produces the largest $\beta$, the scaling factor of the red noise relative to the white noise errors, is used to inflate the errors in the fitted parameters (Section \ref{sec:final}).

 In the residual permutation method (\citealt{Jenkins2002}; \citealt{Southworth2008}; \citealt{Bean2008}; \citealt{Winn2008b}) the best-fit model is subtracted from the data and the residuals are circularly shifted and then added to the data points. A new fit is found, and then the residuals are shifted again, with those at the end wrapped around to the start of the data. In this way, every new synthetic data set will have the same noise characteristics as the actual data but only translated in time. Usually this process continues until the residuals have cycled back to where they originated (e.g. \citealt{Todorov2012}). We perform two different residual permutation procedures to determine the effect of red noise in the precision of our derived parameters. 

Our first residual permutation ($res1$) method uses a procedure very similar to \cite{Todorov2012} where the shifting process continues until the residuals have cycled back to where they originated (one full circular permutation). The resulting parameter values may have non-Gaussian distributions if red noise is present. Consequently, we set the 1$\sigma$ error bars of each parameter as half the range that covers 68\% of the total number of the data points, centered on the best-fit value from either the DE-MCMC or LM analysis. For each fitted parameter we then define $\beta_{res1}$ (the scaling factor of the errors relative to white noise using the $res1$ method) as $\sigma_{w}$/$\sigma_{res1}$, where $\sigma_{w}$ are the error bars derived from the bootstrap Monte Carlo technique or the DE-MCMC technique and the $\sigma_{res1}$ are the error bars derived from the first residual permutation method. 

For the second residual permutation ($res2$) method we update this procedure by allowing for the error bars of the residuals to be taken into account. 
This is similar to step (2), (3), and (4) in the bootstrap procedure described above, however, in step (3) $\sigma_{n}$ is added to the residuals and in step (4) the residuals are added to the data points and a new fit is found. We repeat this process 10000 times and on each step the residuals are circularly shifted. This procedure results in a distribution of fitted values for each parameter from which its uncertainty is estimated using the standard deviation of a Gaussian fit. For each fitted parameter we then define $\beta_{res2}$ (the scaling factor of the errors relative to white noise using the $res2$ method) as $\sigma_{w}/\sigma_{res2}$, where $\sigma_{res2}$ are the error bars derived from the second residual permutation method. The second residual permutation method is limited by the fact that we assume a Gaussian distribution for the errors. 

The next red noise estimation we implement is the time-averaging method. This is done in a similar fashion to the procedure described by \cite{Winn2008b}.  For each light curve we find the best-fitting model and calculate the residuals between the observed and calculated fluxes. Next, the residuals are separated into bins of $N$ points and we calculate the standard deviation, $\sigma_{N}$, of the binned residuals. In our analysis, $N$ ranges from $1$ to $n$, where $n$ is the total number of data points in each respective transit. Using the set of $\sigma_{N}$ and $N$ values we then use a LM non-linear least squares minimization algorithm to find the RMS of red noise ($\sigma_{red}$) and the RMS of white noise ($\sigma_{white}$) using the following equation from \citealt{Pont2006}:
\begin{equation}
	\sigma_{N} =  \sqrt{  \frac{\sigma_{white}^2 }{N}  + \sigma_{red}  } .
\end{equation}
Using $\sigma_{white}$ and $\sigma_{red}$ we estimate $\beta_{time}$, the scaling factor of the errors relative to white noise using the time-averaging method, with the following equation from \cite{Carter2009}: 
\begin{equation}
	\beta_{time} = \sqrt{1 + \left( \frac{\sigma_{red}}{\sigma_{white}} \right)^2} . \label{eq:beta_time}
\end{equation}

Finally, we use the wavelet technique (solveredwv; \citealt{Carter2009}) as a third check of the importance of red noise in the light curve fitting process. In this method the total noise of the transit is assumed to be formed as an additive combination of noise with power spectral density proportional to $1/f^{\alpha}$ (the red noise) and Gaussian white noise.  A downhill simplex method (AMOEBA; \citealt{Nelder1965}; \citealt{Press1992}) algorithm is used to maximize the likelihood that a function of $\sigma_{red}$ and $\sigma_{white}$ is related to the standard deviations of the $1/f^{\alpha}$ and white noise, respectively. A more thorough description of the wavelet model can be found in \citet{Carter2009}. Again, $\beta_{wave}$, the scaling factor of the errors relative to white noise using the wavelet technique, is estimated by using equation (\ref{eq:beta_time}).

\subsubsection{Final error bars on the fitted parameters}\label{sec:final}

To get the final error bars for the fitted parameters we multiply $\sigma_{w}$ by the largest $\beta$ ($\beta_{time}$, $\beta_{res1}$, $\beta_{res2}$, or $\beta_{wave}$) from the residual permutation, time-averaging, and wavelet red noise calculations to account for underestimated error bars due to red noise (\citealt{Winn2008b}). To remain conservative this multiplication step is only done if the largest $\beta$ is greater than one. Finally, in cases where the reduced chi-squared ($\chi^{2}_{r}$) of the data (Table \ref{tb:obs_new}) to the best-fit model is greater than unity we multiply the error bars above by $\sqrt{\chi^{2}_{r}}$ to compensate for the underestimated observational errors (\citealt{Bruntt2006}; \citealt{Southworth2007a};  \citealt{Southworth2007b};  \citealt{Southworth2008}; \citealt{Barnes2013}).

\subsubsection{Additional features of EXOMOP} \label{sec:add}

We calculate the transit duration, $\tau_{t}$, of each of our transit model fits with the following equation \citep{Carter2008}:
\begin{equation}
	\tau_{t} = t_{egress}-t_{ingress},
\end{equation}
where $t_{egress}$ is the best-fitting model time of egress (4th contact), and $t_{ingress}$ is the best-fitting model time of ingress (1st contact). The error on $\tau_{t}$ is set to the $\sqrt{2}$ times the cadence of our observations \citep{Carter2009}.

\texttt{EXOMOP} performs an asymmetry test on each transit. We subtract each light curve by the mirror image of itself about the calculated mid-transit time. This same technique is used in \citet{Turner2013a} and \citet{Pearson2014} to search for asymmetries caused by a possible bow shock in TrES-3b and HAT-P-16b, respectively. This technique is useful for possible bow shock detection because bow shock models of WASP-12b \citep{Llama2011} and HD 189733b \citep{llama2013} predict a distinct asymmetry between the two halves of the transit (\citealt[see fig. 2]{Llama2011}; \citealt[see fig. 3]{llama2013}).

\subsection{EXOMOP model comparison}  \label{sec:exomop_compare}
Using artificial data, we perform several different comparison tests of \texttt{EXOMOP} with two different publicly-available modeling software packages: the Transit Analysis Package\footnote{http://ifa.hawaii.edu/users/zgazak/IfA/TAP.html} (\texttt{TAP}; \citealt{Mandel2002}; \citealt{Carter2009}; \citealt{Gazak2011}; \citealt{Eastman2013}) and \texttt{JKTEBOP}\footnote{http://www.astro.keele.ac.uk/jkt/codes/jktebop.html} \citep{Southworth2004a,Southworth2004b}. We also test if the errors we calculate using \texttt{EXOMOP} are reliable by comparing the errors to analytic estimates. 

We briefly discuss these two modeling packages below. \texttt{TAP} fits the transit light curves with a standard \citet{Mandel2002} model using Markov Chain Monte Carlo techniques and the parameter uncertainties are found with a wavelet likelihood function \citep{Carter2009}. \texttt{JKTEBOP} was adapted from the EBOP program written for eclipsing binary star systems (\citealt{Etzel1981}; \citealt{Popper1981}) and implements the Nelson-Davis-Etzel eclipsing binary model \citep{Nelson1972}. \texttt{JKTEBOP} is useful because it uses biaxial spheroids to model the planet \citep{Southworth2010} and therefore allows for departures from sphericity (whereas \texttt{TAP} is a spherical model). Therefore, any potentially observed non-spherical asymmetries in our data can be modeled with \texttt{JKTEBOP} to determine if the observed asymmetry is caused by a non-spherical planet instead of a bow shock. In addition, \texttt{JKTEBOP} uses a Monte Carlo simulation algorithm to compute errors (\citealt{Southworth2004a,Southworth2004b}; \citealt{Southworth2010}; \citealt{Hoyer2011}). 

We create a synthetic model transit using the analytic equations of \citet{Mandel2002} with a planet-to-star radius ($R_{p }/R_{*}$) = 0.1173,  the scaled semi-major axis ($a/R_{*}$) = 3.033, inclination ($i$) = 82.96$^{\degree}$, period ($P_{p}$) = 1.0914209 d, the linear limb darkening coefficient ($\mu{_1}$) = 0.61797203, the quadratic limb darkening coefficient ($\mu{_2}$) = 0.20813438, eccentricity ($e$) = 0, and argument of periastron ($\omega$) = 0$^{\circ}$. These parameters are chosen because they match the parameters of WASP-12b observed in the near-UV. Next, three sets of different white and red noise parameters are added to the synthetic \citet{Mandel2002} model to explore the effects of noise. The first set of models include only random Gaussian white noise with a standard deviation of 1, 2, 4, and 5 mmag. For the second and third set we create white noise and $1/f^{\alpha}$ red noise both with a standard deviation of 1 mmag where $\alpha$ is equal to 0.33 and 0.66, respectively. In total, we ran 6 models.

For the \texttt{EXOMOP} analysis we use 10000 iterations for the LM fit and 20 chains and 20$^{6}$ links for the DE-MCMC fit. With \texttt{TAP}, we model each transit individually using 5 chains with lengths of $10^{5}$ links each. \texttt{JKTEBOP} is implemented using the Monte Carlo algorithm and residual permutation method described in \citet{Southworth2008}. During the analysis for each model, the time of mid-transit (${T_{c}}$) and  $R_{p }/R_{*}$ are allowed to float. We only model these two parameters for the comparison tests because the errors on them are analytically tractable (see below; \citealt{Carter2008}). The ${i}$, $e$, $\omega$, $\mu{_1}$, $\mu{_2}$, $a/R_{*}$, and the $P_{p}$ of the planet are fixed. In addition, for \texttt{TAP} the white and red noise are left as free parameters. Since \texttt{TAP} does not automatically ensure chain convergence, we perform the Gelman-Rubin statistic (\citealt{Gelman1992}; \citealt{Ford2006}) manually to ensure convergence. In addition, \texttt{TAP} does not take into account the individual error bars on each transit point, whereas both the \texttt{EXOMOP} and \texttt{JKTEBOP} models do take them into account.

\begin{table*}
\centering
\caption{White Gaussian noise model tests with \texttt{EXOMOP}, \texttt{TAP}, and \texttt{JKTEBOP} using synthetic light curves}
\begin{tabular}{cccccccccccc}
\hline
\hline
Model	&
Noise            & 
$R_p/R_\ast$   &
Mid-transit & 
 Red$^{1}$  &
White$^{1}$ &
$\beta_{res2}$ &
$\beta_{res1}$ &
$\beta_{res2}$ &
$\beta_{res1}$ &
Red$^{2}$ &
White$^{2}$ \\ 
  &
 (mmag) & 
& 
(HJD)& 
 (mmag) & 
 (mmag)  &
R$_p$/R$_\ast$ &
R$_p$/R$_\ast$ &
Mid &
Mid &
(mmag)&
(mmag) \\
\hline 
\hline  
\texttt{TAP}		&1	&	0.11785$^{+0.00065}_{-0.00068}$ 	&	 0.00000$^{+0.00024}_{-0.00024}$ 	&	2.8$^{+1.7}_{-1.6}$ 		&	0.94$^{+0.03}_{-0.03}$ 	&  ---	&	---	&---		&---	&---		&---\\
\texttt{TAP}		&2	&	0.1168$^{+0.0012}_{-0.0012}$ 	&	-0.00027$^{+0.00045}_{-0.00044}$ &	3.8$^{+3.1}_{-2.5}$ 		&	1.96$^{+0.07}_{-0.06}$ 	&  ---	&	---	&---		&---&---		&---\\
\texttt{TAP}		&4	&	0.1170$^{+0.0027}_{-0.0027}$ 	&	-0.0007$^{+0.0010}_{-0.0010}$  	&	11.4$^{+7.0}_{-6.6}$ 	&	3.79$^{+0.13}_{-0.13}$  &  ---	&	---	&---	&---&---		&---	\\
\texttt{TAP}		&5	&	0.1165$^{+0.0036}_{-0.0034}$ 	&	-0.0007$^{+0.0012}_{-0.0012}$ 	&	14.6$^{+9.9}_{-8.8}$ 	&	5.06$^{+1.18}_{-0.18}$  &  ---	&	---	&---	&---	&---		&---\\
\hline
\texttt{JKTEBOP} &1	&     0.11775$^{+0.00025}_{-0.00025}$   & 0.00003$^{+0.00013}_{-0.00013}$       &    ---					& ---							  &		& 1.78 & ---	& 0.99&---		&--- \\
\texttt{JKTEBOP} &2	&     0.11743$^{+0.00052}_{-0.00052}$   &-0.00032$^{+0.00028}_{-0.00028}$       &    ---					&  ---							  & 		& 0.79 & --- 	& 0.88&---		&---\\
\texttt{JKTEBOP} &4	&	0.1174$^{+0.0010}_{-0.0010}$		&-0.00070$^{+0.00053}_{-0.00053}$	&    ---					& ---							  &		& 0.31& ---	& 0.80&---		&---\\
\texttt{JKTEBOP} &5	&	0.1160$^{+0.0013}_{-0.0013}$		&-0.00056$^{+0.00069}_{-0.00069}$	&    ---					& ---							  &		& 0.54& ---	& 0.79&---		&---\\
\hline		
\texttt{EXOMOP}	&1	&	0.11769$^{+0.00025}_{-0.00025}$ 	& 0.00001$^{+0.00013}_{-0.00013}$	&	0.0 	&	0.94 	& 0.97	& 0.58	&0.96	&0.85  &  0.00 	& 0.93$^{+0.33}_{-0.33}$ \\
\texttt{EXOMOP}	&2	&	0.11738$^{+0.00051}_{-0.00050}$ 	&-0.00031$^{+0.00028}_{-0.00028}$	&	0.0 	&	1.84	  	& 0.45	& 0.90	& 0.84	&0.88 & 0.00 	& 1.75$^{+0.70}_{-0.70}$ \\
\texttt{EXOMOP}	&4	&	0.1173$^{+0.0011}_{-0.0011}$ 	&-0.00067$^{+0.00059}_{-0.00057}$		&	0.0 	&	3.84	 	&0.91	&0.72	&0.93	&0.74 &0.00 	&  3.44$^{+1.60}_{-1.60}$\\
\texttt{EXOMOP}	&5	&	0.1159$^{+0.0013}_{-0.0013}$ 	& -0.00057$^{+0.00062}_{-0.00062}$	&	0.0 	&	5.12 	 &0.89	&0.65	&0.89	&0.76 &0.00	& 4.80$^{+2.22}_{-2.22}$ 	\\	
\hline			                 
\end{tabular}	
\vspace{-2em}
\tablenotetext{1}{The red and white noise are calculated using the wavelet likelihood technique \citep{Carter2009} described in Section \ref{sec:red_noise}. }
\tablenotetext{2}{The red and white noise are calculated using the Time-Averaging method \citep{Pont2006} described in Section \ref{sec:red_noise}.}
\label{tb:white_compare}
\end{table*}

\begin{table*}
\centering
\caption{Red noise model tests with \texttt{EXOMOP}, \texttt{TAP}, and \texttt{JKTEBOP} using synthetic light curves}
\begin{tabular}{cccccccccccc}
\hline
\hline
Model	&
$\alpha$            & 
R$_p$/R$_\ast$   &
Mid-transit & 
 Red$^{1}$  &
White$^{1}$ &
$\beta_{res2}$ &
$\beta_{res1}$ &
$\beta_{res2}$ &
$\beta_{res1}$ &
Red$^{2}$ &
White$^{2}$ \\ 
  &
Added  & 
& 
(HJD)& 
 (mmag) & 
 (mmag)  &
R$_p$/R$_\ast$ &
R$_p$/R$_\ast$ &
Mid &
Mid &
(mmag)&
(mmag) \\
\hline 
\hline  
\texttt{TAP}& 	0.66 & 0.1197$^{+0.0022}_{-0.0023}$& -0.00045$^{+0.00061}_{-0.00062}$		& 13.2$^{+1.4}_{-1.4}$	&0.618$^{+0.053}_{-0.059}$	&  ---	&	---	&---	&---	&---&---	\\
\texttt{TAP}& 	0.33& 0.1176$^{+0.0016}_{-0.0016}$ & -0.00054$^{+0.00046}_{-0.00047}$		& 9.1$^{+1.5}_{-1.4}$ 	& 0.856$^{+0.043}_{-0.043}$	&  ---	&	---	&---	&---	&---&---\\
\hline 
\texttt{JKTEBOP} & 0.66 &	0.11889$^{+0.00067}_{-0.00067}$ & -0.000036$^{+0.00028}_{-0.00028}$ &    ---& ---		  	&	2.59	& --- & ---	&2.18&---		&--- \\
\texttt{JKTEBOP}&0.33  &	0.11693$^{+0.00052}_{-0.00052}$	&-0.00038$^{+0.00022}_{-0.00022}$ & 	 ---& ---		  	&	1.99	& ---& ---      &1.61&---		&--- 	\\
\hline
\texttt{EXOMOP}&  0.66 & 0.1177$^{+0.0015}_{-0.0029}$ & -0.00004$^{+0.00029}_{-0.00029}$		& 0.46				 & 0.70		&  1.41	& 4.53		&1.43		&3.61     	& 0.55$^{+0.19}_{-0.19}$ & 1.41$^{+0.65}_{-0.65}$\\
\texttt{EXOMOP}&  0.33 & 0.11817$^{+0.00081}_{-0.00081}$ & -0.00018$^{+0.00040}_{-0.00059}$	& 0.16				 & 0.91	&  1.44	&1.67	  	&1.42	         &1.74      & 0.07$^{+0.34}_{-0.07}$& 1.53$^{+0.46}_{-0.46}$\\
\hline
\end{tabular}
\vspace{-2em}
\tablenotetext{1}{The red and white noise are calculated using the wavelet likelihood technique \citep{Carter2009} described in Section \ref{sec:red_noise}. }
\tablenotetext{2}{The red and white noise are calculated using the Time-Averaging method \citep{Pont2006} described in Section \ref{sec:red_noise}. }
\label{tb:red_compare}
\end{table*}

\begin{table*}
\centering
\caption{Parameters fixed for the light curve fitting using EXOMOP}
\begin{tabular}{ccccccc}
  \hline
  \hline
Planet	&
Period&
a/$R_\ast$	&
Inclination & 
Eccentricity &
Omega &
Source \\
 & 
 &
 &(\degree) &
 &(\degree)  
 &
\\  
  \hline
  \hline
   CoRoT-1b 	  	& 1.5089686	& 5.259		& 85.66	&0.071		&276.70	 &1	\\
 GJ436b		 	&2.6438986	&14.41		&86.774 	&0.15		&351	&2 \\
  HAT-P-1b 	  	& 4.46529976	& 9.853		& 85.634	& 0.00		&0.00	& 3	\\
  HAT-P-13b 	 	& 2.9162383	&  ---			& 81.93	& 0.00		&0.00	& 4	\\	
    HAT-P-16b 	  	 & 2.7759690	&7.17		& 86.6	& 0.034		&214	& 5 \\
 HAT-P-22b 	  	&  3.212220	& 8.55		& 86.90	& 0.016		&156.00	&  6	\\
  TrES-2b 			&  2.4706132	& 7.8957		&  83.8646& 0.0002		&143.13	& 7	\\
  TrES-4b 		& 3.5539268	& 6.08		& 82.81	& 0.00		&0.00	& 8	\\
  WASP-1b	&2.5199449 &	5.64			&88.65	& 0.00			&0.00 &9 \\
 WASP-12b 	& 1.09142166	& ---		& 82.72	& 0.0447		&274.44	&10	\\
  WASP-33b 	& 1.2198709	& 3.69		& 86.2	& 0.00		&0.00	& 11	\\	
   WASP-36b 	&1.5373653	&5.977		&83.61	&0.00		&0.00	&12\\	
  WASP-44b 	& 2.4238133	& 8.562		& 86.59	& 0.00		&0.00	&13	\\	
  WASP-48b 	& 2.143634	& 4.23		& 80.09	& 0.00		&0.00	&14	\\	
  WASP-77Ab & 1.3600309	& ---			& 89.4	& 0.00		&0.00	&15	 \\
  \hline
\end{tabular}
\vspace{-2em}
\tablerefs{(1) \citealt{Gillon2009}; (2) \citealt{Knutson2014}; (3) \citealt{Nikolov2014}; (4) \citealt{Southworth2012a}; (5) \citealt{Pearson2014}; (6) \citealt{Bakos2011}; (7) \citealt{Esteves2013}; (8) \citealt{Chan2011}; (9) \citealt{Maciejewski2014}; (10) \citealt{Sing2013}; (11) \citealt{Kovcs2013}; (12) \citealt{Smith2012}; (13) \citealt{Mancini2013}; (14) \citealt{Enoch2011}; (15) \citealt{Maxted2013} }
\label{tb:fit_pars}
\end{table*}

\begin{table*}
\centering
\caption{Limb darkening coefficients for the light curve fitting using \texttt{EXOMOP}}
\begin{tabular}{ccccccc}
\hline
\hline
Planet	&
Filter           & 
Linear coefficient$^{1}$ &
Quadratic coefficient$^{1}$   &
T$_{eff}$ [K]&
[Fe/H]	&
$\log{g}$ [cgs] \\
\hline   
\hline
   CoRoT-1b$^{a}$ 	&Bessell U 	&0.66547		&0.17302	& 5950	& -0.30	& 4.25\\
 GJ436b		&Bessell U	&0.926888	&-0.120646&3350$^{b}$	&-0.15$^{c}$     & 4.427$^{d}$		\\
  HAT-P-1b$^{e}$ 	&Bessell U 	&0.73417		&0.11238	&5980	&+0.130  &4.382	\\
  HAT-P-13b$^{f}$ 	& Bessell U 	&0.89273	       	&-0.02940&5653	& +0.410	& 4.130 	\\	
    HAT-P-16b$^{g}$ 	& Bessell U 	&0.65720		&0.17653 &6158	&+0.170	& 4.340  \\
 HAT-P-22b$^{h}$ 	&Bessell U 	&1.00392	       	&-0.14338&  5302	&+0.24	&4.36	\\
  TrES-2b$^{i}$ 		& Bessell U 	&0.74742		&0.10232	& 5850	&-0.15	& 4.427	\\
  TrES-4b$^{j}$ 		& Bessell U 	&0.61810		&0.20952	&6200	&0.140	& 4.064	\\
 WASP-1b 		& Bessell U 	&0.151543	&0.687788&6110$^{k}$	&0.26$^{d}$	&4.190$^{d}$		\\
"		& Harris B		&0.198846	&0.599307&6110	&0.26	&4.190\\
 WASP-12b$^{l}$  	&Harris R 		&0.61797		&0.20813 	&  6300 	& 0.30	& 4.38	\\
 "	&Bessell U  	&0.30070		&0.31983 	& 6300 	& 0.30	& 4.38\\
  WASP-33b$^{m}$  	& Bessell U 	&0.31668	 	&0.38643	& 7430	&+0.10	&4.30\\	
      "& Harris B 	&0.37146	 	&0.35168	& 7430	&+0.10	&4.30\\	
   WASP-36b$^{n}$  	&Harris R	 	&0.32106		&0.30131	&5880	& -0.31	&4.498\\	
   	"	&Bessell U	&0.70503		&0.13979 &5880	& -0.31	&4.498\\
  WASP-44b$^{o}$ 	& Bessell U 	&0.93916	       	&-0.06561 &5410 	& +0.06	&4.481	\\	
 "		& Harris V	&0.550120	&0.199928&5410 	& +0.06	&4.481 \\
	"	& Harris B	&0.758312	&0.0728504&5410 	& +0.06	&4.481 \\
  WASP-48b$^{p}$ 	&Bessell U 	&0.70217		&0.14181	&5920	&-0.12	& 4.03 \\	
  WASP-77Ab$^{q}$ & Bessell U 	&0.92696	       	&-0.06241&5500	&	0.00	&4.33	\\
\hline   
\end{tabular}
\vspace{-2em}
\tablenotetext{1}{The limb darkening coefficients are taken from \citet{Claret2011} and interpolated to the stellar parameters of their host star}
\tablerefs{(a) \citealt{Barge2008}; (b) \citealt{Moses2013}; (c) \citealt{Bean2006}; (d) \citealt{Torres2008}; (e) \citealt{Torres2008}; (f) \citealt{Bakos2009}; (g) \citealt{Buchhave2010}; (h)  \citealt{Bakos2011}; (i) \citealt{Torres2008}; (j) \citealt{Torres2008}; (k) \citealt{Simpson2011}; (l) \citealt{Hebb2009}; (m) \citealt{Cameron2010}; (n) \citealt{Smith2012}; (o) \citealt{Anderson2012}; (p) \citealt{Enoch2011}; (q)  \citealt{Maxted2013} }
\label{tb:fit_pars_limb}
\end{table*}

The results of the white noise analysis can be found in Table \ref{tb:white_compare} and the red noise analysis in Table \ref{tb:red_compare}. As expected, \texttt{EXOMOP} finds no red noise in the pure white noise tests and red noise in the red noise tests. In every case, the \texttt{EXOMOP} $R_{p }/R_{*}$ values are within 1$\sigma$ to the true $R_{p }/R_{*}$. We find that \texttt{TAP} overestimates the amount of red noise in every test we ran (by 2--14 $\sigma$) including the set of models with only white noise. Consequently, \texttt{TAP} is overestimating the error bars to their fitted parameters because of this excess red noise. Since both \texttt{EXOMOP} and \texttt{TAP} use the wavelet likelihood technique \citep{Carter2009} it is not clear why \texttt{TAP}  is overestimating the amount of red noise in these tests. Our results confirm the need to account for red noise using a variety of methods. Each of the methods used find red noise in the red noise tests but at slightly varying degrees. \citet{Turner2013a} and \citet{Hoyer2012} both conclude that \texttt{JKTEBOP} may be underestimating the errors in its transit fits when compared to \texttt{TAP}. However, neither of these studies conduct a thorough red and white noise test study. Therefore, we believe that \texttt{TAP} is overestimating the error bars in the fitted parameters compared to \texttt{JKTEBOP} and \texttt{EXOMOP} due its incorrect red noise calculation. The \texttt{EXOMOP} and \texttt{JKTEBOP} results are in very good agreement with each other.

To get an general idea if the error estimation in \texttt{EXOMOP} is behaving as expected, we compare our white noise tests (Table \ref{tb:white_compare}) to analytic estimations for the uncertainty in the flux drop, $\delta = (R_{p}/R_{*})^{2}$, and mid-transit time. 
\citet{Carter2008} derive an analytic estimate for the 1$\sigma$ uncertainty in $\delta$ ($\sigma_{\delta}$) to be
\begin{equation}
\sigma_{\delta} =  \frac{\sigma_{g}}{\sqrt{n}}, \label{eq:sigma_d}
\end{equation}
where $\sigma_{g}$ are the Gaussian errors in the relative flux (in our case the noise added) and $n$ is the number of data points. Additionally, the analytic estimate of the 1$\sigma$ uncertainty ($\sigma_{t}$) in the mid-transit time is \citep{Carter2008}:
\begin{equation}
\sigma_{t} =  \frac{\sigma_{g}}{\sqrt{n} \delta}  \left( \tau_{t} - \tau\right) \sqrt{\frac{\tau}{2(\tau_{t} - \tau)}}, \label{eq:sigma_t}
\end{equation}
 where $\tau$ is the ingress/egress duration. Limb darkening and red noise cause the error estimation in equations (\ref{eq:sigma_d}) and (\ref{eq:sigma_t}) to increase \citep{Seager2011}. The error estimations we find using \texttt{EXOMOP} have the same behavior as the analytic estimates by \citet{Carter2008} exactly for both $\sigma_{\delta}$ and $\sigma_{t}$. For example, if the noise doubles in our white noise tests then the error estimates on $R_{p}/R_{*}$ also double (Table \ref{tb:white_compare}). The \texttt{JKTEBOP} error bars also mimic this analytic behavior but the \texttt{TAP} error bars do not. Due to this result we believe the error estimation in \texttt{EXOMOP} is reliable.

\subsection{EXOMOP analysis of the systems}\label{sec:EXOMOP_analysis}

Each individual transit is modeled with \texttt{EXOMOP} using 10000 iterations for the LM model and 20 chains and 20$^{6}$ links for the DE-MCMC model. During the analysis $T_{c}$ and $R_{p }/R_{*}$ are always left as free parameters for each transit. We then systematically fit every combination with $a/R_{*}$, $i$, ${T_{c}}$, and $R_{p }/R_{*}$ set as free parameters. The BIC is used to assess over-fitting of the data and the model that produces the lowest BIC value is always chosen. For every planet except HAT-P-13b, WASP-12b, WASP-44b, and WASP-77Ab the BIC is higher when fitting for $a/R_{*}$ and $i$. The $a/R_{*}$, $e$, $\omega$, $i$, and $P_{p}$ of each of the planets are fixed to their values listed in Table \ref{tb:fit_pars}. The linear and quadratic limb darkening coefficients in each filter are taken from \citet{Claret2011} and interpolated to the stellar parameters of the host stars (see Table \ref{tb:fit_pars_limb}) using the \texttt{EXOFAST} applet\footnote{http://astroutils.astronomy.ohio-state.edu/exofast/limbdark.shtml}\citep{Eastman2013}. In addition, a linear or quadratic least squares fit is modeled to the OoT baseline simultaneously with the \cite{Mandel2002} model. The BIC is also used to determine whether to include a linear or quadratic OoT baseline fit in the best-fit model. 

The fitted parameters from either the LM or DE-MCMC best-fitting model that produce the highest error bars are reported. In every case both models find results within 1$\sigma$ of each other. The light curve parameters obtained from the \texttt{EXOMOP} analysis and the derived transit durations are summarized in Tables \ref{tb:1_light}--\ref{tb:5_light}. The modeled light curves can be found in Figs. \ref{fig:light_1}--\ref{fig:light_7}. The physical parameters for our targets are derived as outlined in Section \ref{sec:physical_properites} (Tables \ref{tb:parms_1}--\ref{tb:parms_2}). A thorough description of the modeling and results of each system can be found in Section \ref{sec:indiv_systems}. We also perform the asymmetry test (described in Section \ref{sec:add}) for each transit to search for any non-spherical asymmetries. 

\begin{table*}
\centering
\caption{Parameters derived in this study for the CoRoT-1b, GJ436b, HAT-P-1b, HAT-P-16b, HAT-P-22b, TrES-2b, TrES-4b, and WASP-1b light curves using \texttt{EXOMOP} }
\begin{tabular}{cccccc}
\hline
\hline
Planet 	&	CoRoT-1b 	&	GJ436b 	&	GJ436b 	&	GJ436b 	&	HAT-P-1b 	\\
Date 	&	  2012 December 06 	&	2012 March 23 	&	2012 April 07 	&	All 	&	2012 October 02 	\\
Filter$^{1}$ 	&	U 	&	U 	&	U 	&	U 	&	U 	\\
$T_{c}$ (HJD-2450000) 	&	6268.98963$^{+0.00070}_{-0.0013}$ 	&	6009.8889$^{+0.0019}_{-0.0020}$ 	&	6025.7322$^{+0.0073}_{-0.0068}$ 	&	--- 	&	6203.64907$^{+0.00084}_{-0.00095}$  	\\
R$_p$/R$_\ast$ 	&	0.1439$^{+0.0020}_{-0.0018}$ 	&	0.0930$^{+0.0083}_{-0.0048}$ 	&	0.0703$^{+0.0099}_{-0.0071}$ 	&	0.0758$^{+0.0086}_{-0.0075}$ 	&	0.1189$^{+0.0010}_{-0.0014}$ 	\\
Duration (min)   	&	149.9$^{+1.9}_{-1.9}$ 	&	59.6$^{+2.5}_{-2.5}$ 	&	58.30$^{+1.45}_{-1.45}$ 	&	59.55$^{+1.07}_{-1.07}$   	&	172.12$^{+0.95}_{-0.95}$ 	\\
$\beta_{res2}^{a}$  (R$_p$/R$_\ast$) 	&	0.74 	&	0.79 	&	1.17	&	1.14 	&	1.35 	\\
$\beta_{res2}^{a}$ (Mid) 	&	0.74 	&	0.75	&	1.15	&	--- 	&	1.30 	\\
$\beta_{res1}^{b}$  ($R_{p}/R_{\ast}$) 	&	$+1.04\thinspace -0.70$ 	&	$+1.92  \thinspace-0.59$ 	&	$+2.35\thinspace -1.67$ 	&	$+2.64\thinspace -2.28$ 	&	$+1.33\thinspace -1.94$ 	\\
$\beta_{res1}^{b}$  (Mid) 	&	   $+0.70  \thinspace-2.02$ 	&	$+1.16  \thinspace-1.48$ 	&	$+2.15\thinspace -2.00$ 	&	--- 	&	+2.13  \thinspace-2.43 	\\
$\beta_{time}^{c}$ 	&	   1.00 	&	1.00 	&	1.03 	&	1.01 	&	1.03 	\\
White Noise$^{d}$ (mmag) 	&	  2.47$^{+1.04}_{-1.04}$ 	&	1.87$^{+0.66}_{-0.66}$ 	&	2.52$^{+1.41}_{-1.41}$ 	&	2.86$^{+1.38}_{-1.38}$ 	&	1.50$^{+0.64}_{-0.64}$ 	\\
Red Noise$^{d}$ (mmag) 	&	   0.00 	&	0.00 	&	0.66$^{+1.05}_{-0.66}$ 	&	0.40$^{+0.75}_{-0.40}$ 	&	0.35$^{+0.29}_{-0.29}$ 	\\
$\beta_{wavelet}^{e}$ 	&	  1.01 	&	1.004	&	1.03 	&	1.01 	&	1.02   	\\
White Noise$^{f}$ (mmag) 	&	  3.47  	&	2.19 	&	2.08 	&	1.95 	&	  1.28   	\\
 Red Noise$^{f}$ (mmag) 	&	  0.00    	&	0.20	&	0.49 	&	0.32 	&	0.25 	\\
OoT Baseline Function  	&	None	&	None 	&	None 	&	--- 	&	 None \\	
\hline 											
Planet 	&	HAT-P-16b 	&	HAT-P-16b 	&	HAT-P-22b 	&	HAT-P-22b	&	HAT-P-22b	\\
Date 	&	2013 November 02 	&	All 	&	2013 February 22      	&	2013 March 22	&	All	\\
Filter$^{1}$ 	&	U 	&	U 	&	U     	&	U	&	U	\\
$T_{c}$ (HJD-2450000) 	&	6598.79110$^{+0.00060}_{-0.00059}$ 	&	--- 	&	6346.8144$^{+0.0013}_{-0.0014}$ 	&	6738.70864$^{+0.00061}_{-0.00063}$  	&	---	\\
R$_p$/R$_\ast$ 	&	0.1115$^{+0.0011}_{-0.0011}$ 	&	0.10645$^{+0.00067}_{-0.00067}$ 	&	0.1151$^{+0.0021}_{-0.0022}$    	&	0.1072$^{+0.0013}_{-0.0012}$    	&	0.10797$^{+0.00086}_{-0.00094}$    	\\
Duration (min) 	&	185.44$^{+1.28}_{-1.28}$ 	&	181.78$^{+3.06}_{-3.06}$ 	&	172.46$^{+1.66}_{-1.66}$     	&	172.89$^{+2.38}_{-2.38}$  	&	170.50$^{+3.05}_{-3.05}$  	\\
$\beta_{res2}$$^{a}$  (R$_p$/R$_\ast$) 	&	0.67	&	1.08	&	0.50        	&	0.71	&	1.18	\\
$\beta_{res2}$$^{a}$ (Mid) 	&	0.88	&	--- 	&	0.51     	&	0.83	&	---	\\
$\beta_{res1}$$^{b}$  (R$_{p}/R_{\ast}$) 	&	$+0.16\thinspace -0.24$ 	&	$+0.29\thinspace -0.38$ 	&	0.10\thinspace -0.70     	&	0.23\thinspace -0.15    	&	0.53\thinspace -0.56    	\\
$\beta_{res1}$$^{b}$  (Mid) 	&	$+0.75  \thinspace-0.54$ 	&	--- 	&	0.59  \thinspace-0.53     	&	0.53  \thinspace-0.32	&	---	\\
$\beta_{time}$$^{c}$ 	&	1.00 	&	1.00 	&	1.00    	&	1.00	&	1.00	\\
White Noise$^{d}$ (mmag) 	&	1.23$^{+0.61}_{-0.61}$ 	&	0.97$^{+0.25}_{-0.25}$ 	&	2.23$^{+0.65}_{-0.65}$    	&	1.11$^{+0.52}_{-0.52}$    	&	1.19$^{+0.36}_{-0.36}$    	\\
Red Noise$^{d}$ (mmag)   	&	0.00 	&	0.00 	&	0.00     	&	0.00	&	0.00	\\
$\beta_{wavelet}$$^{e}$ 	&	1.00 	&	1.00 	&	1.00           	&	1.00	&	1.00	\\
White Noise$^{f}$ (mmag) 	&	2.07  	&	0.86	&	3.06     	&	2.07	&	1.42	\\
Red Noise$^{f}$ (mmag) 	&	0.00 	&	0.00 	&	0.20	&	0.00	&	0.00	\\
OoT Baseline Function  	&	Linear 	&	--- 	&	None	&	Linear	&	---	\\
\hline 											
Planet				&	TrES-2b    	&	TrES-4 	&	TrES-4b 	&	WASP-1b 	&	WASP-1b 	\\
Date 				&	  2012 October 29 	&	2011 July 26	&	2011 July 26 	&	2013 September 19 	&	2013 September 19 	\\
Filter$^{1}$ 			&	U    			&	R 		&	U 	&	U 	&	B 	\\
$T_{c}$ (HJD-2450000) 	&	6230.5980$^{+0.00059}_{-0.00060}$ 	&	5769.7536$^{+0.0040}_{-0.0040}$ 	&	5769.7532$^{+0.0036}_{-0.0037}$ 	&	6555.9381$^{+0.0038}_{-0.0025}$ 	&	6555.9393$^{+0.0027}_{-0.0027}$ 	\\
R$_p$/R$_\ast$ 		&	0.1243$^{+0.0022}_{-0.0024}$   	&	0.0880$^{+0.0055}_{-0.0055}$      	&	0.1094$^{+0.0052}_{-0.0052}$ 	&	0.0938$^{+0.0023}_{-0.0023}$ 	&	0.1018$^{+0.0040}_{-0.0040}$ 	\\
Duration (min) 		&	106.68$^{+1.19}_{-1.19}$   	&	205.35$^{+2.73}_{-2.73}$ 	&	216.96$^{2.64}_{2.64}$ 	&	219.49$^{+3.15}_{-3.15}$ 	&	224.17$^{+3.18}_{-3.18}$ 	\\
$\beta_{res2}$$^{a}$  (R$_p$/R$_\ast$) 	&	0.72	&	1.6	&	2.23 	&	1.28 	&	1.89 	\\
$\beta_{res2}$$^{a}$ (Mid) 	&	0.97	&	1.46	&	2.25 	&	1.19 	&	1.68 	\\
$\beta_{res1}$$^{b}$  (R$_p$/R$_\ast$) 	&	+0.26\thinspace -0.27      	&	+0.54\thinspace -0.85	&	+0.67\thinspace -0.74 	&	+0.43\thinspace -0.58 	&	+0.45\thinspace -0.67 	\\
$\beta_{res1}$$^{b}$  (Mid) 	&	  +0.32  \thinspace-0.86   	&	+1.30  \thinspace-1.24 	&	+1.14  \thinspace-1.15 	&	+1.96\thinspace -1.28 	&	+1.03\thinspace -1.10 	\\
$\beta_{time}$$^{c}$ 	&	1.00 	&	1.00    	&	1.00 	&	1.01 	&	1.00 	\\
White Noise$^{d}$ (mmag) 	&	1.67$^{+0.53}_{-0.53}$   	&	1.78$^{+1.05}_{-1.05}$ 	&	3.79$^{+1.40}_{-1.40}$ 	&	3.12$^{+1.76}_{-1.76}$ 	&	2.09$^{+0.90}_{-0.90}$ 	\\
Red Noise$^{d}$ (mmag)   	&	0.00       	&	0.00 	&	0.00 	&	0.32$^{+0.32}_{-0.32}$ 	&	0.00 	\\
$\beta_{wavelet}$$^{e}$ 	&	1.00 	&	1.00 	&	  1.00 	&	1.00 	&	1.00 	\\
White Noise$^{f}$ (mmag) 	&	2.23	&	3.08	&	3.32 	&	2.37 	&	1.69 	\\
Red Noise$^{f}$ (mmag) 	&	0.00 	&	0.001  	&	0.001 	&	0.00 	&	0.00 	\\
OoT Baseline Function  	&	Linear 	&	 Quad	&	Linear	&	Linear 	&	Linear 	\\
\hline 											
 \end{tabular}
 \vspace{-2em}
\tablenotetext{1}{Filter: B is the Harris B (330--550 nm), R is the Harris R (550--900 nm), V is the Harris V (473--686 nm) and U is the Bessell U (303--417 nm)  }
\tablenotetext{a}{$\beta_{res2}$ is found by using the second residual permeation method (Section \ref{sec:red_noise}) }
\tablenotetext{b}{$\beta_{res1}$ is found by using the first residual permeation method (Section \ref{sec:red_noise})}
\tablenotetext{c}{$\beta_{time}$ is the scaling factor for the Time-Averaging method \citep{Pont2006} (Section \ref{sec:red_noise}) }
\tablenotetext{d}{The red and white noise are calculated using the Time-Averaging method \citep{Pont2006} (Section \ref{sec:red_noise}) }
\tablenotetext{e}{$\beta_{wave}$ is the scaling factor for the wavelet likelihood technique \citep{Carter2009} (Section \ref{sec:red_noise}) }
\tablenotetext{f}{The red and white noise are calculated using the wavelet likelihood technique \citep{Carter2009} (Section \ref{sec:red_noise})}
\label{tb:1_light}
\end{table*}

\begin{table*}
\centering
\caption{Parameters derived in this study for the WASP-1b, WASP-33b, WASP-36b, and WASP-44b light curves using \texttt{EXOMOP}  }
\begin{tabular}{cccccc}
\hline
 \hline
Planet 	&	WASP-1b 	&	WASP-1b 	&	WASP-1b 	&	WASP-1b 	&	WASP-33b \\
Date 	&	2013 October 22 	&	2013 October 22 	&	All 	&	All 	&	2012 December 01 \\
Filter$^{1}$           	&	U 	&	B 	&	U 	&	B 	&	U \\
$T_{c}$ (HJD-2450000) 	&	6588.69666$^{+0.00090}_{-0.00082}$ 	&	6588.6961$^{+0.0008}_{-0.0012}$ 	&	--- 	&	--- 	&	6263.8434$^{+0.0022}_{-0.0029}$ \\
R$_p$/R$_\ast$                         	&	0.09630$^{+0.00092}_{-0.00092}$ 	&	0.10096$^{+0.00097}_{-0.00097}$ 	&	0.0964$^{+0.0010}_{-0.00010}$ 	&	0.1013$^{+0.0018}_{-0.0018}$ 	&	0.1125$^{+0.0047}_{-0.0097}$ \\
Duration (min) 	&	222.01$^{+3.22}_{-3.22}$ 	&	213.71$^{+3.22}_{-3.22}$ 	&	223$^{+1.83}_{-1.83}$ 	&	224.17$^{+1.83}_{-1.83}$ 	&	164.77$^{+2.15}_{-2.15}$ \\
$\beta_{res2}$$^{a}$  (R$_p$/R$_\ast$) 	&	1.03 	&	1.26 	&	1.41 	&	1.73 	&	2.81 \\
$\beta_{res2}$$^{a}$ (Mid) 	&	1.00 	&	1.20 	&	--- 	&	-- 	&	2.70 \\
$\beta_{res1}$$^{b}$ (R$_{p}$/R$_\ast$) 	&	+0.69\thinspace -0.81 	&	+0.81\thinspace -0.30 	&	+0.64\thinspace -0.57 	&	+1.63\thinspace -0.73 	&	+3.88\thinspace -8.40 \\
$\beta_{res1}$$^{b}$ (Mid) 	&	+1.53\thinspace -1.40 	&	+1.50\thinspace -2.20 	&	--- 	&	--- 	&	+3.60 \thinspace-4.55\\
$\beta_{time}$$^{c}$ 	&	1.03 	&	1.06 	&	1.002 	&	1.01 	&	  1.10  \\
White Noise$^{d}$ (mmag) 	&	1.71$^{+0.96}_{-0.96}$ 	&	1.12$^{+0.70}_{-0.70}$ 	&	1.28$^{+0.72}_{-0.72}$ 	&	1.81$^{+0.94}_{-0.94}$ 	&	  2.73$^{+1.31}_{-1.31}$ \\
Red Noise$^{d}$ (mmag) 	&	0.39$^{+0.62}_{-0.39}$ 	&	0.41$^{+0.30}_{-0.30}$ 	&	0.17$^{+0.90}_{-0.17}$ 	&	0.29$^{+0.33}_{-0.29}$ 	&	  1.23 $^{+2.82}_{-1.23}$ \\
$\beta_{wave}$$^{e}$ 	&	1.00 	&	1.04 	&	1.004 	&	1 	&	  1.54  \\
Wavelet White Noise$^{f}$ (mmag) 	&	1.32 	&	0.79 	&	1.91 	&	1.36 	&	1.28 \\
Wavelet Red Noise$^{f}$ (mmag) 	&	0.00 	&	0.22 	&	0.18 	&	0.00 	&	  1.52 \\
OoT Baseline Function  	&	None	&	 Linear 	&	 ---  	&	 --- 	&	Linear\\
\hline										
   Planet 	&	WASP-33b 	&	WASP-33b 	&	WASP-33b 	&	WASP-36b 	&	WASP-36b \\
Date 	&	2012 December 01 	&	2012 October 01 	&	All 	&	2012 December 29 	&	2013 March 15 \\
Filter$^{1}$           	&	B 	&	U 	&	U 	&	R 	&	U \\
$T_{c}$ (HJD-2450000) 	&	6263.8419$^{+0.0036}_{-0.0076}$ 	&	6202.84778$^{+0.00067}_{-0.00069}$ 	&	--- 	&	6290.86129$^{+0.00034}_{-0.00026}$ 	&	6367.72898$^{+0.00095}_{-0.00058}$ \\
R$_p$/R$_\ast$                         	&	0.1127$^{+0.0054}_{-0.0056}$ 	&	0.1017$^{+0.0027}_{-0.0027}$ 	&	0.1086$^{+0.0022}_{-0.0007}$ 	&	0.13850$^{+0.00071}_{-0.00082}$ 	&	0.1316$^{+0.0018}_{-0.0018}$ \\
Duration (min) 	&	167.31$^{+2.15}_{-2.15}$ 	&	165.50$^{+0.71}_{-0.71}$ 	&	166.94$^{+0.55}_{-0.55}$ 	&	109.46$^{+0.72}_{-0.72}$ 	&	107.96$^{+1.41}_{-1.41}$ \\
$\beta_{res2}$$^{a}$  (R$_p$/R$_\ast$) 	&	1.44 	&	1.21 	&	2.33 	&	1.24 	&	0.80 \\
$\beta_{res2}$$^{a}$ (Mid) 	&	1.197 	&	1.16 	&	--- 	&	1.18 	&	0.75 \\
$\beta_{res1}$$^{b}$ (R$_p$/R$_\ast$) 	&	+0.85\thinspace -1.67 	&	2.34\thinspace -2.70 	&	+6.95\thinspace -3.68 	&	+0.95\thinspace -0.41 	&	+0.31\thinspace -0.25 \\
$\beta_{res1}$$^{b}$ (Mid) 	&	+3.31\thinspace -3.33 	&	+4.11 \thinspace-2.70 	&	--- 	&	+1.59\thinspace -0.50 	&	+0.68\thinspace -0.43 \\
$\beta_{time}$$^{c}$ 	&	1.05 	&	1.01 	&	1.02 	&	1.002 	&	1.00 \\
White Noise$^{d}$ (mmag) 	&	6.45$^{+3.50}_{-3.50}$ 	&	3.03$^{+1.31}_{-1.31}$ 	&	3.21$^{+1.35}_{-1.35}$ 	&	1.99$^{+0.86}_{-0.86}$ 	&	2.30$^{+1.02}_{-1.02}$ \\
Red Noise$^{d}$ (mmag) 	&	1.96$^{+1.71}_{-1.71}$ 	&	0.28$^{+0.28}_{-0.28}$ 	&	0.67$^{+0.24}_{-0.24}$ 	&	0.13$^{+0.68}_{-0.13}$ 	&	0.00 \\
$\beta_{wave}$$^{e}$ 	&	1.05 	&	1.39   	&	1.14 	&	1.00 	&	1.00 \\
 White Noise$^{f}$ (mmag) 	&	5.13 	&	1.18 	&	1.84 	&	1.80 	&	3.47 \\
 Red Noise$^{f}$ (mmag) 	&	1.58 	&	1.13 	&	  1.00 	&	0.01 	&	0.00 \\
OoT Baseline Function 	&	None	&	Quadratic 	&	 --- 	&	 Linear 	&	 Linear\\
\hline
\end{tabular}
\vspace{-2em}
\tablenotetext{1}{Filter: B is the Harris B (330--550 nm), R is the Harris R (550--900 nm), V is the Harris V (473--686 nm) and U is the Bessell U (303--417 nm)  }
\tablenotetext{a}{$\beta_{res2}$ is found by using the second residual permeation method (Section \ref{sec:red_noise}) }
\tablenotetext{b}{$\beta_{res1}$ is found by using the first residual permeation method (Section \ref{sec:red_noise})}
\tablenotetext{c}{$\beta_{time}$ is the scaling factor for the Time-Averaging method \citep{Pont2006} (Section \ref{sec:red_noise}) }
\tablenotetext{d}{The red and white noise are calculated using the Time-Averaging method \citep{Pont2006} (Section \ref{sec:red_noise}) }
\tablenotetext{e}{$\beta_{wave}$ is the scaling factor for the wavelet likelihood technique \citep{Carter2009} (Section \ref{sec:red_noise}) }
\tablenotetext{f}{The red and white noise are calculated using the wavelet likelihood technique \citep{Carter2009} (Section \ref{sec:red_noise})}
\label{tb:2_light}
\end{table*}

\begin{table*}
\centering
\caption{Parameters derived in this study for the WASP-48b light curve using \texttt{EXOMOP}  }
\begin{tabular}{cccccc}
\hline
 \hline

\hline 										
Planet 	&	WASP-48b 								\\
Date 	&	  2011 October 09 								\\
Filter$^{1}$           	&	U 								\\
$T_{c}$ (HJD-2450000) 	&	5844.7249$^{+0.0019}_{-0.0017}$ 		\\						
R$_p$/R$_\ast$                         	&	0.0916$^{+0.0017}_{-0.0017}$ 	\\							
Duration (min) 	&	192.20$^{+1.73}_{-1.73}$ 					\\			
$\beta_{res2}$$^{a}$  (R$_p$/R$_\ast$) 	&	1.16 					\\			
$\beta_{res2}$$^{a}$ (Mid) 	&	1.19 							\\	
$\beta_{res1}$$^{b}$ (R$_p$/R$_\ast$) 	&	+0.11\thinspace -0.98 	\\							
$\beta_{res1}$$^{b}$ (Mid) 	&	+1.35  \thinspace-1.24 			\\					
$\beta_{time}$$^{c}$ 	&	1.00 								\\
White Noise$^{d}$ (mmag) 	&	2.16$^{+0.96}_{-0.96}$ 			\\					
Red Noise$^{d}$ (mmag) 	&	0.00 								\\
$\beta_{wave}$$^{e}$ 	&	1.00 								\\
 White Noise$^{f}$ (mmag) 	&	2.22 							\\	
 Red Noise $^{f}$(mmag) 	&	0.02 								\\
 OoT Baseline Function 	&	 None							\\	
\hline
\end{tabular}
\vspace{-2em}
\tablenotetext{1}{Filter: B is the Harris B (330--550 nm), R is the Harris R (550--900 nm), V is the Harris V (473--686 nm) and U is the Bessell U (303--417 nm)  }
\tablenotetext{a}{$\beta_{res2}$ is found by using the second residual permeation method (Section \ref{sec:red_noise}) }
\tablenotetext{b}{$\beta_{res1}$ is found by using the first residual permeation method (Section \ref{sec:red_noise})}
\tablenotetext{c}{$\beta_{time}$ is the scaling factor for the Time-Averaging method \citep{Pont2006} (Section \ref{sec:red_noise}) }
\tablenotetext{d}{The red and white noise are calculated using the Time-Averaging method \citep{Pont2006} (Section \ref{sec:red_noise}) }
\tablenotetext{e}{$\beta_{wave}$ is the scaling factor for the wavelet likelihood technique \citep{Carter2009} (Section \ref{sec:red_noise}) }
\tablenotetext{f}{The red and white noise are calculated using the wavelet likelihood technique \citep{Carter2009} (Section \ref{sec:red_noise})}
\label{tb:3_light}
\end{table*}

\begin{table*}
\centering
\caption{Parameters derived in this study for the HAT-P-13b, WASP-12b, and WASP-44b light curves using \texttt{EXOMOP}}
\begin{tabular}{ccccc}
\hline
\hline
Parameter	&
Value             & 
Value &
 Value  &
 Value \\
 \hline
 \hline
Planet 							&	HAT-P-13b 					&	 WASP-12b 						&	 WASP-12b 						&	 WASP-12b 		\\
Date 							&	 2013 March 02 				&	  2011 November 15 					&	2011 November 15 					&	  2012 March 22 		\\
Filter$^{1}$           					&	 U 							&	   U 								&	R 								&	U 		\\
$T_{c}$ (HJD-2450000) 				&	6354.6974$^{+0.0014}_{-0.0014}$	&	   5881.98375$^{+0.00047}_{-0.00078}$	&	5881.98229$^{+0.00080}_{-0.00080}$ 	&	6009.67929$^{+0.00060}_{-0.00057}$ 		\\
R$_p$/R$_\ast$   					&	0.0850$^{+0.0022}_{-0.0014}$ 		&	   0.11963$^{+0.00082}_{-0.00082}$ 		&	0.1153$^{+0.0016}_{-0.0016}$  		&	0.12313$^{+0.00087}_{-0.00087}$ 		\\
a/R$_\ast$   						&	5.280$^{+0.065}_{-0.065}$ 		&	3.189$^{+0.021}_{-0.021}$ 				&	3.057$^{+0.052}_{-0.051}$ 			&	 3.202$^{+0.025}_{-0.036}$ 		\\
Duration (min) 					&	202.44$^{+1.38}_{-1.38}$ 			&	  176.44$^{+3.08}_{-3.08}$ 			&	180.57$^{+3.08}_{-3.08}$			 	&	179.58$^{+1.39}_{-1.39}$ 		\\
$\beta_{res2}$$^{a}$  (R$_p$/R$_\ast$) 	&	1.48 							&	 0.99 							&	1.50 	&	0.70 		\\
$\beta_{res2}$$^{a}$ (Mid) 			&	1.35 							&	 0.91 							&	1.46 	&	0.70 		\\
$\beta_{res2}$$^{a}$ (a/R$_\ast$) 		&	1.348 						&	 0.91 							&	1.47 	&	0.73 		\\
$\beta_{res1}$$^{b}$ (R$_p$/R$_\ast$) 	&	$+1.55  \thinspace-0.63$ 			&	  $+0.49\thinspace -1.01$ 				&	 $+1.20\thinspace -1.26$ 	&	$+0.31\thinspace -0.71$ 		\\
$\beta_{res1}$$^{b}$ (Mid) 			&	+0.98  \thinspace-1.35 			&	  $+0.64  \thinspace-1.70$ 			&	 $+1.23  \thinspace-1.99$ 	&	$+0.23 \thinspace-0.89$ 		\\
$\beta_{res1}$$^{a}$  (a/R$_\ast$)  		&	+0.87 $ \thinspace-0.99$ 			&	  $+1.29\thinspace$ -0.80 				&	 $+1.75\thinspace -2.01$ 	&	$+1.30\thinspace -0.78$ 		\\
$\beta_{time}$$^{c}$ 				&	1.00 							&	  1.00  	&	1.00 	&	1.00 		\\
White Noise$^{d}$ (mmag) 			&	1.67$^{+0.46}_{-0.46}$ 	&	  1.59$^{+0.74}_{-0.74}$ 	&	1.42$^{+0.65}_{-0.65}$ 	&	 2.31$^{+0.65}_{-0.65}$ 		\\
Red Noise$^{d}$ (mmag) 				&	 0.00 	&	  0.00 	&	 0.00 	&	  0.00   		\\
$\beta_{wave}$$^{e}$ 				&	 1.00 	&	  1.00   	&	 1.00   	&	 1.00    		\\
 White Noise$^{f}$ (mmag) 			&	 1.20 	&	  1.22 	&	   1.11 	&	1.62 		\\
 Red Noise$^{f}$ (mmag) 				&	 0.00 	&	  0.00  	&	  0.00 	&	0.00 		\\
 OoT Baseline Function 				&	 None 	&	Linear 	&	Linear 	&	Linear 		\\
 \hline										\\
Planet 							&	 WASP-12b 					&	WASP-12b 						&	WASP-12b 					&	WASP-44b 		\\
Date 							&	 2012 October 02 				&	 2012 November 30 					&	 All 							&	  2011 October 13 		\\
Filter$^{1}$           					&	U 							&	      U 							&	 U 							&	U 		\\
$T_{c}$ (HJD-2450000) 				&	6202.95339$^{+0.00045}_{-0.00055}$ &	6262.88831$^{+0.00068}_{-0.00068}$ 	&	--- 							&	5848.8477$^{+0.0013}_{-0.0013}$ 		\\
R$_p$/R$_\ast$   					&	0.11660$^{+0.00077}_{-0.00077}$ 	&	0.1193$^{+0.0014}_{-0.0014}$  		&	0.12016$^{+0.00076}_{-0.00065}$ 	&	0.1228$^{+0.0028}_{-0.0028}$ 		\\
a/R$_\ast$   						&	3.096$^{+0.023}_{-0.046}$ 		&	3.313$^{+0.046}_{-0.051}$ 			&	 3.217$^{+0.038}_{-0.026}$ 	&	8.31$^{+0.30}_{-+0.30}$    		\\
Duration (min) 					&	 179.43$^{+1.43}_{-1.43}$ 		&	170.78$^{+1.31}_{-1.31}$ 				&	171.26$^{+2.17}_{-2.17}$ 	&	135.81$^{+1.60}_{-1.60}$ 		\\
$\beta_{res2}$$^{a}$  (R$_p$/R$_\ast$) 	&	0.97 							&	1.14 	&	1.87  	&	1.15 		\\
$\beta_{res2}$$^{a}$ (Mid) 			&	0.92 	&	1.02 	&	--- 	&	1.07 		\\
$\beta_{res2}$$^{a}$ (a/R$_\ast$) 		&	0.92 	&	1.02 	&	1.27 	&	1.25		\\
$\beta_{res1}$$^{b}$ (R$_p$/R$_\ast$) 	&	$+0.77\thinspace -0.41$ 	&	$+1.01\thinspace -1.00$ 	&	$+1.00\thinspace -1.40$ 	&	+0.25\thinspace -0.28 		\\
$\beta_{res1}$$^{b}$ (Mid) 			&	$+0.92  \thinspace-1.23$ 	&	$+0.95  \thinspace-0.91$ 	&	--- 	&	0.86 \thinspace-0.72 		\\
$\beta_{res1}$$^{a}$  (a/R$_\ast$)  		&	$+0.56  \thinspace-1.97$ 	&	$+1.15\thinspace -1.16$ 	&	$+2.03\thinspace -1.61$ 	&	0.87  \thinspace-0.53 		\\
$\beta_{time}$$^{c}$ 				&	1.02 	&	 1.00 	&	1.00 	&	1.00 		\\
White Noise$^{d}$ (mmag) 			&	2.07$^{+0.91}_{-0.91}$ 	&	 3.35$^{+1.32}_{-1.32}$ 	&	2.51 $^{+1.31}_{-1.31}$ 	&	5.05$^{+1.70}_{-1.70}$ 		\\
Red Noise$^{d}$ (mmag) 				&	 0.15$^{+0.15}_{-0.15}$ 	&	 0.00 	&	0.00 	&	  0.00 		\\
$\beta_{wave}$$^{e}$ 				&	 1.00 	&	1.00 	&	 1.00 	&	1.00 		\\
Wavelet White Noise$^{f}$ (mmag) 		&	1.56 	&	3.23 	&	2.06 	&	5.25  		\\
Wavelet Red Noise$^{f}$ (mmag) 		&	0.001 	&	 0.00 	&	0.00 	&	0.00 		\\
OoT Baseline Function 				&	 Linear 	&	 Linear 	&	--- 	&	Linear		\\										\\
\hline
\end{tabular}
\vspace{-2em}
\tablenotetext{1}{Filter: B is the Harris B (330--550 nm), R is the Harris R (550--900 nm), V is the Harris V (473--686 nm) and U is the Bessell U (303--417 nm)  }
\tablenotetext{a}{$\beta_{res2}$ is found by using the second residual permeation method (Section \ref{sec:red_noise}) }
\tablenotetext{b}{$\beta_{res1}$ is found by using the first residual permeation method (Section \ref{sec:red_noise})}
\tablenotetext{c}{$\beta_{time}$ is the scaling factor for the Time-Averaging method \citep{Pont2006} (Section \ref{sec:red_noise}) }
\tablenotetext{d}{The red and white noise are calculated using the Time-Averaging method \citep{Pont2006} (Section \ref{sec:red_noise}) }
\tablenotetext{e}{$\beta_{wave}$ is the scaling factor for the wavelet likelihood technique \citep{Carter2009} (Section \ref{sec:red_noise}) }
\tablenotetext{f}{The red and white noise are calculated using the wavelet likelihood technique \citep{Carter2009} (Section \ref{sec:red_noise})}
\label{tb:4_light}
\end{table*}

\begin{table*}
\centering
\caption{Parameters derived in this study for the WASP-44b and WASP-77Ab light curves using \texttt{EXOMOP}}
\begin{tabular}{cccc}
\hline
\hline
Parameter	&
Value             & 
Value &
 Value  \\
 Value \\
 \hline
 \hline
Planet 	&	WASP-44b 	&	WASP-44b 	&	WASP-77Ab 				\\
Date 	&	2013 October 19 	&	2013 October 09 	&	 2012 December 06  				\\
Filter$^{1}$           	&	B 	&	V 	&	U 				\\
$T_{c}$ (HJD-2450000) 	&	6585.68580$^{+0.00063}_{-0.00063}$ 	&	6585.68618$^{+0.00077}_{-0.00053}$ 	&	6271.65804$^{+0.00032}_{-0.00035}$ 			\\
R$_p$/R$_\ast$   	&	0.1236$^{+0.0018}_{-0.0019}$ 	&	0.1164$^{+0.0017}_{-0.0017}$ 	&	0.12612$^{+0.00098}_{-0.00094}$ 			\\
a/R$_\ast$   	&	8.59$^{+0.11}_{-+0.12}$    	&	8.33$^{+0.09}_{-+0.14}$    	&	5.396$^{+0.054}_{-0.054}$ 				\\
Duration (min) 	&	126.21$^{+2.73}_{-2.73}$ 	&	129.10$^{+2.84}_{-2.84}$ 	&	129.67$^{+1.61}_{-1.61}$ 				\\
$\beta_{res2}$$^{a}$  (R$_p$/R$_\ast$) 	&	1.04 	&	1.23	&	1.69 				\\
$\beta_{res2}$$^{a}$ (Mid) 	&	1.04 	&	1.26	&	1.55 				\\
$\beta_{res2}$$^{a}$ (a/R$_\ast$) 	&	1.01	&	1.21	&	1.47 				\\
$\beta_{res1}$$^{b}$ (R$_p$/R$_\ast$) 	&	+0.11\thinspace -0.14 	&	+0.92\thinspace -0.75	&	0.36  \thinspace-0.53 				\\
$\beta_{res1}$$^{b}$ (Mid) 	&	+0.58\thinspace -0.49 	&	+1.62\thinspace -1.03 	&	1.07  \thinspace-0.87 				\\
$\beta_{res1}$$^{a}$  (a/R$_\ast$)  	&	+0.44\thinspace -0.32 	&	+0.96\thinspace -1.77 	&	1.00  \thinspace-1.33 				\\
$\beta_{time}$$^{c}$ 	&	1.00 	&	1.00 	&	1.00 				\\
White Noise$^{d}$ (mmag) 	&	2.18$^{+1.06}_{-1.06}$ 	&	2.04$^{+1.02}_{-1.02}$ 	&	1.53 $^{+0.51}_{-0.51}$ 				\\
Red Noise$^{d}$ (mmag) 	&	0.00 	&	0.00 	&	0.00 				\\
$\beta_{wave}$$^{e}$ 	&	1.00 	&	1.00 	&	1.00 				\\
 White Noise$^{f}$ (mmag) 	&	2.45 	&	2.18 	&	1.34 				\\
 Red Noise$^{f}$ (mmag) 	&	0.00 	&	0.00 	&	0.00     				\\
 OoT Baseline Function 	&	Linear	&	Linear	&	 Linear				\\
\hline
\end{tabular}
\vspace{-2em}
\tablenotetext{1}{Filter: B is the Harris B (330--550 nm), R is the Harris R (550--900 nm), V is the Harris V (473--686 nm) and U is the Bessell U (303--417 nm)  }
\tablenotetext{a}{$\beta_{res2}$ is found by using the second residual permeation method (Section \ref{sec:red_noise}) }
\tablenotetext{b}{$\beta_{res1}$ is found by using the first residual permeation method (Section \ref{sec:red_noise})}
\tablenotetext{c}{$\beta_{time}$ is the scaling factor for the Time-Averaging method \citep{Pont2006} (Section \ref{sec:red_noise}) }
\tablenotetext{d}{The red and white noise are calculated using the Time-Averaging method \citep{Pont2006} (Section \ref{sec:red_noise}) }
\tablenotetext{e}{$\beta_{wave}$ is the scaling factor for the wavelet likelihood technique \citep{Carter2009} (Section \ref{sec:red_noise}) }
\tablenotetext{f}{The red and white noise are calculated using the wavelet likelihood technique \citep{Carter2009} (Section \ref{sec:red_noise})}
\label{tb:5_light}
\end{table*}

\begin{table*}
\centering
\caption{Physical Properties of CoRoT-1b, GJ436b, HAT-P-1b, HAT-P-13b, HAT-P-22b, TrES-2b, TrES-4b, and WASP-1b derived from the light curve modeling}
\begin{tabular}{ccccccc}
\hline
\hline
Parameter (units)	&
Value			&	
Source			&
Value			&
Source			&
Value		&	
Source			\\
\hline 
\hline 				
Planet 					&CoRoT-1b					& --	&GJ436b							&	---	&HAT-P-1b					&---\\
M$_{b}$ (M$_{Jup}$)  		& 1.07$\pm$0.17				&1	&0.0728$\pm$0.0024				&	1	&0.529$\pm$0.020		  	 	&1	 	\\
Near-UV R$_p$/R$_\ast$		&0.1439$^{+0.0020}_{-0.0018}$	&1	&0.0758$^{+0.0086}_{-0.0075}$		&	1	&0.1189$^{+0.0010}_{-0.0015}$	&1 \\
Optical R$_p$/R$_\ast$		&0.1381$^{+0.0007}_{-0.0015}$	&2	&0.08310$\pm$0.00027				&	3	&0.11802$\pm$0.00018			&4 \\
Near-UV R$_{b}$ (R$_{Jup}$)	&  1.48$\pm$0.13				&1	&0.342$\pm$0.041					&	1	&1.358$\pm$0.036			 	& 1		\\
Optical R$_{b}$ (R$_{Jup}$)	&  1.42$\pm$0.24				& 2	&0.3739$\pm$0.0097				&	3	&1.319$\pm$0.019				& 4	\\
$\rho_{b}$ ($\rho_{Jup}$)		&0.33$\pm$0.10				&1	&1.30$\pm$0.11					&	1	&0.269$\pm$0.040				& 1 		\\
$\log{g_{b}}$ (cgs)			&3.12$\pm$0.20				&1	&3.180$\pm$0.032					&1		&2.912$\pm$0.048				& 1	\\
 T$^{'}_{eq}$ (K)			&1834$\pm$46					&2	&686 $\pm$10						&3	& 1322$\pm$15				& 4		 \\
H (km)					& 705$\pm$320					& 1  & 230	$\pm$17						&1	&1008$\pm$73						&1	\\
 $\Theta$					&      0.039$\pm$0.013			&1	&0.0267$\pm$0.0015				&1	&0.0403$\pm$0.0032				 & 1\\
 Orbital inclination			&	85.66$^{+0.62}_{-0.48}$		& 2	&86.774$\pm$	0.030				&3	& 85.634$\pm$0.056			& 4	\\
 Orbital eccentricity			&	0.071$^{+0.62}_{-0.48}$        	&2	&0.150$\pm$0.012					&3	& 0.00					 	 & 4		\\
  a (au)					& 0.0259$^{+0.0011}_{-0.0020}$	& 2	&0.03109$\pm$0.00074				&3	&0.05561$\pm$0.00083		  &4		\\
 Period (d)					&1.508976552$\pm$0.000000097	& 1	&2.64389788$\pm$0.00000010		&1	&4.4652968$\pm$0.0000018	  &1	\\
 $T_{c}(0)$ (BJD)			&2454138.303971$\pm$0.000036	&1	&2454238.479958$\pm$0.000039		&1	&2453979.93165$\pm$0.00025 &1	\\
  \hline
 Planet 					&HAT-P-13b				&---		&HAT-P-16b					& ---	&HAT-P-22b					&--- \\
 M$_{b}$ (M$_{Jup}$) 		&0.906$\pm$0.023			&1		&4.189$\pm$0.092				&  1	&2.148$\pm$0.062				&1\\
 Near-UV R$_p$/R$_\ast$		&0.0850$^{+0.0022}_{-0.0014}$&1		    &0.10645$\pm$0.00067		&1    &0.1079$\pm$0.00094	 		&1 \\
Optical R$_p$/R$_\ast$		&0.0870498$\pm$0.0024		&5		&0.1071$\pm$0.0014			&7	&0.1065$\pm$0.0017			&9 \\
 Near-UV R$_{b}$ (R$_{Jup}$)	&1.452$\pm$0.052			&1		&1.28$\pm$0.056 				&  1	&1.092$\pm$0.047				&1\\
 Optical R$_{b}$ (R$_{Jup}$)	&1.487$\pm$0.038			&5		&1.190$\pm$0.035				&  7	&1.080$\pm$0.058				&9\\
 $\rho_{b}$ ($\rho_{Jup}$)	&0.272$\pm$0.021			&1		&1.86$\pm$0.24					&  1	& 1.61$\pm$0.21				&1\\
 $\log{g_{b}}$ (cgs)			&3.008$\pm$0.032			&1		&3.858$\pm$0.053				&  1	&3.691$\pm$0.063				&1	\\
 T$^{'}_{eq}$ (K)			&1740$\pm$27				&1		&1571$\pm$21					&  7	&1463$\pm$19					&9\\
 H (km)					&863$\pm$65				&1		& 109$\pm$13					& 1	&150$\pm$22					&1	\\
 $\Theta$					&0.0405$\pm$0.0023		&1		&0.237$\pm$0.017				&  1	&0.186$\pm$0.017				&1	\\		
 Orbital inclination ($\degree$)	&81.93$\pm$0.26			&5		&87.74$\pm$0.59				&  7	&86.9$^{+0.6}_{-0.5}$			&9	\\
  Orbital eccentricity			&0.0133$\pm$0.0041		&6		&0.034$\pm$0.003.				&  8	&0.016$\pm$0.009				&9\\
   a (au)					&0.0431$\pm$0.0012			&1		&0.04130$\pm$0.00047			&  7	&0.0414$\pm$0.0005			&9\\
 Period (d)				&2.9162382$\pm$0.0000016	&1		&2.775970244$\pm$0.00000066	& 1	&3.2122312$\pm$0.0000012     		&1\\	
$T_{c}(0)$ (BJD)			&2455176.53864$\pm$0.00023&1		&2455027.592939$\pm$0.00019	& 1	&2454930.22296$\pm$0.00025		&1	\\
\hline
Planet						&TrES-2b						&---  		&TrES-4b						&---		&   WASP-1b					& ---\\	
 M$_{b}$ (M$_{Jup}$)  			& 1.44$\pm$0.21				&	10 	& 	0.917$\pm$0.070			& 1		&0.846$\pm$0.054				& 1\\
Near-UV R$_p$/R$_\ast$			&0.1243$\pm$0.0024		 	&1		&0.1094$^{+0.0052}_{-0.0052}$			&	1	&0.0964$^{+0.0010}_{-0.00010}$	&1 \\
Optical R$_p$/R$_\ast$			&0.125358$^{+0.000019}_{-0.000024}$&10	&0.09745$\pm$0.00076			&	11	&0.1013$\pm$0.0018			&1 \\
 Near-UV R$_{b}$ (R$_{Jup}$)		&1.215$\pm$0.049			&	1 	 &	1.91$\pm$0.11				& 1		&1.379$\pm$0.033				& 1 \\
 Optical R$_{b}$ (R$_{Jup}$)		&1.245$^{+0.045}_{-0.041}$		&	10 	&	1.706$\pm$0.056			& 11		&1.449$\pm$0.041				& 1 \\
 $\rho_{b}$ ($\rho_{Jup}$)			& 1.82$\pm$0.23			&	10 	&   0.173$\pm$0.022					& 1		&0.26022$\pm$0.028		 	& 1	\\
$\log{g_{b}}$ (cgs)				& 3.798$\pm$0.046				&	10 	&   2.89$\pm$0.055				& 1		&2.998$\pm$0.039				& 1	\\
 T$^{'}_{eq}$ (K)				& 1472$\pm$12					&	10 	&	1778$\pm$22				& 11		& 1812$\pm$14				& 1 \\
H (km)						& 118$\pm$12					&      1	&1373$\pm$167				&1		& 920$\pm$82					&1 \\
$\Theta$						& 0.216$\pm$0.020				&	10 	&   0.0393$\pm$0.0038			& 1		&0.0366$\pm$0.0034			& 1	\\
 a (au)						&0.0367$^{+0.0013}_{-0.0012}$	&	10 	&	0.05084$\pm$0.00050		&  11	&0.03889$^{+0.00053}_{-0.00073}$& 12	\\
Orbital inclination ($\degree$)		&83.8646$^{+0.0041}_{-0.0036}$	&	10 	&	82.81$\pm$0.37			&  11	&88.65$\pm$0.55				& 13	\\
Orbital eccentricity				&0  0.0002$^{+0.0010}_{-0.0002}$	&	10 	&	0						&  11	&0							& 13	\\
Period (d)					&2.4706132$\pm$0.0000001		&	10 	&3.5539246$\pm$0.0000014 		&  1		&2.51994529$\pm$0.00000056 	& 1	 \\
$T_{c}(0)$ (BJD)				&2454969.39661$\pm$0.0048		&	10 	&2454223.79850$\pm$0.00032	&  1		&2453912.51504 $\pm$0.00035	& 1	\\
\hline 
 \end{tabular}
 \vspace{-2em}
\tablerefs{(1) Our Study; (2) \citealt{Gillon2009}; (3) \citealt{Knutson2014}; (4) \citealt{Nikolov2014}; (5) \citealt{Southworth2012a}; (6) \citealt{Winn2010}; (7) \citealt{Ciceri2013}; (8) \citealt{Husnoo2012}; (9) \citealt{Bakos2011}; (10) \citealt{Barclay2012}; (11) \citealt{Chan2011}; (12) \citealt{Maciejewski2014}; (13) \citealt{Stempels2007}}
\label{tb:parms_1}
\end{table*}

\begin{table*}
\centering
\caption{Physical Properties of WASP-12b, WASP-33b, WASP-36b, WASP-44b, WASP-48b, and WASP-77Ab derived from the light curve modeling}
\begin{tabular}{ccccccc}
\hline
\hline
Parameter (units)	&
Value			&	
Source			&
Value			&
Source			&
Value		&	
Source	\\
\hline
\hline
Planet						&   WASP-12b					& ---	&  WASP-33b					& ---	&	WASP-36b				&---			\\	
 M$_{b}$ (M$_{Jup}$)  			&2.01$\pm$0.14				& 1	&3.28$\pm$0.73				& 1	&2.286$\pm$0.066				&1				\\
 Near-UV R$_p$/R$_\ast$			&0.12016$^{+0.00076}_{-0.00065}$ &1	&0.1086$^{+0.0022}_{-0.0007}$	&1	&0.1316$^{+0.0018}_{-0.0018}$		&1 \\
  Optical R$_p$/R$_\ast$			&0.1173$\pm$0.0005			&2	&0.1143$\pm$0.0002			&3	&0.13850$^{+0.00071}_{-0.00082}$	&1 \\
 Near-UV R$_{b}$ (R$_{Jup}$)		&1.835$\pm$0.08					& 1 	&1.594$^{+0.043}_{-0.043}$		& 1	&1.218$\pm$0.028				&1		\\
 Optical R$_{b}$ (R$_{Jup}$)		&1.860$\pm$0.090				& 2 &1.679$^{+0.019}_{-0.030}$		& 3	&1.281$\pm$0.026				&1		\\
 $\rho_{b}$ ($\rho_{Jup}$)			&0.326$\pm$ 0.049	 			& 1	&0.65$\pm$0.14				& 1	&1.017$\pm$0.068				&1				\\
$\log{g_{b}}$ (cgs)				&3.210$\pm$0.057				& 1	& 3.459$\pm$0.098				& 1	&3.538$\pm$0.028				&1		\\
 T$^{'}_{eq}$ (K)				& 2483$\pm$79				& 1	&2723$\pm$37					& 4	&1724$\pm$39					&5	\\
H (km)						&773$\pm$103					&1	&477$\pm$108					&1	&252$\pm$17					&1\\
$\Theta$						&0.0389$\pm$0.0055			& 1	&0.065$\pm$0.015				& 1	&0.0905$\pm$0.0047			&1	\\
 a (au)						&0.0235$\pm$0.0011				& 1	&0.0259$^{+0.0005}_{-0.0005}$	&3 &0.02643$\pm$0.00026			&5	\\
Orbital inclination ($\degree$)		&82.96$^{+0.50}_{-0.44}$	& 2	&86.2$\pm$0.2					&3	&83.61$\pm$0.21 				&5	\\
Orbital eccentricity				&0.0447$\pm$0.0043				& 2&0							& 3	&0							&5		\\
Period (d)					&1.09142119$\pm$0.00000021	& 1		&1.21987016$\pm$0.00014		& 1	&1.53736423$\pm$0.00000057 	&1	 \\
$T_{c}(0)$ (BJD)				&2455147.45820$\pm$0.00013	& 1	&24552984.82964$\pm$0.00030	& 1	&2455569.83817$\pm$0.00010	&1	\\
\hline 
Planet						& WASP-44b						& --- &WASP-48b 			&---		&WASP-77Ab						&--- \\
M$_{b}$ (M$_{Jup}$)  			& 0.867$\pm$0.064					&1	&0.984$\pm$0.085			&1		&1.76$\pm$0.057				&1\\	
Near-UV R$_p$/R$_\ast$			&0.1228$\pm$0.0028				&1	&0.0916$\pm$0.0017 		&1		&0.1305$^{+0.0010}_{-0.0010}$		&1$^{a}$ \\
Optical R$_p$/R$_\ast$			&0.1164$\pm$0.0017				&1	&0.0980$\pm$0.0010		&7		&0.13012$\pm$0.00065			&8 \\
Near-UV R$_{b}$ (R$_{Jup}$)		& 1.03$\pm$0.038					&1	&1.560$\pm$0.088		&1		&1.21$\pm$0.02				&1	\\
Optical R$_{b}$ (R$_{Jup}$)		&0.98$\pm$0.032					&1	&1.67$\pm$0.10					&7		&1.21$\pm$0.02					&8	\\
$\rho_{b}$ ($\rho_{Jup}$)			&0.86$\pm$0.11					&1	&0.198$\pm$0.039			&1		&0.928$\pm$0.055				&1		\\
$\log{g_{b}}$ (cgs)				&3.35$\pm$0.05					&1	&2.941$\pm$0.092			&1		&3.471$\pm$0.022				&1		\\
 T$^{'}_{eq}$ (K)				&1304$\pm$36						&6	&2035$\pm$52				&7		&1674$\pm$24					&1	\\
H (km)						&292$\pm$32						&1	&1178$\pm$415			&1		&	286$\pm$59		&1	\\
$\Theta$						&0.0664$\pm$0.0068				&1	&0.0340$\pm$0.0046		&1		&0.0694$\pm$0.0043				&1	\\
 a (au)						&0.03443$\pm$0.00099				&6	&0.0344$\pm$0.0026		&7		&0.02396$\pm$0.00043				&1\\
Orbital inclination ($\degree$)		&86.59							&6	&80.09$\pm$0.55 			&7		&89.40$\pm$0.7					&8\\
Orbital eccentricity				&0								&6	&0						&7		&0								&8	\\
Period (d)					&2.4238120$\pm$0.0000012			&1 	&2.14363592$\pm$0.0000046		&1		&1.3600306$\pm$0.0000012			&1	\\
$T_{c}(0)$ (BJD)				& 2455434.37655$\pm$0.00020		&1	&2455364.55217$\pm$0.00020&1	&2455870.44977$\pm$0.00014	&1	\\
\hline
\end{tabular}	
 \vspace{-2em}
\tablerefs{(1) Our Study; (2) \citealt{Maciejewski2013}; (3) \citealt{Kovcs2013}; (4) \citealt{Cameron2010}; (5) \citealt{Smith2011}; (6) \citealt{Anderson2012}; (7) \citealt{Enoch2011}; (8) \citealt{Maxted2013}}
\tablenotetext{a}{The near-UV R$_p$/R$_\ast$ of WASP-77Ab is corrected for the dilution of the companion stars (Section \ref{sec:wasp77b}) }
\label{tb:parms_2}
\end{table*}

\section{Calculated Physical Properties of the Systems}\label{sec:physical_properites}


We use the results of our light curve modeling with \texttt{EXOMOP} to calculate the planetary and geometrical parameters of our targets(mass, radius, density, surface gravity, equilibrium temperature, Safronov number, atmospheric scale height). The physical parameters of all our systems can be found in Tables \ref{tb:parms_1}--\ref{tb:parms_2}. The planetary mass, $M_{p}$, can be calculated using the following equation (\citealt{Winn2010b}; \citealt{Seager2011}):
\begin{equation}
M_{p} =  \left( \frac{\sqrt{1-e^2}}{28.4329} \right) \left( \frac{ K_{*} } {\sin{i}} \right) \left( \frac{P_{b}}{1 yr} \right)^{1/3} \left( \frac{M_{*}}{\msun} \right)^{2/3} M_{jup},
\end{equation}
where K$_{*}$ is the radial velocity amplitude of the host star and $P_{p}$ is the orbital period of the planet. We adopt the formula by \citet{Southworth2007a} to calculate the surface gravitational acceleration, $g_{p}$:
\begin{equation}
g_{p} = \frac{2 \pi} {P_{p}} \left(\frac{a}{R_{p}} \right)^{2} \frac{ \sqrt{1-e^{2} }} { \sin{i} } K_{*}. 
\end{equation}
The equilibrium temperature, $T_{eq}$, is derived using the relation \citep{Southworth2010}: 
\begin{equation}
T_{eq} = T_{eff} \left( \frac{1-A}{4 F} \right)^{1/4} \left(  \frac{R_{*}} {2 a } \right)^{1/2},
\end{equation}
where $T_{eff}$ is the effective temperature of the host star, $A$ is the Bond albedo, and $F$ is the heat redistribution factor. This formula is simplified by making the assumption, as done in \citet{Southworth2010}, that $A = 1- 4F$; the resulting equation is the modified equilibrium temperature, T$^{'}_{eq}$: 
\begin{equation}
T^{'}_{eq} = T_{eff} \left(  \frac{R_{*}} {2 a} \right)^{1/2}.
\end{equation}
The Safronov number, $\Theta$, is a measure of the ability of a planet to gravitationally scatter or capture other bodies in nearby orbits \citep{Safronov1972}. We calculate $\Theta$ using the equation from \citet{Southworth2010}: 
\begin{equation}
\Theta = \frac{M_{p} a}{M_{*} R_{p}}. 
\end{equation}
Differences between Safronov numbers could point to differences in migration or stopping mechanisms \citep{Seager2011}. As defined by \citet{Hansen2007}, Class I hot Jupiters have $\Theta$ = 0.07$\pm$0.01 and Class II have $\Theta$ = 0.04$\pm$0.01. However, \citet{Southworth2012c} find that this devision of hot Jupiters into two classes was not evident when using a greater sample of planets. The atmospheric scale height, $H,$ is calculated using \citep{deWit2013}
\begin{equation}
H = \frac{k_{B} T^{'}_{eq} }{\mu g_{b}},
\end{equation}
where $k_{B}$ is Boltzmann's constant and $\mu$ is the mean molecular weight in the planet's atmosphere (set to 2.3; \citealt{deWit2013}).

\vspace{-1.5em}
\subsection{Period Determination}\label{sec:period}

By combining our \texttt{EXOMOP} derived mid-transit times with previously published mid-transit times, we refine the orbital period of our targets. When necessary, the mid-transit times are transformed from HJD, which is based on UTC time, into BJD, which is based on Barycentric Dynamical Time (TDB), using the online converter\footnote{http://astroutils.astronomy.ohio-state.edu/time/hjd2bjd.html} by \citet{Eastman2010}.  We derive an improved ephemeris for each target by performing a weighted linear least-squares analysis using the following equation:
\begin{equation}
			T_{c} =  T_{c}(0) + P_{p}\times{E},	
\end{equation}
where $T_{c}(0)$ is the mid-transit time at the discovery epoch in $BJD_{TDB}$, $P_{p}$ is the orbital period of the target, and $E$ is the integer number of cycles after their discovery paper. See Tables \ref{tb:parms_1}--\ref{tb:parms_2} for an updated T$_{c}(0)$ and P$_{p}$ for each system. The results of the transit timing analysis for all our targets can be found in Table \ref{tb:mr_timing} (the entire table can be found online).

\begin{table*}
\centering
\caption{Results of the transit timing analysis$^{1}$}
\begin{tabular}{ccccccc}
\hline
\hline
 Planet Name            	& 
$T_{c}$ ($BJD_{TDB}$) 	&
$T_{c}$ error (d)			&
Epoch				&
O-C (d)				&
O-C error (d)			&
Source 				\\
\hline
\hline
CoRoT-1b				&	 2456268.990397	& 0.00013 & 1412	& 0.0000060		&0.000135 		& This paper \\
CoRoT-1b				&	2454138.328594	&0.00039	 & 0		& 0.000222		&0.000392		& \citealt{Csizmadia2010}\\
\hline
\end{tabular}
\vspace{-2em}
\tablenotetext{1}{This table is available in its entirety in machine-readable form in the online journal.  A portion is shown here for guidance regarding its form and content. }	
\label{tb:mr_timing}	
\end{table*}
 \section{Individual Systems} \label{sec:indiv_systems}

\subsection{CoRoT-1b}

CoRoT-1b is the first transiting exoplanet discovered by the CoRoT satellite (\citealt{Baglin2003}; \citealt{Barge2008}). Several follow-up primary transit photometry studies of the system find no signs of a changing period (\citealt{Bean2009}; \citealt{Gillon2009}; \citealt{Csizmadia2010}; \citealt{Rauer2010}; \citealt{Southworth2011}; \citealt{Sada2012}; \citealt{Ranjan2014}). CoRoT-1b's atmosphere may have a temperature inversion (\citealt{Snellen2009}; \citealt{Alonso2009}; \citealt{Rogers2009}; \citealt{Gillon2009}; \citealt{Zhao2012}) or an isothermal profile \citep{Deming2011}. Infrared transmission spectroscopy observations by \citet{Schlawin2014} disfavor a TiO/VO-rich spectrum for CoRoT-1b, suggesting the temperature inversion is caused by another absorber in the atmosphere or that flat spectrum is due to clouds or a haze layer. \citet{Pont2010} observed the Rossiter-McLaughlin effect \citep{Winn2011} for this planet and found that the projected spin-orbit angle is not aligned with the stellar spin axis with $\lambda = 77^{\circ}\pm11^{\circ}$. The Rossiter-McLaughlin effect is important because planets that are not orbiting coplanar with their host stars may exhibit bow shock variability (See Section \ref{sec:bow_var}, \citealt{Vidotto2011c}; \citealt{llama2013}).

We observed CoRoT-1b on 2012 December 07 using the U filter (Table \ref{tb:obs_new}), which is the first published near-UV light curve of this planet (Fig.~\ref{fig:light_1}). Our derived physical parameters (Table \ref{tb:parms_1}) agree with previous studies and reduce the uncertainty on the period by a factor of 5 compared to \citet{Gillon2009}. We also find a near-~UV R$_p$/R$_\ast$ = 0.1439$^{+0.0020}_{-0.0018}$ which is 2.3$\sigma$ larger that its optical R$_p$/R$_\ast$ = 0.1381$^{+0.0007}_{-0.0015}$ \citep{Gillon2009}. An early near-UV or any non-spherical asymmetries are not seen in this transit of CoRoT-1b.

\subsection{GJ436b}
GJ436b, a hot Neptune, was discovered through radial velocity measurements \citep{Butler2004} and later confirmed to be a transiting exoplanet \citep{Gillon2007b}. There have been extensive ground-based and space-based photometry and spectral studies of the GJ436b (e.g. \citealt{Maness2007}; \citealt{Deming2007}; \citealt{Demory2007}; \citealt{Gillon2007a,Gillon2007b}; \citealt{Alonso2008}; \citealt{Bean2008,Bean2008b}; \citealt{Coughlin2008}; \citealt{Ribas2008}; \citealt{Southworth2008,Southworth2010}; \citealt{Shporer2009}; \citealt{Figueira2009}; \citealt{Caceres2009}; \citealt{Ballard2010}; \citealt{Gibson2010}; \citealt{Pont2010}; \citealt{Knutson2011,Knutson2014}; \citealt{Ehrenreich2011}; \citealt{Shabram2011}; \citealt{Stevenson2012}; \citealt{Line2013}; \citealt{Moses2013}; \citealt{Gaidos2014}; \citealt{Kulow2014}; \citealt{Knutson2014}; \citealt{Lanotte2014}). The host star is found to be inactive (e.g. \citealt{Bean2006}; \citealt{Wright2007}; \citealt{Torres2007}; \citealt{Madhusudhan2009}; \citealt{Ballerini2012}; \citealt{Braun2012}; \citealt{Albrecht2012}; \citealt{Kislyakova2013}) and there are two other transiting planets in the system (\citealt{Ribas2008}; \citealt{Ballard2010}; \citealt{Stevenson2012}; \citealt{Knutson2013}). The host star being inactive reduces the possibility of bow shock variability in our near-UV observations (\citealt{Vidotto2011c}; \citealt{llama2013}). In our sample, GJ436b has the lowest planetary mass, is the only hot Neptune, and the only planet orbiting an M-dwarf. \\
\indent We observed the first near-UV light curve of GJ436b on 2012 March 23 and subsequently on 2012 April 07 (Table \ref{tb:obs_new}, Fig.~\ref{fig:light_1}). The light curves obtained for this object are noisy because the observations are reaching the precision limit for the 1.55-m Kuiper telescope due to the small transit depth and the faintness of the M-dwarf in the near-UV (these observations are $\sim 2 \times$ the photon limit). However, there are no non-spherical asymmetries in the near-UV lights of GJ436b. Our physical parameters (Table \ref{tb:parms_1}) and light curve solution (Table \ref{tb:1_light}) are consistent with previous studies. We find a near-UV R$_p$/R$_\ast$ = 0.0758$^{+0.0086}_{-0.0075}$ which is consistent within 1$\sigma$ of its optical R$_p$/R$_\ast$ = 0.08310$\pm$0.00027 \citep{Knutson2014}.

\subsection{HAT-P-1b}
HAT-P-1b is the first planet discovered by the HATNet project (\citealt{Bakos2002}; \citealt{Bakos2007}) and the planet orbits one of the stars in a visual binary (\citealt{Bakos2007}; \citealt{Liu2014}). There have been many follow-up transit observations of HAT-P-1b (e.g. \citealt{Winn2007}; \citealt{Johnson2008}; \citealt{Todorov2010}; \citealt{Sada2012}; \citealt{Wilson2015}). Secondary eclipse measurements by \citet{Beky2013} found a $2\sigma$ upper limit of 0.64 for HAT-P-1b's geometric albedo between 577 and 947 nm. \citet{Nikolov2014} found a conclusive detection of both sodium and water in the transmission spectra using the Space Telescope
Imaging Spectrograph onboard the \textit{HST}. Rossiter-McLaughlin effect measurements of the system found that HAT-P-1b is aligned (3.7$^{\circ}\pm2.1^{\circ}$) with the host star's equator (\citealt{Johnson2008}). HAT-P-1b has the longest orbital period of all the planets in our study. 

The first near-UV light curve of HAT-P-1b was observed on 2012 October 02 (Table \ref{tb:obs_new}, Fig.~\ref{fig:light_1}). The stellar binary was used as the main reference star in our light curve analysis since the two stars are nearly identical in their stellar parameters (\citealt{Bakos2007}; \citealt{Liu2014}) and will experience similar variations due to the atmosphere (the stars are only separated by 11$\arcsec$). Our light curve solution (Table \ref{tb:1_light}) and derived planetary parameters (Table \ref{tb:parms_1}) agree with previous studies. We find a near-UV R$_p$/R$_\ast$ = 0.1189$^{+0.0010}_{-0.0015}$ which is within 1$\sigma$ of the optical R$_p$/R$_\ast$ =  0.11802$\pm$0.00018 \citep{Nikolov2014}. We do not observe an early ingress or any non-spherical asymmetries in the light curve of HAT-P-1b.

\subsection{HAT-P-13b}
HAT-P-13b, is an inflated hot Jupiter in a nearly circular orbit (\citealt{Bakos2009}) and the system also has a massive outer planet (M$_{p,c} \sin{i_{c}}$ = 14.3 $M_{Jup}$ ; \citealt{Winn2010}; \citealt{Knutson2013}), on a highly eccentric orbit \citep{Bakos2009}. Follow-up photometry studies have refined the planetary parameters of HAT-P-13b and searched for possible transit timing variations (\citealt{Winn2010}; \citealt{Szabo2010}; \citealt{Pal2011}; \citealt{nascimbeni2011}; \citealt{Fulton2011}; \citealt{Southworth2012a}; \citealt{Sada2016}). In addition, \citet{Winn2010} performed Rossiter-McLaughlin effect measurements of the system and found that HAT-P-13b is likely aligned (1.9$^{\circ}\pm8.6^{\circ}$) with its host star's equator. HAT-P-13 has the highest metallicity of all the host stars in our sample.
 
We observed the first near-UV transit of HAT-P-13b on 2013 March 02 (Table \ref{tb:obs_new}, Fig.~\ref{fig:light_1}). Our light curve (Table \ref{tb:4_light}) and physical parameters (Table \ref{tb:parms_1}) agree with previous studies and the error on our period is improved by a factor of 1.6 over the error found by \citet{Southworth2012a}. We find a near-UV R$_p$/R$_\ast$ = 0.0850$^{+0.0022}_{-0.0014}$, which is consistent with its optical R$_p$/R$_\ast$ = 0.0871$\pm$0.0024 \citep{Southworth2012a}. \citet{Turner2013a} suggest that their non-detection of a bow shock around TrES-3b could have caused by the low metallicity of the host star. Therefore, HAT-P-13b is an important target to test this suggestion since it has a high metallicity. Despite HAT-P-13 having a high metallicity, we do not observe an early near-UV ingress.

\subsection{HAT-P-16b}
HAT-P-16b is a hot Jupiter with a radius of 1.289$\pm$0.066 $R_{Jup}$ and an abnormally large mass of 4.193$\pm$0.094 $M_{Jup}$ \citep{Buchhave2010}. Spectroscopic and photometric studies have confirmed and improved upon the discovery values (\citealt{Husnoo2012}; \citealt{Ciceri2013}; \citealt{Pearson2014}; \citealt{Sada2016}). It was found through Rossiter-McLaughlin observations (\citealt{Moutou2011}) that HAT-P-16b's projected spin-orbit angle of $\lambda = -10^{\circ}\pm16^{\circ}$ is aligned with the stellar spin axis. HAT-P-16b has the highest planetary mass in our sample. 

We observed the second near-UV transit of HAT-P-16b on 2013 November 02 using the near-UV filter (Table \ref{tb:obs_new}, Fig.~\ref{fig:light_2}). This near-UV transit is observed to follow-up the observations done by \citet{Pearson2014}. We perform a combined analysis with our near-UV transit and the near-UV transit presented by \citet{Pearson2014} since they used the same telescope/filter and the data reduction pipeline (\texttt{ExoDRPL}) as we do in this study. This combined light curve is binned by 2 min to minimize the contribution of red noise. Our light curve solution (Table \ref{tb:1_light}) and derived planetary parameters (Table \ref{tb:parms_1}) agree with previous studies. The error on our period improved by a factor of 2 over that presented by \citet{Pearson2014}. We also find a near-UV radius of R$_p$/R$_\ast$ = 0.10645$\pm$0.00067, which is consistent within 1$\sigma$ of its optical radius R$_p$/R$_\ast$ = 0.1071$\pm$0.0014 \citep{Ciceri2013}. The near-UV light curves used in this study are stable (the R$_p$/R$_\ast$ values are constant) over the $\sim$1 year time period observed. 

A very extended planetary magnetosphere (\citealt{Vidotto2011b}, see fig. 9) or a clumpy magnetosheath could cause a double transit to occur if the material absorbing the near-UV radiation is concentrated in a small area. Specifically, if the absorbing material does not fill the entire planetary magnetosphere then there will be a gap between the absorbing material and the planetary radius (thus causing a double transit). The early ingress scenario described in the introduction assumes a filled planetary magnetosphere (constant absorption from the planet to the bow-shock) resulting in a blended absorption light curve. \citet{Pearson2014} suggest they may have observed a double transit in their 2012 December 29 near-UV data of HAT-P-16b at a phase of -0.0305 or $\sim$26 minutes before the start of ingress (see their fig. 1). These authors cautioned that this $2 \sigma$ feature requires follow-up observations. Our observations of HAT-P-16b do not reproduce this characteristic. Therefore, we believe the feature seen by \citet{Pearson2014} may have been an unknown systematic in their dataset or is temporal.

\subsection{HAT-P-22b}
HAT-P-22b, a hot Jupiter, was discovered by \citet{Bakos2011} around a G5 star that is part of a binary system with a distant M-dwarf companion (\citealt{Bakos2011}; \citealt{Knutson2013}). This planet is a pL class exoplanet as defined by the \citet{Fortney2008} due to a low incoming flux impinging on its atmosphere. The host star of HAT-P-22b has the lowest mass of the hot Jupiter hosting stars in our sample. 

We observed the first  follow-up light curves of HAT-P-22b on 2013 February 22 and 2013 March 22 using the U filter (Table \ref{tb:obs_new}, Fig.~\ref{fig:light_2}). We combined the near-UV data and binned it by 2 mins (this time was chosen to minimize the contribution of red noise). The derived planetary parameters agree with the discovery values and the error on the period is improved by a factor of 7.5 (Table \ref{tb:parms_1}). We also find a near-UV radius of R$_p$/R$_\ast$= 0.1079$\pm$0.00094, which is consistent with its optical R$_p$/R$_\ast$~=~0.1065$\pm$0.0017 \citep{Bakos2011}. Any non-spherical asymmetries are not seen in our data.

\subsection{TrES-2b} 

The hot Jupiter TrES-2b was the first transiting planet discovered in the Kepler field \citep{ODonovan2006}. Follow-up transit observations have confirmed and refined the planetary parameters of this system (\citealt{Holman2007}; \citealt{Colon2010}; \citealt{Mislis2010}; \citealt{Gilliland2010}; \citealt{Croll2010}; \citealt{ODonovan2010}; \citealt{Scuderi2010}; \citealt{Southworth2011};\citealt{Kipping2011a}; \citealt{Kipping2011b}; \citealt{Christiansen2011}; \citealt{Schroter2012}; \citealt{Barclay2012}; \citealt{Esteves2013}; \citealt{Ranjan2014}). In addition, Rossiter-McLaughlin effect measurements of the system found that TrES-2b is aligned with its host star's equator (-9$^{\circ}\pm12^{\circ}$) and orbits in a prograde orbit (\citealt{Winn2008a}). TrES-2b has the lowest albedo of any exoplanet currently known \citep{Kipping2011b}. 

We observed the first near-UV light curve of TrES-2b on 2012 October 29 (Table \ref{tb:obs_new}, Fig.~\ref{fig:light_2}). There is non clear evidence for any non-spherical asymmetries in TrES-2b. The TrES-2 system parameters were measured by \citet{Esteves2013} using 3 years of observations by the \textit{Kepler} spacecraft. Due to their extensive analysis, we choose to only derive the near-UV radius of the planet (Table \ref{tb:parms_1}). We find a near-UV $R_{p}/R_{\ast}$ = 0.1243$\pm$0.0024, which is consistent with its optical $R_{p}/R_{\ast}$ = 0.125358$^{+0.000019}_{-0.000024}$ \citep{Esteves2013}.

\subsection{TrES-4b} 
The hot Jupiter TrES-4b has a very low density and is one of the most highly inflated transiting giant planets known to date \citep{Mandushev2007}. Primary transit follow-up studies have refined these planetary parameters and searched for transit timing variations (\citealt{Torres2008}; \citealt{Sozzetti2009}; \citealt{Southworth2010}; \citealt{Chan2011}; \citealt{Sada2012}; \citealt{Ranjan2014}; \citealt{Sozzetti2015}). TrES-4b was found to be aligned (6.3$^{\circ}\pm4.7^{\circ}$) with its host star's equator using measurements of the Rossiter-McLaughlin effect (\citealt{Narita2010}). This system has the largest planetary radius and largest host star mass and radius in our sample.

Our observations of TrES-4b were conducted on 2011 July 26 using the Bessell U and Harris R filters (Table \ref{tb:obs_new}). We present the only published near-UV light curve of TrES-4b (Fig.~\ref{fig:light_5}, Table \ref{tb:1_light}). Our planetary parameters agree with the discovery values and improve the error on the period by a factor of 2.3 (Table \ref{tb:parms_1}). We also find a near-UV R$_p$/R$_\ast$ = 0.1094$^{+0.0052}_{-0.0052}$, which is larger by 2$\sigma$ of its optical R$_p$/R$_\ast$ = 0.09745$\pm$0.00076 \citep{Chan2011}.  
We do not observe any non-spherical asymmetries in our data due to the presence of an optically thick bow shock.

\subsection{WASP-1b}

WASP-1b is the first exoplanet discovered by the SuperWASP survey (\citealt{Pollacco2006}; \citealt{Cameron2007}).  Several follow-up photometry studies that have refined these planetary parameters and searched for transit timing variations (\citealt{Charbonneau2007b}; \citealt{Shporer2007}; \citealt{Southworth2008}; \citealt{Szabo2010}; \citealt{Southworth2012c}; \citealt{Sada2012}; \citealt{Maciejewski2014}; \citealt{Granata2014}). \citet{Wheatley2010} observed the secondary transit of WASP-1b and found a strong temperature inversion in its atmosphere and ineffective day-night energy redistribution. Rossiter-McLaughlin effect measurements found WASP-1b to be misaligned with the equator of its host star (\citealt{Stempels2007}; \citealt{Albrecht2011}; \citealt{Simpson2011}). WASP-1 is the only F star (F7V) in our sample. 

Here we present the first near-UV light curves of WASP-1b (Table \ref{tb:obs_new}; Fig.~\ref{fig:light_3}, Table \ref{tb:1_light}, Table \ref{tb:2_light}). The light curve solution (Table \ref{tb:1_light}, Table \ref{tb:2_light}) and the derived planetary parameters (Table \ref{tb:parms_1}) are in agreement to previous studies. We also find a near-UV radius of R$_p$/R$_\ast$ = 0.0964$\pm$0.0010, which is smaller by 3.5 $\sigma$ with its optical radius of R$_p$/R$_\ast$ = 0.1048$\pm$0.0014 \citep{Granata2014}. We do not see an early ingress or any non-spherical asymmetries in our near-UV transits. Our near-UV light curves are stable over the 1 month time period observed.

\subsection{WASP-12b}
WASP-12b is a hot Jupiter orbiting a G0 star with a short orbital period \citep{Hebb2009}. There have been extensive  photometric and spectroscopic  studies  of WASP-12b (\citealt{Lopez-Morales2010}; \citealt{Maciejewski2011,Maciejewski2013}; \citealt{Chan2011}; \citealt{Campo2011}; \citealt{Croll2011}; \citealt{Madhusudhan2011};  \citealt{Husnoo2011}; \citealt{Cowan2012}; \citealt{Crossfield2012}; \citealt{Haswell2012}; \citealt{Sokov2012}; \citealt{Southworth2012b}; \citealt{Sada2012}; \citealt{Zhao2012}; \citealt{Bechter2014}; \citealt{Copperwheat2013}; \citealt{Crossfield2013}; \citealt{Fohring2013}; \citealt{Fossati2013}; \citealt{Mandell2013}; \citealt{Sing2013}; \citealt{Swain2013}; \citealt{Teske2014}; \citealt{Stevenson2014}; \citealt{Stevenson2014b}; ; \citealt{Croll2015}; \citealt{Burton2015}; \citealt{Kreidberg2015}; \citealt{Collins2015}). WASP-12 is a triple star system with a binary M dwarf system in orbit around the G0 star (\citealt{Crossfield2012}; \citealt{Bergfors2013}; \citealt{Bechter2014}). Previous studies by \citealt{Fossati2010b}, \citet{Haswell2012}, and \citep{Nichols2015} observed an early near-UV ingress with \emph{HST} using the Cosmic Origins Spectrograph. However, these studies have a low number of data points and therefore follow-up near-UV studies are needed. Ground-based near-UV observations \citep{Copperwheat2013} and additional space-based UV observations of WASP-12b using the Space Telescope Imaging Spectrograph instrument on \textit{HST} \citep{Sing2013} all do not observe any asymmetries in their near-UV light curves. Finally, WASP-12b has the closest orbital distance and planetary radius in our study and is the top candidate predicted by VJH11a to exhibit an early near-UV ingress.  

Our observations were conducted from 2011 November to 2012 November (Table \ref{tb:obs_new}, Table \ref{tb:4_light}; Fig.~\ref{fig:light_4}). These observations were performed to follow-up the previous near-UV observations and to confirm the detection of an early ingress. We didn't account for the M-dwarf companions in the our analysis, because they contribute a negligible amount of flux at the wavelengths observed (\citealt{Copperwheat2013}). We combined all the near-UV transits and binned the light curve by 1 min and 30 s (this time was chosen to minimize the dominance of red noise). The derived planetary parameters (Table \ref{tb:parms_2}) are in agreement to previous studies. Our near-UV radius is within $1\sigma$ of that found by \citet{Copperwheat2013} and \citet{Sing2013}. We also find a near-UV R$_{p}/R_{\ast}$ = 0.12016$^{+0.00076}_{-0.00065}$, which is 2.5$\sigma$ larger than optical radius of R$_p$/R$_\ast$ = 0.1173$\pm$0.0005 \citep{Maciejewski2013}. The larger near-UV radius is consistent with Rayleigh scattering (Section \ref{sec:atmo}). We do not observe an early ingress in any of our near-UV light curves. Our near-UV light curves are stable over the $\sim$1 year time period observed.

\subsection{WASP-33b} 
WASP-33b is a hot Jupiter \citep{Cameron2010} that orbits a bright (V-mag = 8.3) $\delta$ Scuti variable host star (\citealt{Herrero2011}). It is the first planet discovered to orbit an A-type star (\citealt{Herrero2011}). This system has been extensively studied with photometry and spectroscopy (\citealt{Herrero2011}; \citealt{Moya2011}; \citealt{Smith2011}; \citealt{Deming2012}; \citealt{Sada2012}; \citep{deMooij2013}; \citealt{vonEssen2014}; \citealt{Haynes2015}; \citealt{Johnson2015}; \citealt{vonEssen2015}; \citealt{Hardy2015}). Secondary eclipse measurements indicate that WASP-33b has a low albedo \citep{deMooij2013} and inefficient heat-transport from the day-side to the night-side (\citealt{Smith2011}; \citealt{Deming2012}; \citealt{Madhusudhan2012}; \citealt{deMooij2013}; \citealt{Haynes2015}). In our study, WASP-33b has the highest highest planetary equilibrium temperature and is the only planet around an A star. 

We observed the first near-UV light curve of WASP-33b on 2012 October 01 and subsequently in the B and U bands on 2012 December 01 (Table \ref{tb:obs_new}, Fig.~\ref{fig:light_5}). We did not take into account the pulsations in our modeling because it was found by \citealt{vonEssen2014} that taking them into account did not change their final parameter results. We see the variability of the host star in our transits, residuals, and asymmetry test very clearly. The light curve solution (Table \ref{tb:2_light}) and derived physical parameters (Table \ref{tb:parms_2}) are consistent with previous studies (e.g. \citealt{Cameron2010}; \citealt{Kovcs2013}; \citealt{vonEssen2014}). We find a near-UV $R_{p}/R_{\ast}$ = 0.1086$^{+0.0022}_{-0.0007}$, which is consistent with its optical $R_{p}/R_{\ast}$ = 0.1066$\pm$0.0009 \citep{Cameron2010}. There are non-spherical asymmetries in our light curves, however, the amplitude and shape of the variability in the residuals are due to host star's variability. Our near-UV transits are stable over the several months observed.

\subsection{WASP-36b}
The hot Jupiter WASP-36b was discovered around a G2 dwarf \citep{Smith2012}. The host star shows low levels of stellar activity and has undergone little or no tidal spin-up due to the planet \citep{Smith2012}. WASP-36 has the lowest metallicity of all the hot Jupiter host stars in our sample.

We observed the first near-UV light curve of WASP-36b on 2012 December 29 and an additional R band transit on 2013 March 15 (Table \ref{tb:obs_new}, Table \ref{tb:2_light}, Fig.~\ref{fig:light_6}). The derived physical parameters (Table \ref{tb:parms_2}) agree with the discovery values and the error on the period is improved by a factor of 4.7. We also find a near-UV radius of $R_{p}/R_{\ast}$ = 0.1316$^{+0.0018}_{-0.0018}$ which is 2.6$\sigma$ smaller than the optical radius of R$_p$/R$_\ast$ = 0.13850$^{+0.00071}_{-0.00082}$.

\subsection{WASP-44b} 
The hot Jupiter WASP-44b is a highly inflated planet in orbit around a G8V star \citep{Anderson2012}. The host star, WASP-44, is found to be inactive based off observations of weak Ca II H$\&$K emission and no rotational modulation (\citealt{Anderson2012}). The first follow-up light curve of WASP-44b \citep{Mancini2013} indicates a constant radius from the optical to NIR wavelengths. This system has the smallest host star radius of all hot Jupiter systems in our study. 

We observed the first near-UV light curve of WASP-44b on 2012 October 13 using the U filter and subsequently on 2013 October 19 with the B and V filter (Table \ref{tb:obs_new}, Table \ref{tb:4_light}, Table \ref{tb:5_light}, Fig.~\ref{fig:light_6}). The light curve solution (Table \ref{tb:2_light}) and planetary parameters (Table \ref{tb:parms_2}) are consistent with the discovery value and the error on the period is improved by a factor of 1.2 \citep{Mancini2013}. We also find a near-UV radius of R$_p$/R$_\ast$ = 0.1228$\pm$0.0028, which is larger by 1.4$\sigma$ with its optical radius of R$_p$/R$_\ast$ =  0.1164$\pm$0.0017. An early near-UV or any non-spherical asymmetries are not observed in the data.

\subsection{WASP-48b} 
WASP-48b is a typical inflated hot Jupiter orbiting a slightly evolved F star (\citealt{Enoch2011}). These parameters were confirmed by follow-up J-band primary transit observations by \citet{Sada2012}. Secondary eclipse measurements indicate that WASP-48b has a weak temperature inversion and moderate day/night recirculation \citep{ORourke2014}. \citet{Ciceri2015} find that the spectrum of WASP-48b is flat from the optical to near-IR, which suggests that the atmosphere is not affected by large Rayleigh scattering. WASP-48 is the oldest system in our study with an age of 7.9$^{2.0}_{-1.6}$ Gyr and may have undergone synchronization of its stellar rotation with the planetary orbital period due to interactions with WASP-48b (\citealt{Enoch2011}). 

We observed WASP-48b on 2012 October 9 using the U filter (Table \ref{tb:obs_new}, Table \ref{tb:3_light}, Fig.~\ref{fig:light_6}). The derived planetary parameters (Table \ref{tb:parms_2}) agree with the discovery values. We find a near-UV R$_p$/R$_\ast$= 0.0916$\pm$0.0017 which is 2.4$\sigma$ smaller than its optical R$_p$/R$_\ast$ = 0.0980$\pm$0.0010 \citep{Enoch2011}. We do not observe an early ingress in our near-UV transit.

\subsection{WASP-77Ab} \label{sec:wasp77b}

WASP-77Ab is a hot Jupiter orbiting a G8 star in a double-star system (\citealt{Maxted2013}). The host star exhibits moderate chromospheric activity determined by emission in the cores of the Ca II H $\&$ K lines and rotational modulation with a period of 15.3 days \citep{Maxted2013}. WASP-77 is the only multi-star system in our sample where both companions are solar-like stars (G8 and K).\\ 
\indent On 2012 December 06 using the U filter we observed the first follow-up light curve of WASP-77Ab (Table \ref{tb:obs_new}, Fig.~\ref{fig:light_7}). The light curve solution is in Table \ref{tb:5_light}.\\
\indent We make sure to correct for the dilution due to the companion star being in our aperture using the procedure described below. The separation of the stars are 3.3$\arcsec$ (our seeing was 2.31--6.93$\arcsec$) and the magnitude differences
between the components of the binary in the near-UV are $\Delta m_{u} = 2.961\pm$0.015 \citep{Maxted2013}. WASP-77Ab orbits around the brighter companion (WASP-77A). We perform the procedure described below to find the corrected R$_p$/R$_\ast$ value and error. (1) We model the light curve with \texttt{EXOMOP} and find $\left(R_{p}/R_{\ast}\right)_{uncor}$= 0.12612$^{+0.00098}_{-0.00094}$ for the uncorrected case. (2) We then calculate the flux of WASP-77B ($F_{2}$) using the following equation:  
\begin{equation}
m_{1} - m_{2} = \Delta m_{u} =  2.5 \log{\left(   \frac{F_{1}}{F_{2}}  \right)},
\end{equation}
where $m_{1}$ is the magnitude of WASP-77A, $m_{2}$ is the magnitude of WASP-77B, $F_{2}$ is the flux measured in the aperture for WASP-77A (set equal to $1 \ erg \ s\ ^{-1} \ cm^{-2} \ \AA^{-1}$), and $F_{1}$ is the flux of WASP-77B (found to be $0.0654034F_{2}$). (3) We then find the corrected $\left( R_{p}/R_{\ast} \right)_{cor}$ value using the equations
\begin{equation}
\left(\frac{R_{p}}{R_{\ast}}\right)_{cor}    = \sqrt{\frac{\Delta F}{F_{cor}}  }
\end{equation}
\begin{equation}
\left(\frac{R_{p}}{R_{\ast}}\right)_{cor}    =  \sqrt{\frac{\Delta F}{F_{2} - F_{1}} } 
\end{equation}
where $\Delta F$ is the change in flux and is equal to $\left(R_p/R_\ast\right)_{uncor}^{2}$ and $F_{cor}$ is the corrected flux for WASP-77A. (4) We propagate all the errors (including $\Delta m_{u}$ and the error from \texttt{EXOMOP} modeling) to find the new error on the $\left( R_{p}/R_{\ast} \right)_{cor}$. Performing this procedure, we find a near-UV radius of $\left(R_{p}/R_{\ast}\right)_{cor}$= 0.1305$\pm0.0010$.\\
\indent We agree with the discovery values for our planetary parameters and the error on the period is improved by a factor of 1.7 (Table \ref{tb:parms_2}). The near-UV radius of WASP-77Ab of $\left(R_p/R_\ast\right)_{cor}$= 0.1305$\pm0.0010$ is consistent with its optical R$_p/R_\ast$ = 0.13012$\pm$0.00065 \citep{Maxted2013}.

\section{Discussion} 
\subsection{Asymmetric Transits} \label{sec:nobowshock}

A large early ingress (Figs. \ref{fig:light_1}--\ref{fig:light_7}) or significant ($>$0.5$\%$) R$_p$/R$_\ast$ difference (Tables \ref{tb:1_light}--\ref{tb:5_light}) is not observed in any of our near-UV light curves. To investigate whether the transit shapes are symmetrical, we perform an asymmetry test where we subtract the mirror image of the transit with itself (See Section \ref{sec:add}). Non-spherical asymmetries do not appear in any of these tests with the exception of WASP-33b, which is potentially the result of its host star's variability (\citealt{Herrero2011}; \citealt{Smith2011}; \citealt{Kovcs2013}). Therefore, within the precision (1.23 -- 5.54 mmag) and timing resolution (61 - 137 s) of our observations no asymmetries are observed. Our results are consistent with the previous non-detections of an early ingress in the ground-based near-UV light curves of HAT-P-5b \citep{Southworth2012b}, HAT-P-16b \citep{Pearson2014}, TrES-3b \citep{Turner2013a}, WASP-12b (\citealt{Copperwheat2013}),  WASP-17b \citep{Bento2014}, and XO-2b \citep{Zellem2015}.

Additionally, the non-detection of asymmetrical transits confirms and expands upon the theoretical modeling done by \citealt{BenJaffel2014} and \citealt{Turner2016}. These theoretical studies concentrated on modeling the corona around solar-like stars. Therefore, since the targets in this study are deliberately chosen to have a variety of planetary and host star parameters (Table \ref{tb:compare}), based on the work in this paper we do not expect to observe near-UV asymmetries caused by an opacity source in the stellar corona in any system regardless of its spectral type.

\subsubsection{Variability in the bow shock} \label{sec:var_bow}

Assuming that the bow shock is sufficiently optically thick to absorb light from the host star during transit, then we need to assess whether shock variability is a key factor in the non-detections. It is predicted that bow shock variations would be common for planets that are not circularized, not in the corotation radius of their host star, and orbiting around active stars (\citealt{Vidotto2011c}; \citealt{llama2013}). Rossiter-McLaughlin effect (\citealt{Mclaughlin1924}; \citealt{Rossiter1924}; \citealt{Winn2011}) measurements and activity indicators can assess whether any of the systems would be prone to bow shock variability.

Measurements of the Rossiter-McLaughlin effect can be used to determine whether our systems are coplanar with their hosts stars. If the coronal material is axisymmetric and if a planet's orbital plane and the stellar equator are coplanar then the planet will move through coronal material of constant density and temperature during transit. In our sample, we have 4 planets (CoRoT-1b, WASP-1b, WASP-12b, WASP-33b) that are not aligned with their stars, 5 planets (HAT-P-1b, HAT-P-13b, HAT-P-16b, TrES-2b, TrES-4b) that are aligned with their stars, and 6 planets (GJ 436b, HAT-P-22b, WASP-36b, WASP-44b, WASP-48b, WASP-77Ab) that are in need of Rossiter-McLaughlin measurements (See Table \ref{tb:compare}). Therefore, it is possible that members of our sample may exhibit shock variability due to the planet moving through coronal material with different densities. However, this phenomenon does not explain all our non-detections since the planets that are aligned with their host stars are moving through coronal material with a similar density and through an environment with a constant stellar magnetic field. \\
\indent Furthermore, if the host stars are active then fluctuations in the stellar wind, flaring, or coronal mass ejections could cause inhomogeneity in the coronal outflow. The $R^{'}_{HK}$ index, the ratio between chromospheric activity to the total bolometric emission of the star, can be used to gauge the amount of stellar activity of a star \citep{Noyes1984} and more active stars exhibit higher $R^{'}_{HK}$ indices. In our sample, there are 6 planets (CoRoT-1b, GJ436b, HAT-P-13b, TrES-4b, WASP-1b, WASP-12b) with a $R^{'}_{HK}$ index lower than the sun ($R^{'}_{HK,\sun}$ = -4.96), 1 planet (HAT-P-16b) with a $R^{'}_{HK} > R^{'}_{HK,\sun}$, 2 planets (HAT-P-1b, TrES-2b) with a $R^{'}_{HK} \sim R^{'}_{HK,\sun}$, and 6 planets (HAT-P-22b, WASP-33b, WASP-36b, WASP-44b, WASP-48b, WASP-77Ab) that do not have a $R^{'}_{HK}$ index measured (Table \ref{tb:compare}). Therefore, some of the non-detections of the planets around the active stars could be caused by their orbits moving through inhomogeneous coronal material. Also, stellar flares can raise the coronal temperature above the maximum temperature allowed for shock formation (VJH11a). HAT-P-16b (the only planet in our sample known to orbit an active host star) is observed more than once and all the observations result in non-detections despite six months between successive observations. Additionally, WASP-12b, WASP-1b, and GJ436b (planets known to orbit non-active host stars) are observed more than once and also result in non-detections. Therefore, variability of the coronal plasma may be causing some of our non-detections but not all of them.\\
\indent However, the interpretation of variability causing some of our non-detections changes significantly if we now consider the theoretical study by \citealt{Turner2016}. \citealt{Turner2016} did an an extensive parameter study to determine if temperature ($3\times10^{4}$ -- $2\times10^{6} \rm{K}$) or density ($10^{4} - 10^{8} \rm{cm^{-4}}$) changes in the coronal outflow would cause variation in the absorption due to the bow shock. They find that under all reasonable conditions for a steady state and varying stellar corona that no absorption occurred in the bow shock. Therefore, we did not observe any asymmetries in our observations because the bow shock does not actually cause any absorption in the first place and not due to bow shock variability.

\subsection{Wavelength dependence on the planetary radius} \label{sec:atmo} 
Observing the primary transit of an exoplanet at multiple wavelengths allows for an investigation into the composition and structure of its atmosphere. The measured R$_p$/R$_\ast$ depends on the opacity of the planetary atmosphere and thus allows for useful insights into the atmosphere's spectral features and composition. If the opacity in our near-UV band is dominated by Rayleigh scattering of molecular hydrogen, it may be possible to place strong upper limits on the planet's 10 bar radius (\citealt{Tinetti2010}; \citealt{Benneke2013}; \citealt{Benneke2012}; \citealt{Griffith2014}). Such constraints can break the degeneracy between an exoplanet's physical radius and atmospheric composition in radiative transfer retrievals (e.g. \citealt{Etangs2008}; \citealt{Tinetti2010}; \citealt{Benneke2012}; \citealt{Griffith2014}).

The R$_p$/R$_\ast$ of 10 exoplanets (GJ436b, HAT-P-1b, HAT-P-13b, HAT-P-16b, HAT-P-22b, TrES-2b, WASP-33b, WASP-44b, WASP-48b, WASP-77Ab) are consistent to within 1$\sigma$ of their optical R$_p$/R$_\ast$ (Tables \ref{tb:parms_1}--\ref{tb:parms_2}). Clouds in the upper atmospheres of these planets are consistent with these observations because clouds reduce the strengths of spectral features (e.g. \citealt{Seager2000}; \citealt{Brown2001}; \citealt{Gibson2013b}; \citealt{Kreidberg2014}). Also, day-side spectral features may be absent due to an isothermal pressure-temperature profile \citep{Fortney2006}. These planets are consistent with other transiting exoplanet observations with flat spectra in optical wavelengths on TrES-3b \citep{Turner2013a}, GJ 3470b (a hot Uranus; \citealt{Biddle2014}), GJ 1214b (\citealt{Bean2011}; \citealt{Kreidberg2014}), WASP-29b \citep{Gibson2013a}, and HAT-P-32b \citep{Gibson2013b}. 

We also find that some of our targets do not exhibit a flat spectrum. The R$_p$/R$_\ast$ of CoRoT-1b, TrES-4b, and WASP-12b are larger than their optical R$_p$/R$_\ast$ by 2.3, 2, and 2.5$\sigma$, respectively (Tables \ref{tb:parms_1}-\ref{tb:parms_2}). This variation corresponds to a difference in the radius of 6 scale heights (H) for both CoRoT-1b and TrES-4b and 3H for WASP-12b. This is consistent with the 6H variation observed in HD 189733b (\citealt{Sing2011}). A larger near-UV radius may indicate non-uniform clouds \citep{Griffith2014} or Rayleigh scattering in their atmospheres (\citealt{Tinetti2010}; \citealt{Benneke2013}; \citealt{Benneke2012}; \citealt{Griffith2014}). Additionally, the near-UV R$_p$/R$_\ast$ of WASP-1b and WASP-36b are smaller than their optical R$_p$/R$_\ast$ by 3.6 and 2.6$\sigma$ (Tables \ref{tb:parms_1}-\ref{tb:parms_2}), respectively. To our knowledge, this is the first time a hot Jupiter has been observed to have a smaller near-UV transit depth than that measured in the optical. Additionally, the near-UV transit depths of WASP-1b and WASP-36b are smaller than every transit depth measurement made on the planets (Table \ref{tb:rp_rstar}). A smaller transit depth could be caused by aerosol absorption \citep{Sing2013}, however, more work is needed to investigate possible opacity sources. The variation corresponds to a difference of 7 and 20H for WASP-1b and WASP-36b, respectively. The large scale height variations are similar with the 13H variation found for WASP-103b \citep{Southworth2015}. These results are consistent with other exoplanets not having flat spectrum (HD 209458b, \citealt{Sing2008}; HAT-P-5b, \citealt{Southworth2012b}; GJ 3470b, \citealt{Nascimbeni2013}; Qatar-2, \citealt{Mancini2014}).  

\begin{figure*}
\center
\begin{tabular}{cc}
\vspace{0.5cm}
\epsfig{file=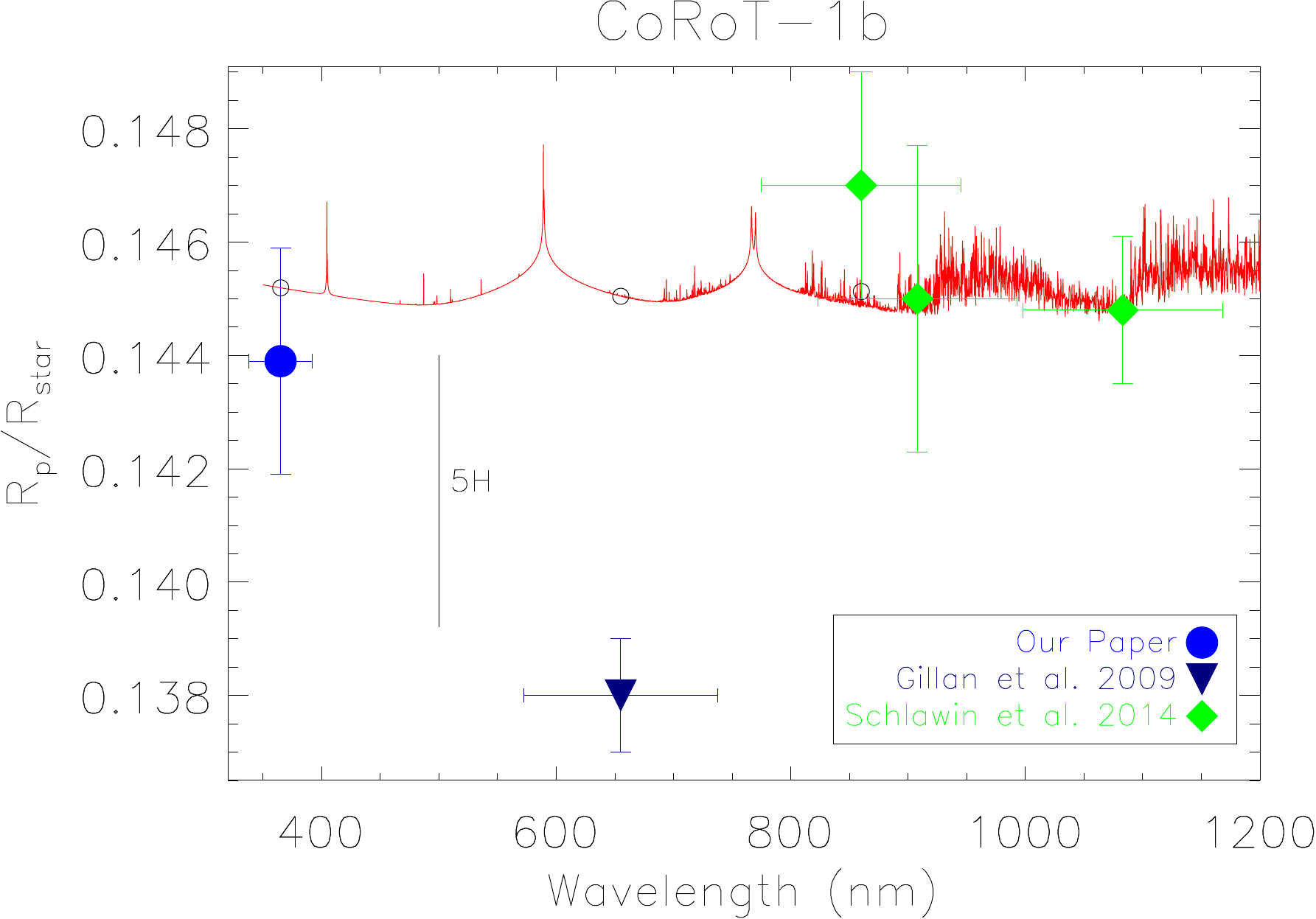,width=0.50\linewidth} 	&  
\epsfig{file=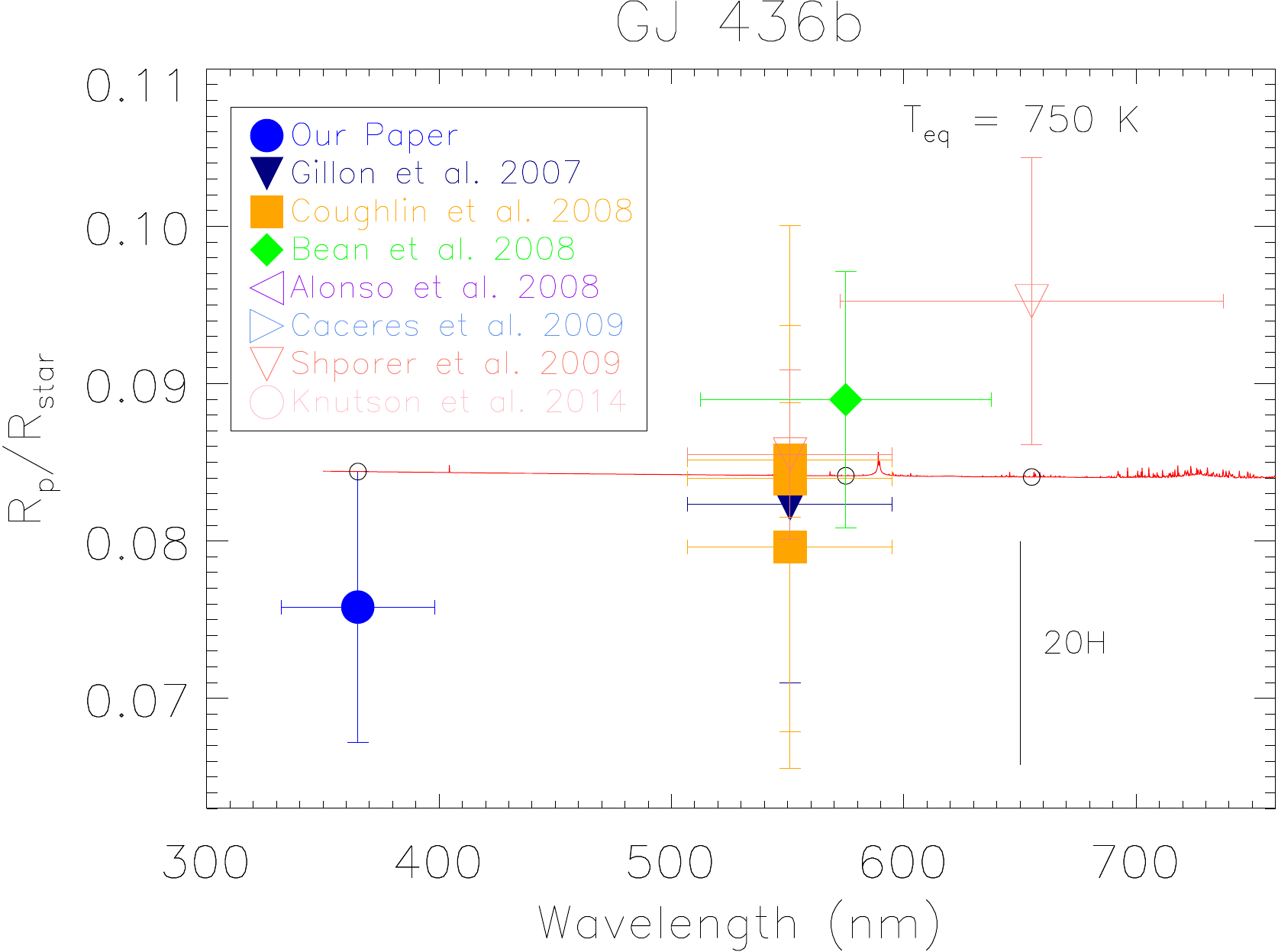,width=0.47\linewidth} \\
\vspace{0.5cm}
\epsfig{file=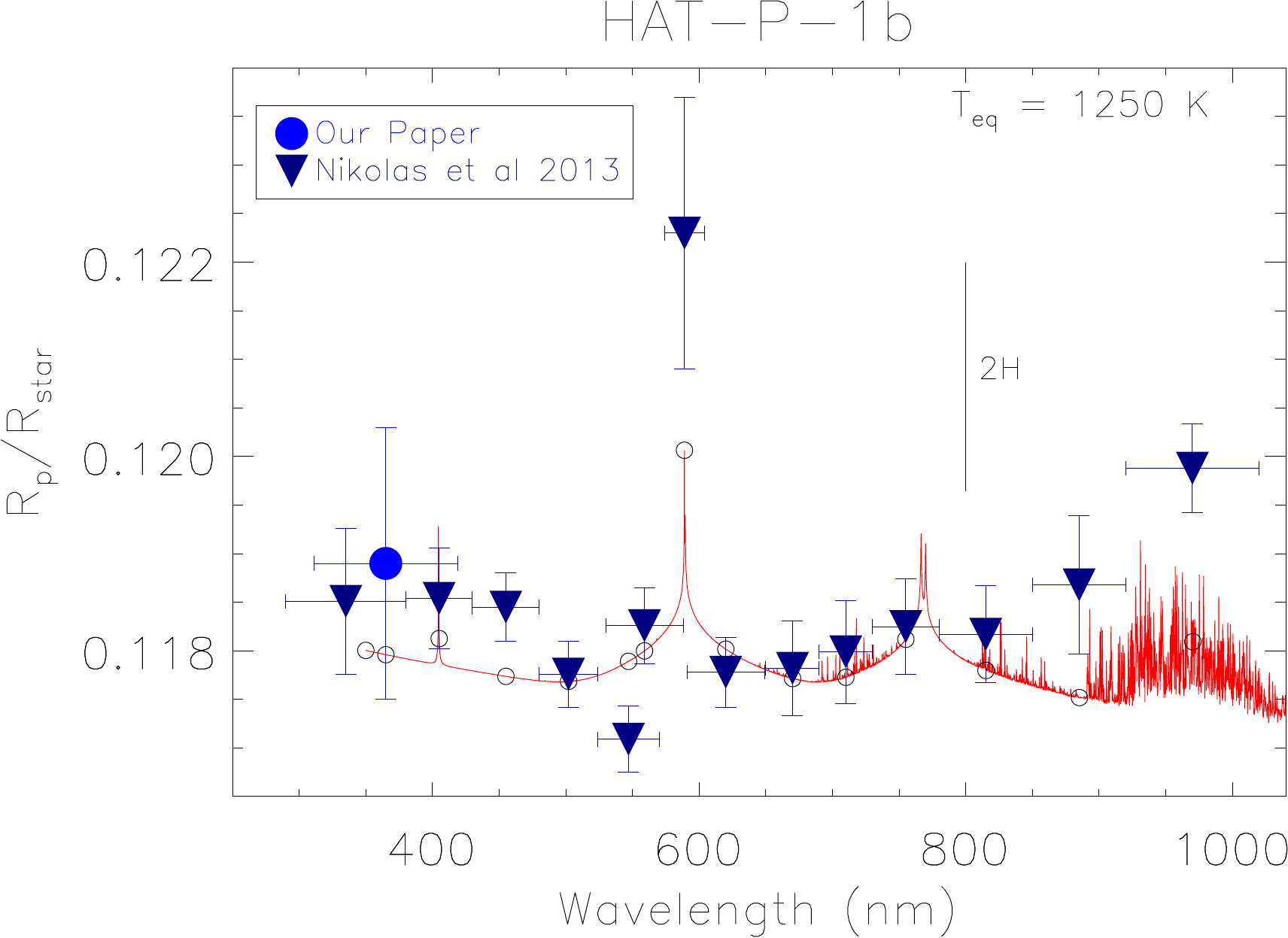,width=0.47\linewidth} 	&  
\epsfig{file=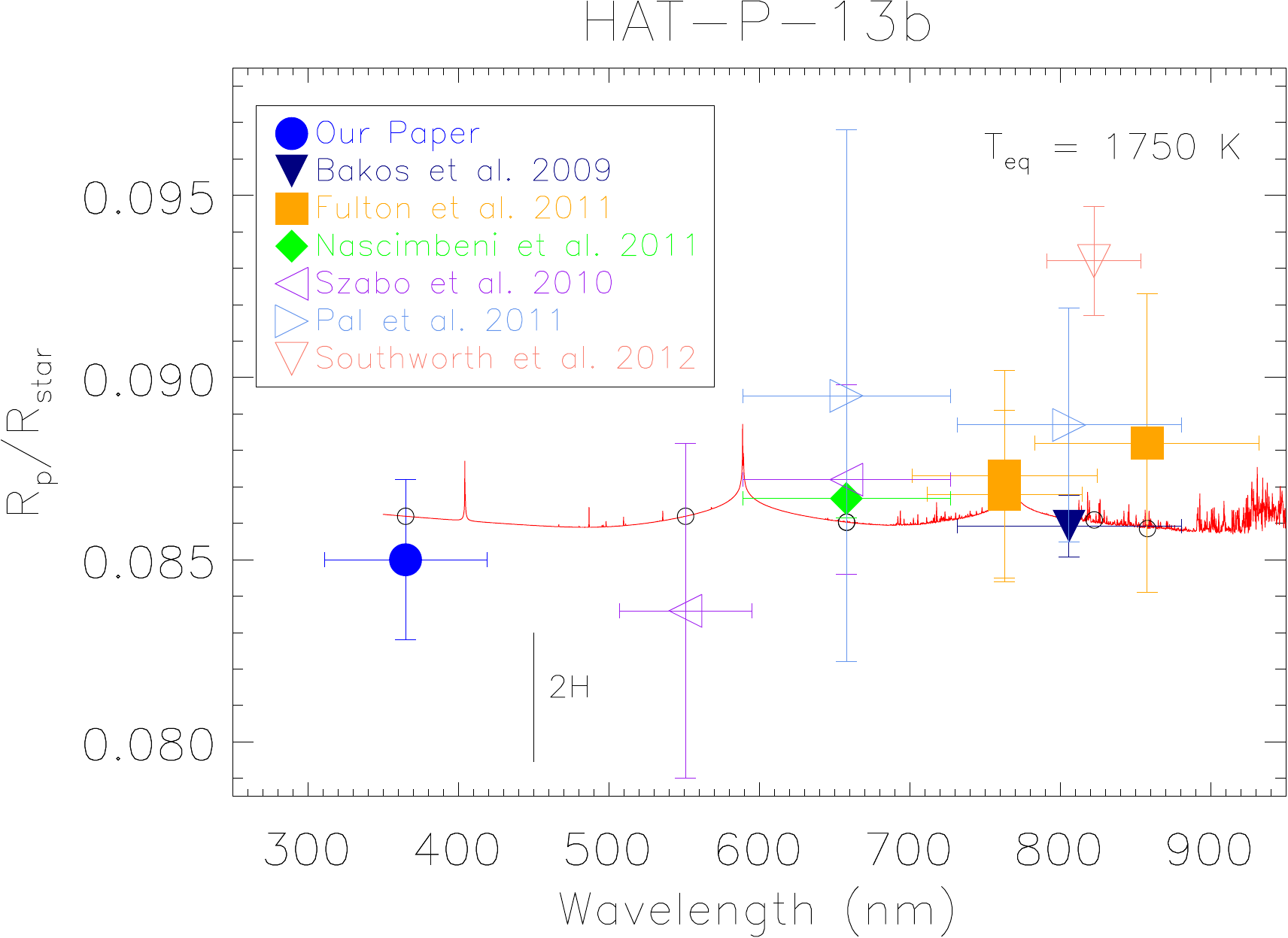,width=0.47\linewidth} \\
\vspace{0.5cm}
\epsfig{file=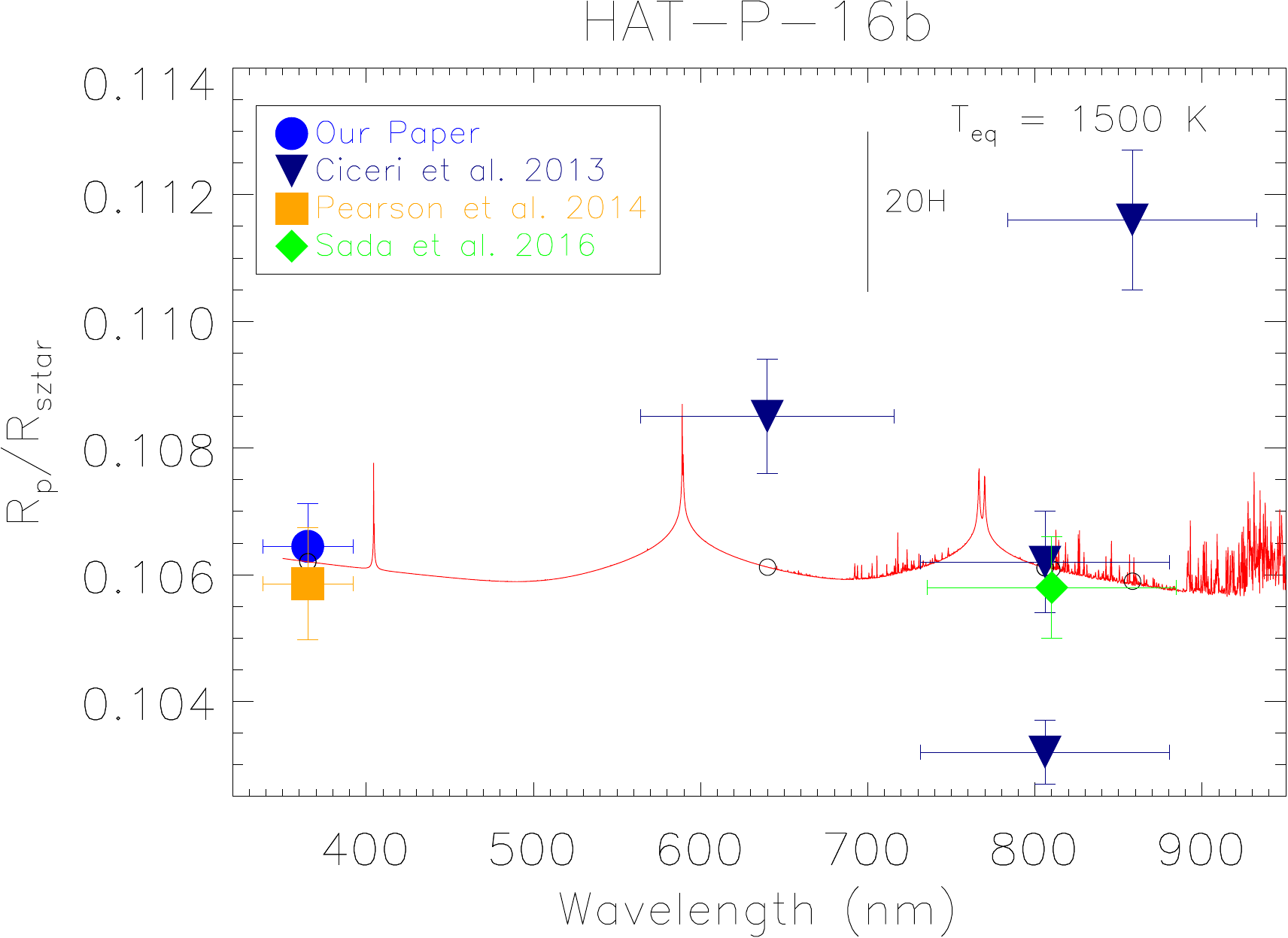,width=0.47\linewidth} 	&  
\epsfig{file=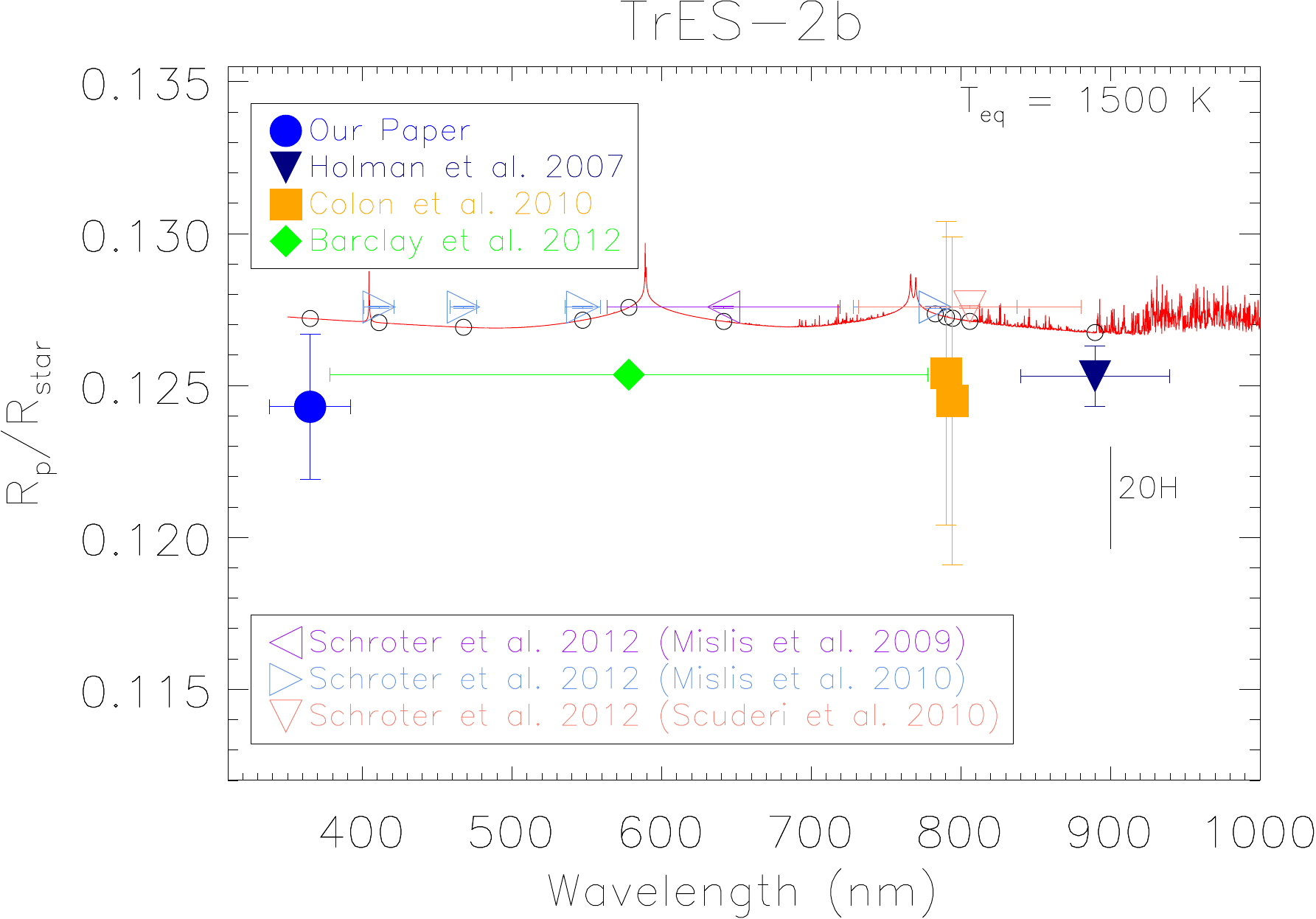,width=0.47\linewidth} \\
\end{tabular}
\caption{Variation of R$_p$/R$_\ast$ vs. wavelength for CoRoT-1b, GJ 436b, HAT-P-1b, HAT-P-13b, HAT-P-16b, and TrES-2b. Over-plotted in red are atmospheric models by \citet{Fortney2010} for planets with a 1 $M_{Jup}$, $g_{p} = 25 m/s$, base radius of 1.25 $R_{Jup}$ at 10 bar, and $T_{eq}$ (specified on plot). The scale height of the planet is also shown on each plot for reference.}
\label{fig:rp}
\end{figure*}

\begin{figure*}
\center
\begin{tabular}{cc}
\vspace{0.5cm}
\epsfig{file=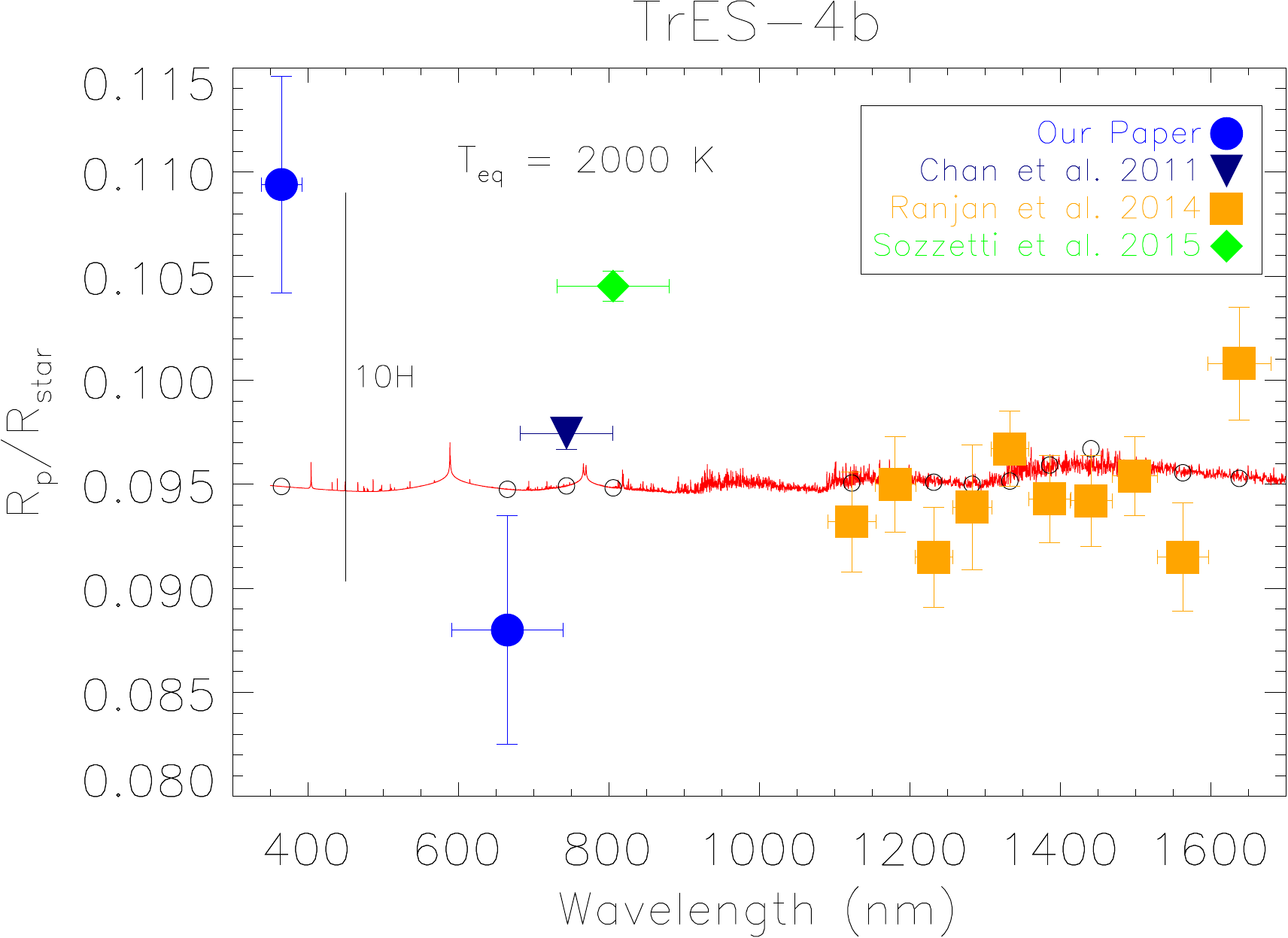,width=0.47\linewidth} 	&  
\epsfig{file=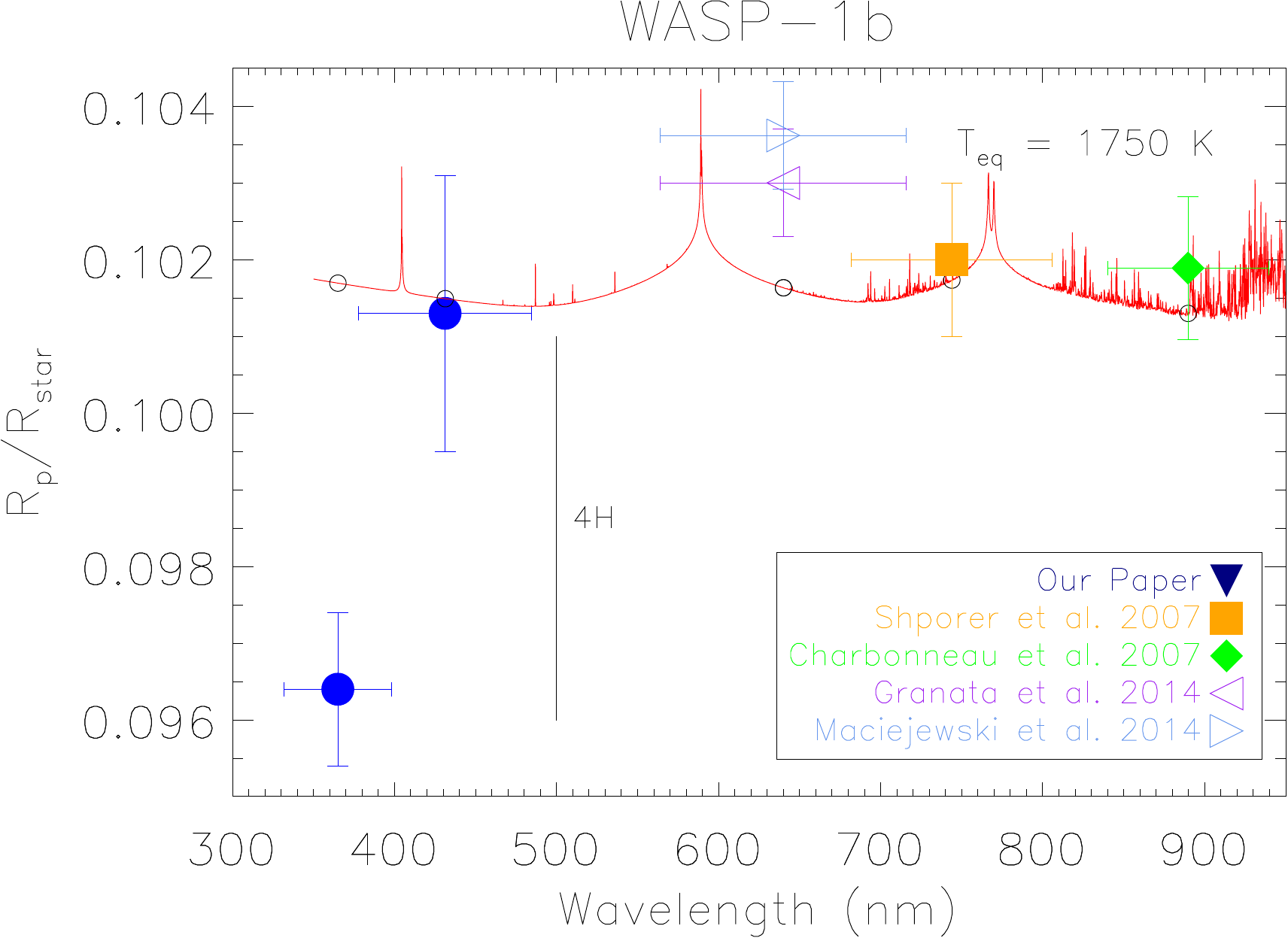,width=0.47\linewidth} \\
\vspace{0.5cm}
\epsfig{file=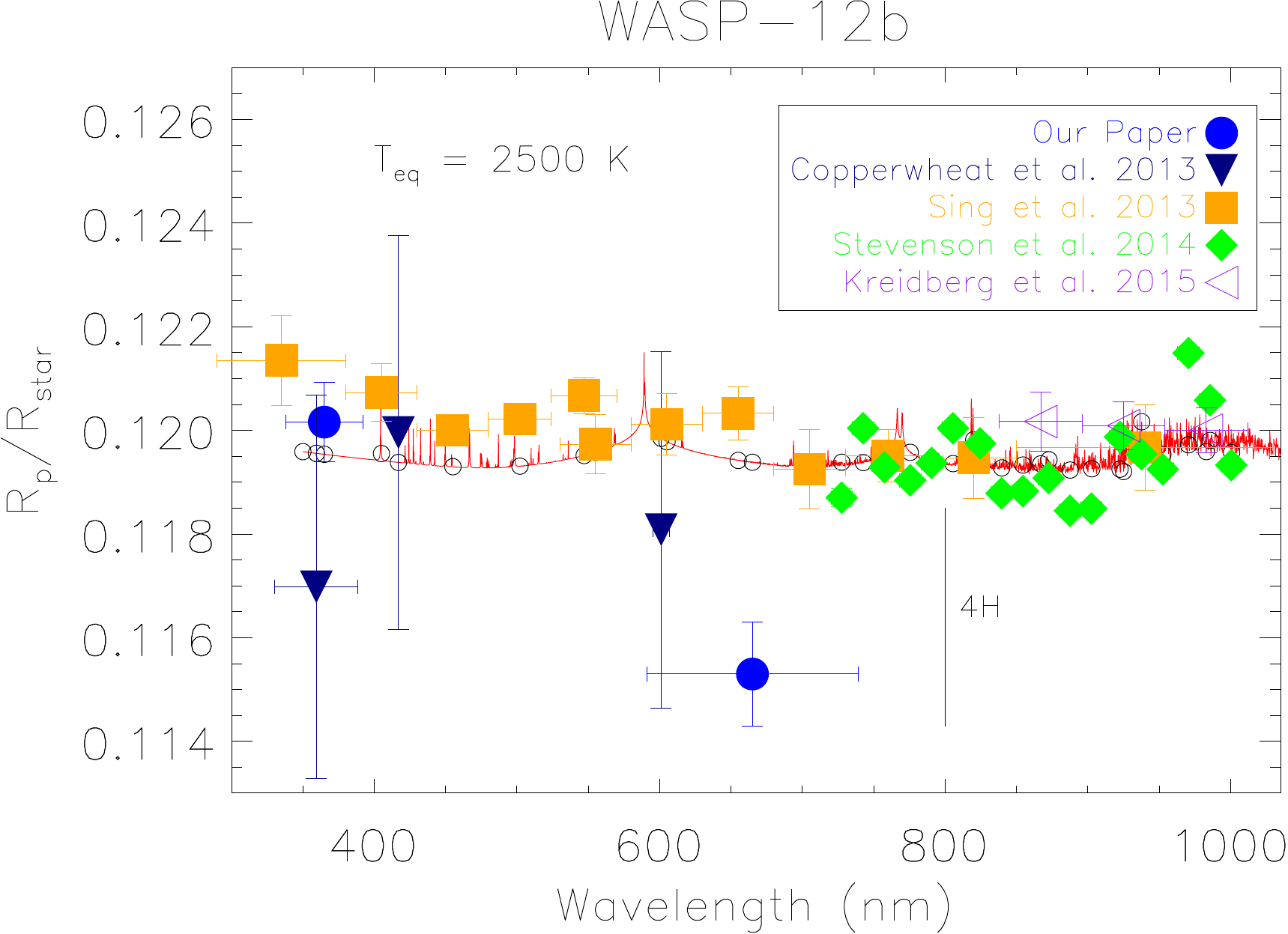,width=0.47\linewidth} 	&    
\epsfig{file=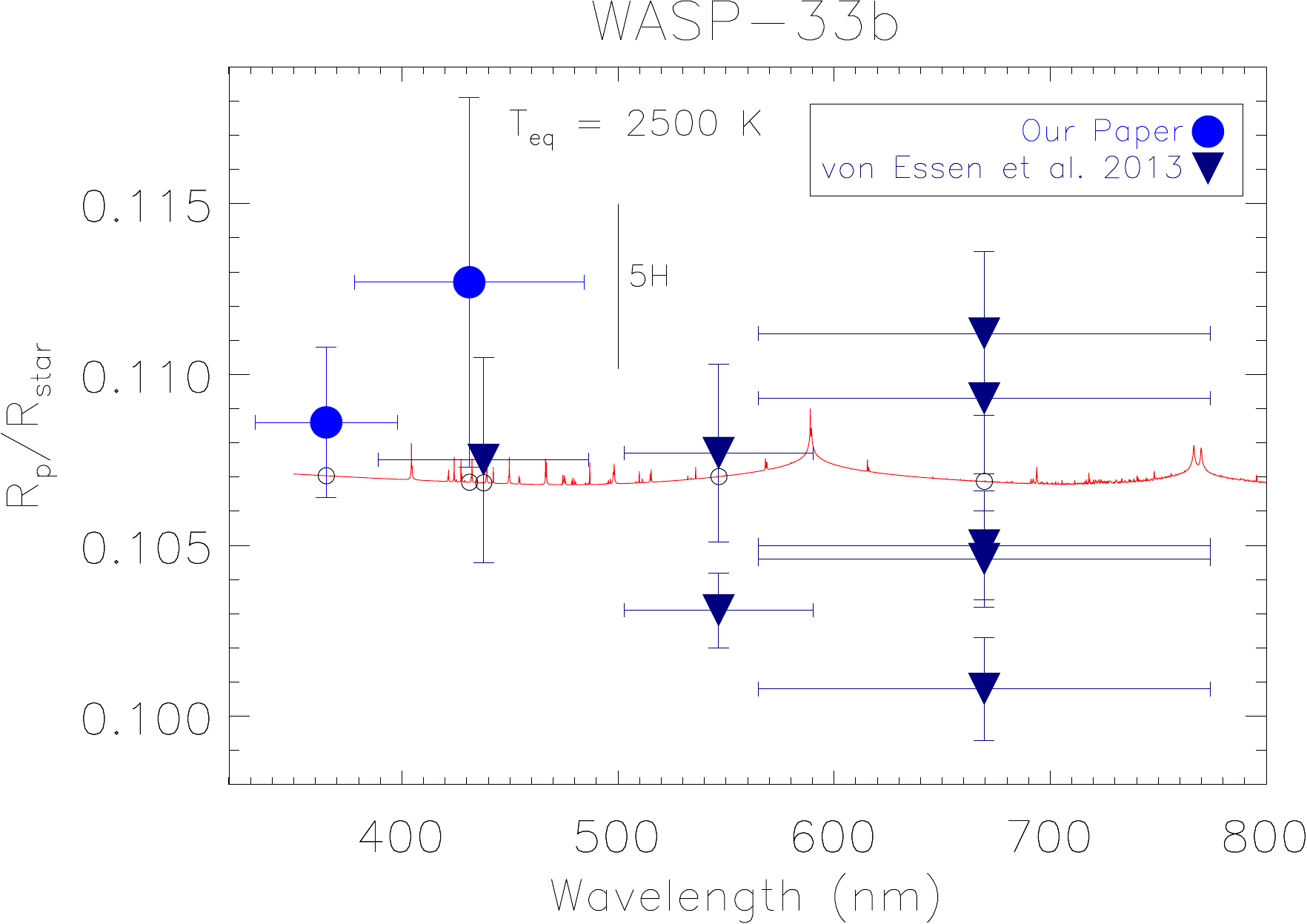,width=0.48\linewidth} \\   
\vspace{0.5cm}
\epsfig{file=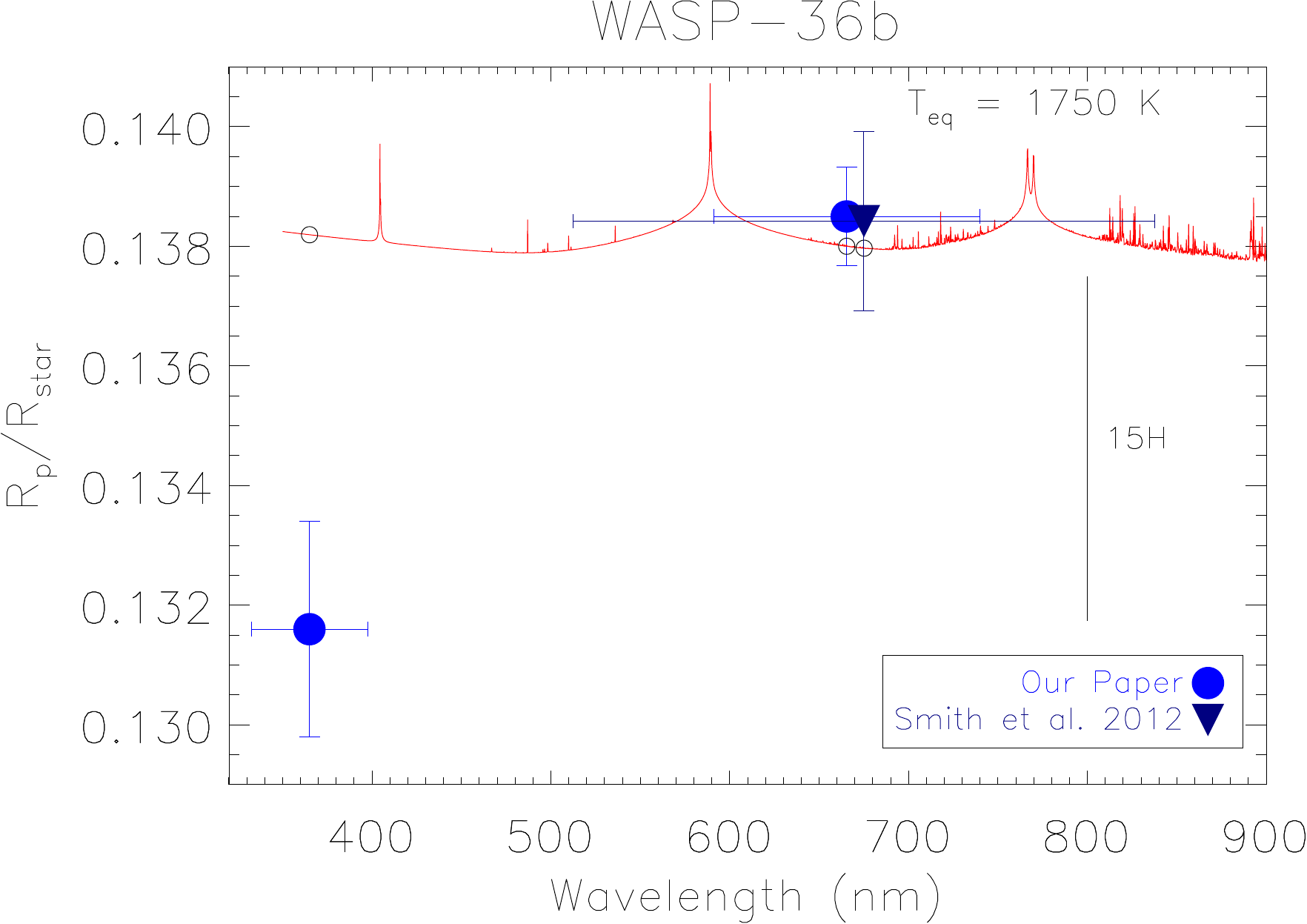,width=0.47\linewidth} 	&    %
\epsfig{file=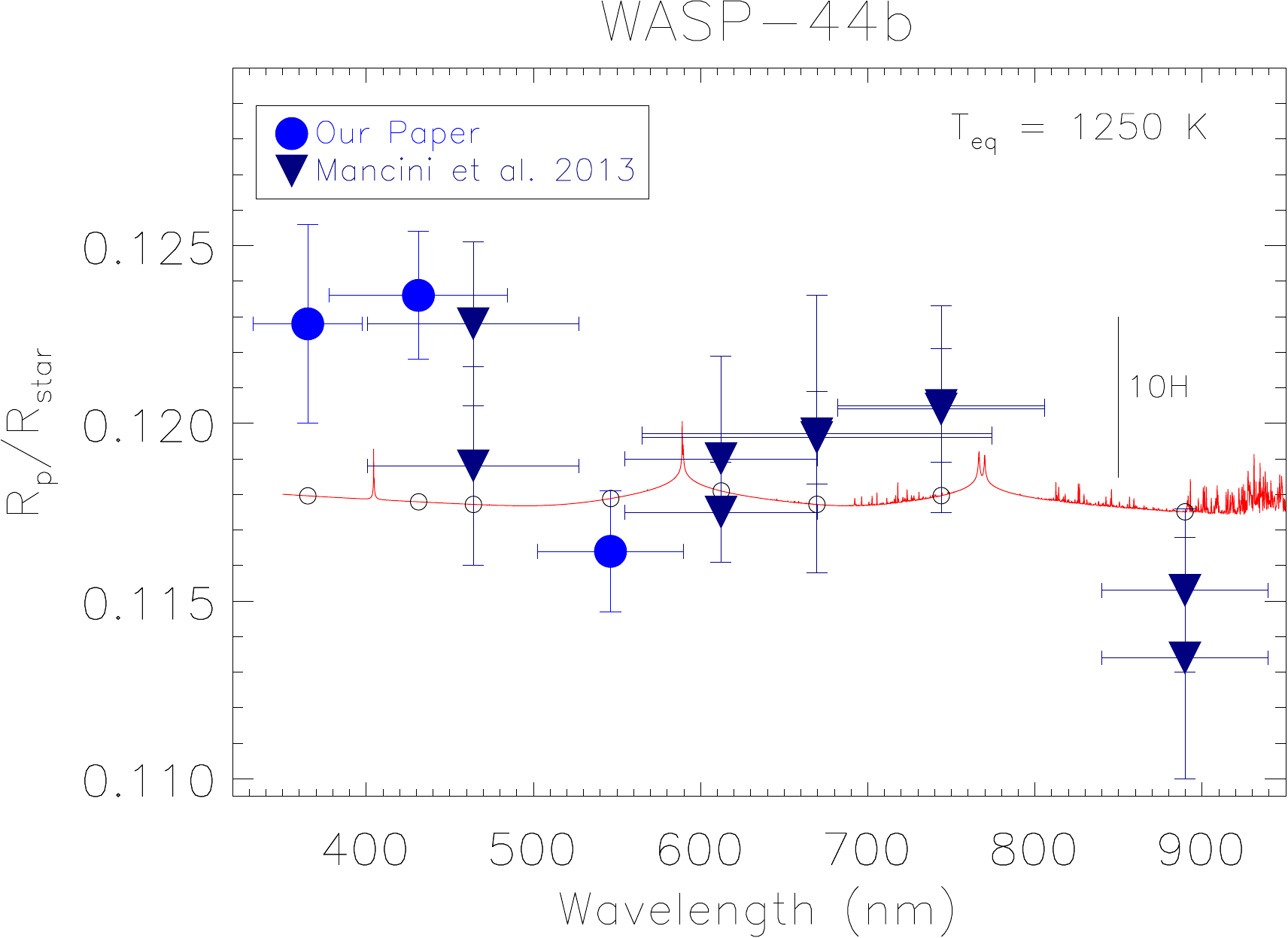,width=0.46\linewidth} \\   
\end{tabular}
\caption{Variation of R$_p$/R$_\ast$ vs. wavelength for TrES-4b, WASP-1b, WASP-12b, WASP-33b, WASP-36, and WASP-44b. The observation of a smaller near-UV than the optical radius on WASP-1b and WASP-36b are the first of such a detection on a hot Jupiter. Other comments are the same as Fig. \ref{fig:rp}. }
\label{fig:rp2}
\end{figure*}

\begin{figure*}
\center
\epsfig{file=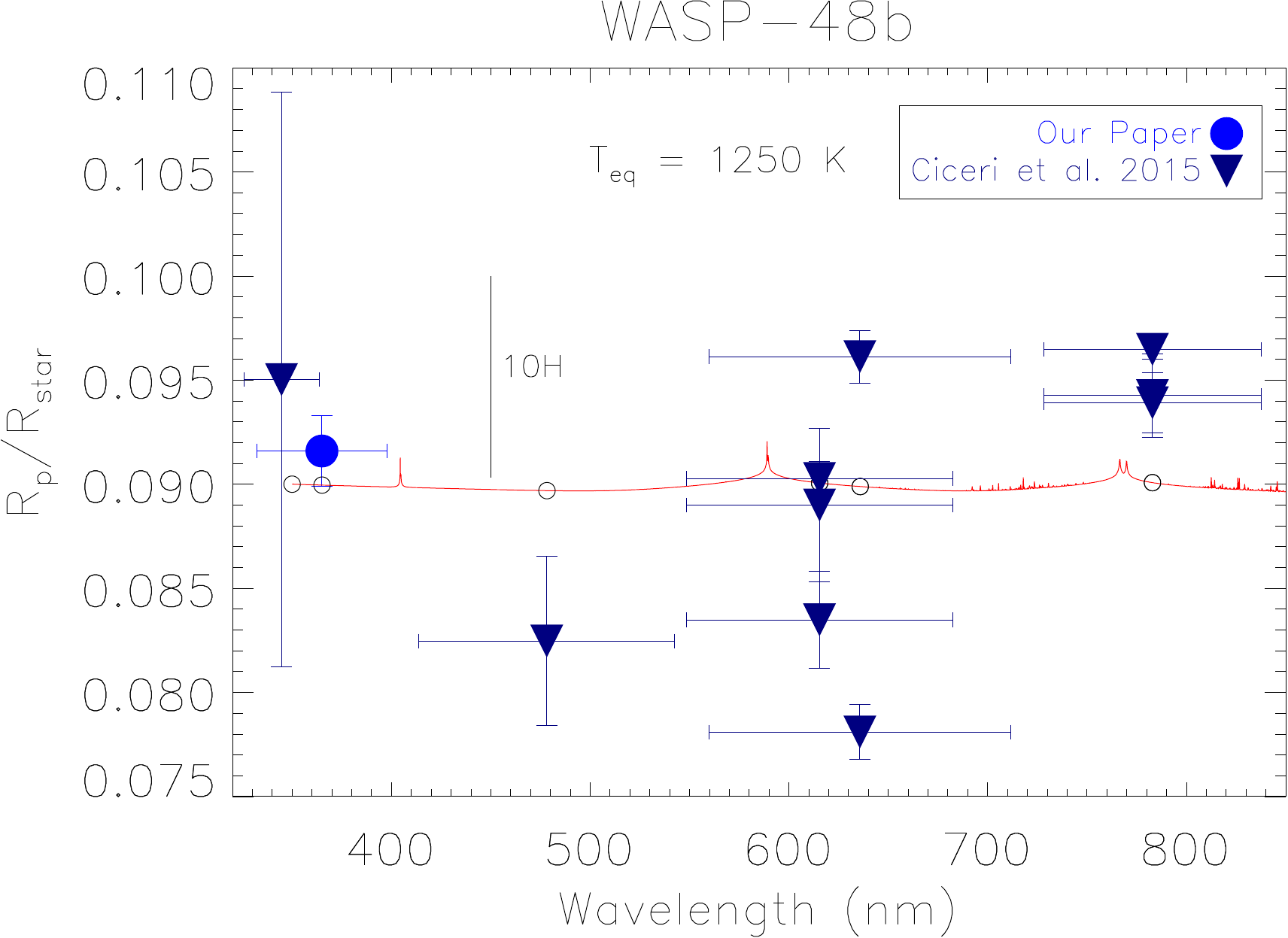,width=0.52\textwidth} 	
\caption{Variation of R$_p$/R$_\ast$ vs. wavelength for WASP-48b. Other comments are the same as Fig. \ref{fig:rp}. }
\label{fig:rp3}
\end{figure*}

For illustration, we compare the observed R$_p$/R$_\ast$ differences with wavelength for each target (Table \ref{tb:rp_rstar}) to theoretical predictions by \citet{Fortney2010} for a model planetary atmosphere (Figure \ref{fig:rp}--\ref{fig:rp3}). The models used were estimated for planets with a 1 $M_{Jup}$, $g_{p} = 25 m/s$, base radius of 1.25 $R_{Jup}$ at 10 bar, solar metallicity, and $T_{eq}$ closest to the measured value for each exoplanet (with model choices of 500, 750, 1000, 1250, 1500, 1750, 2000, 2500 K). Additionally, TiO  and VO  opacity were not included in the synthetic model. A vertical offset was applied to the model to provide the best fit to the spectral changes. This comparison is useful as it is illustrative of the size of variation of the observations compared to what the models predict. However, in-depth radiative transfer models calculated for all the exoplanets are still needed to fully understand their transmission spectra.

Next, we apply the MassSpec concept (\citealt{Etangs2008}, \citealt{deWit2013}) to the spectral slope to determine if the observed radius variations are consistent with Rayleigh scattering. This approximation assumes a well-mixed, isothermal atmosphere in chemical equilibrium, and an effective atmospheric opacity source with an extinction cross section which follows a power-law index, $\alpha$, $\sigma = \sigma_{0} (\lambda/\lambda_{0})^{\alpha}$. The slope of the spectrum is related to $\alpha$ by using the scale height (\citealt{Etangs2008}) 
\begin{equation}
	\alpha H = \frac{d R_{b}(\lambda)}{d  \ln \lambda},	
\end{equation}
where $\lambda$ is the wavelength. An  $\alpha =-4$ would be consistent with Rayleigh scattering \citep{Etangs2008}. In order to calculate $\alpha$, we use our near-UV R$_p$/R$_\ast$ values and the literature values of the nearest wavelength (Table \ref{tb:rp_rstar}). In some cases, the nearest literature wavelength are not in the blue part of the spectra. This lack of measurment can cause a problem in the interpretation of $\alpha$ because the U and B bands are the only bands where strong spectral features are not present (\citealt{Tinetti2010}; \citealt{Benneke2012}; \citealt{Benneke2013}; \citealt{Griffith2014}). The calculation of $\alpha$ also assumes that only a single species is dominant in the the atmosphere, an assumption that may not always hold. We find an $\alpha$ of -17.3$\pm$7.9, -19.1$\pm$2.4 , and -5.9$\pm$0.9 for CoRoT-1b, TrES-4b, and WASP-12b, respectively. The spectral index calculated for WASP-12b (see also \citealt{Sing2013}) and CoRoT-1b are within 2$\sigma$ of Rayleigh scattering. Follow-up observations are needed to confirm this result. It is possible that the index of TrES-4b is caused by scattering from aerosols \citep{Sing2013} but this suggestion needs to be explored in greater detail. Additionally, an $\alpha$ of +11.6$\pm$1.1 and +34.8$\pm$2.7 are found for WASP-1b and WASP-36b, respectively. This is the first time a positive $\alpha$ has been estimated for any exoplanet and more theoretical modeling is needed to identify possible opacity sources.

\subsubsection{Variability due to the host stars} \label{sec:bow_var}
Our interpretation that the observed wavelength variations are due to the planetary atmosphere assumes that the host stars do not vary significantly due to stellar activity. The presence of stellar activity and star spots on the stellar surface can produce variations in the observed transit depth when measured at different times (e.g. \citealt{Czesla2009}; \citealt{Oshagh2013}; \citealt{Oshagh2014}; \citealt{Zellem2015}). This effect is particularly stronger in the near-UV than in the optical and can mimic a Rayleigh scattering signature (e.g. \citealt{Oshagh2014}; \citealt{McCullough2014}). As described in section \ref{sec:var_bow}, for the planets with measured $R^{'}_{HK}$ indices only one (HAT-P-16b) in our sample is known to orbit an active star (Table \ref{tb:compare}). Additionally, no obvious star spot crossing is seen in our data (Figs. \ref{fig:light_1}-\ref{fig:light_5}). The WASP-1b and WASP-36b near-UV and optical observations were done at the same time, thus the influence of stellar activity on the smaller near-UV transit depth result should be minimal.

Next, we estimate how much the transit depth changes due to unocculted spots using the formalization presented by \citet{Sing2011}. The three main assumptions of this method are that the emission spectrum of the spots are treated as a stellar spectrum but with a lower effective temperature, the surface brightness outside the spots does not change, and no facule are present. These assumptions lead to an overall dimming of the star and increase in the transit depth. \citet{Sing2011} find that the change in transit depth due to unocculted spots, $\Delta(R_{p}/R_\ast)$, is
\begin{equation}
\Delta \left(\frac{R_{p}}{R_\ast}\right) =\frac{1}{2}\frac{\Delta d}{d} \frac{R_{p}}{R_\ast},
\end{equation}
where
\begin{equation}
    \frac{\Delta d}{d} = \Delta f(\lambda_{0},t) \left(1 - \frac{ F_{\lambda}^{T_{spot} }}{  F_{\lambda}^{T_{star}}}\right)\Big/\left(1 - \frac{ F_{\lambda_{0}}^{T_{spot} }}{  F_{\lambda_{0}}^{T_{star}} }\right) ,
\end{equation}
$\Delta f(\lambda_{0},t)$ is the total dimming at the reference wavelength ($\lambda_{0}$) over some time scale ($t$), and $F_{\lambda}^{T}$ is the surface brightness of the stellar models at the temperature of the star ($T_{star}$) and the spot ($T_{spot}$). An exact value for $\Delta(R_{p}/R_\ast)$ is beyond that scope of this paper since the $\Delta f(\lambda_{0},t)$ and $T_{spot}$ are unknown for all targets. \citet{Sing2011} find for HD 189733b that $\Delta \left(R_{p}/R_\ast\right) = 2.08\times10^{-3}/2 \left(R_{p}/R_\ast\right)$ between 375--400 $\rm{nm}$ assuming $T_{spot} = 4250 \rm{K}$, $T_{star} = 5000 \rm{K}$, $\Delta f(\lambda_{0})$ = 1$\%$, and $\lambda_{0} = 400 \rm{nm}$. Therefore, unocculted spots have minimal influence (assuming the stars we are observing have spots similar to HD 189733b) on the observed transit depth variations since the influence of these spots are about 10 times smaller (e.g. $\Delta \left[R_{p}/R_\ast\right]$ = 0.00014 for WASP-36b) than our final error bars (Tables \ref{tb:parms_1}-\ref{tb:parms_2}). This result is also consistent with the recent study by \citet{Llama2015} that find that stellar activity similar to that of the sun has minimal effect on the transit depth in the wavelengths explored in our study. Nonetheless, follow-up observations and host star monitoring are encouraged to monitor the effect of stellar activity on the transit depth variations we observe.

\begin{table*}
\centering
\caption{$R_{p}/R_{\ast}$ and central wavelength $\lambda_{eff}$ from this paper and previous literature for all targets$^{1}$ }
\setlength\tabcolsep{4pt}
\centering
\begin{tabular}[t]{ccccc}
\hline
\hline
Planet & Source & Filter & Wavelength (nm) & Rp/$R_{*}$ \\
\hline
CoRoT-1b &This Paper & Bessell U         & 370 & 0.1439$^{+0.0020}_{-0.0018}$\\
CoRoT-1b & \citealt{Gillon2009} & R SPECIAL        & 655 & 0.1381 $\pm$ 0.0007 \\
CoRoT-1b & \citealt{Schlawin2014}             & IRTF &  860		&0.1470 $\pm$ 0.0020\\
\hline
\end{tabular}
\vspace{-2em}
\tablenotetext{1}{This table is available in its entirety in machine-readable form in the online journal.  A portion is shown here for guidance regarding its form and content. }	
\label{tb:mr}
\label{tb:rp_rstar}
\end{table*}

\section{Conclusions}
We investigate the primary transits of the 15 exoplanets (CoRoT-1b, GJ436b, HAT-P-1b, HAT-P-13b, HAT-P-16, HAT-P-22b, TrES-2b, TrES-4b, WASP-1b, WASP-12b, WASP-33b, WASP-36b, WASP-44b, WASP-48b, WASP-77Ab) using ground-based near-UV and optical filters to study their atmospheres (Section \ref{sec:atmo}; Figure \ref{fig:rp}--\ref{fig:rp3}; Table \ref{tb:rp_rstar}). A constant R$_p$/R$_\ast$ from near-UV to optical wavelengths is found for 10 targets (GJ436b, HAT-P-1b, HAT-P-13b, HAT-P-16b,  HAT-P-22b, TrES-2b, WASP-33b, WASP-44b, WASP-48b, WASP-77Ab), suggestive of clouds in their atmospheres. Additionally, the near-UV R$_p$/R$_\ast$ of 3 targets (CoRoT-1b, TrES-4b, WASP-12b) are larger and 2 targets (WASP-1b, WASP-36b) are smaller by at least 2$\sigma$ from their optical R$_p$/R$_\ast$. The atmospheric implications of the transit depth variations are explored (Section \ref{sec:atmo}) and we find that the spectral slope of WASP-12b and CoRoT-1b are consistent with Rayleigh scattering. To our knowledge this is the first time a hot Jupiter has been observed to have a smaller near-UV transit depth than optical and a possible opacity source that can cause such a radius variation is currently unknown. The WASP-1b and WASP-36b near-UV and optical observations were done at the same time, thus limiting the influence of stellar activity on the transit depth variations. Follow-up observations are encouraged to confirm all our results but especially the observation of a smaller near-UV transit depth.  

Additionally, we do not detect any near-UV light curve asymmetries in all of the 15 targets within the precision (~1.23~--~6.22~mmag) and timing resolution (27 -- 137 s) of our observations (Table \ref{tb:obs_new}; Section \ref{sec:nobowshock}). All the non-detections in this study confirm and expand upon the theoretical modeling done by \citet{BenJaffel2014} and \citealt{Turner2016} that near-UV asymmetries cannot be seen from the ground. These findings are consistent with the previous ground-based non-detection of asymmetries in HAT-P-16b (\citealt{Pearson2014}) and WASP-12b (\citealt{Copperwheat2013}) and 4 (HAT-P-5b, TrES-3b, WASP-17b, XO-2b) other exoplanets (\citealt{Southworth2012b}; \citealt{Turner2013a}; \citealt{Bento2014}; \citealt{Zellem2015}). 

Finally, for each target we derive a new set of planetary system parameters and the orbital period and ephemeris are updated to help with follow-up observations (Tables \ref{tb:parms_1}--\ref{tb:parms_2}). Our data includes the first published ground-based near-UV light curves of 12 of the targets (CoRoT-1b, GJ436b, HAT-P-1b, HAT-P-13b, HAT-P-22b, TrES-2b, TrES-4b, WASP-1b, WASP-33b, WASP-36b, WASP-48b, WASP-77Ab) and greatly expands the number of near-UV light curves in the literature. 

\section*{Acknowledgments}

J. Turner, K. Pearson, R. Zellem, J. Teske, and C. Griffith were partially supported by the NASA's Planetary Atmospheres program. J. Turner was also partially funded by the Virginia Space Grant Consortium Graduate Research Fellowship Program and by the National Science Foundation Graduate Research Fellowship under Grant No. DGE-1315231.

We sincerely thank the University of Arizona Astronomy Club, the Steward Observatory TAC, the Steward Observatory telescope day crew, John Bieging, Elizabeth Green, Don McCarthy, Maria Schuchardt, the Lunar and Planetary Laboratory, and the Associated Students of the University of Arizona for supporting this research. 

We also thank John Southworth, Jason Eastman, Ian Crossfield, Josh Carter, and John Johnson for their helpful comments on \texttt{JKTEBOP}, \texttt{EXOFAST}, \texttt{TAP}, and \texttt{EXOMOP}. We also thank Joe Llama for the discussion on his bow shock models and Aline Vidotto, Moira Jardine, and Christiane Helling on the useful discussion of their UV bow shock predictions. Finally, we would like to thank Michael Cushing, Jon Bjorkman, Robert Johnson, Carl Schmidt, Phil Arras, and Andrew Collier Cameron for their insightful comments on this research.

This research has made use of the Exoplanet Orbit Database \citep{Wright2011exo}, Exoplanet Data Explorer at exoplanets.org, Exoplanet Transit Database, Extrasolar Planet Transit Finder, NASA's Astrophysics Data System Bibliographic Services, and the International Variable Star Index (VSX) database, operated at AAVSO, Cambridge, Massachusetts, USA.

We would also like to thank the anonymous referee
for their insightful comments during the publication process. This
manuscript is much improved thanks to their comments.


\bibliographystyle{mn2e}
\bibliography{reference_new.bib}

\label{lastpage}

\end{document}